\documentclass[sigconf,authorversion]{acmart}
\usepackage{enumitem}
\usepackage[disable]{todonotes}

\AtBeginDocument{%
  \providecommand\BibTeX{{%
    \normalfont B\kern-0.5em{\scshape i\kern-0.25em b}\kern-0.8em\TeX}}}

\copyrightyear{2024}
\acmYear{2024}
\setcopyright{rightsretained}
\acmConference[ITiCSE-WGR 2024]{Proceedings of the 2024 Working Group Reports on Innovation and Technology in Computer Science Education}{July 8--10, 2024}{Milan, Italy}
\acmBooktitle{Proceedings of the 2024 Working Group Reports on Innovation and Technology in Computer Science Education (ITiCSE-WGR 2024), July 8--10, 2024, Milan, Italy}\acmDOI{10.1145/XXXXXXX}

\usepackage{color}
\usepackage{todonotes}
\usepackage{graphicx}
\usepackage{relsize}
\usepackage{dirtytalk}
\usepackage{enumitem}
\usepackage{multicol}
\usepackage{todonotes}
\usepackage{rotating}
\usepackage{subcaption}
\usepackage{multirow}
\usepackage{svg}
\usepackage{natbib}
\usepackage{colortbl}
\usepackage{wasysym}
\usepackage{balance}
\RequirePackage[skins]{tcolorbox}
\newtcolorbox{custombox}[1]{
	colback=gray!10,
	colframe=gray!70,
	left=1mm,
	right=1mm,
	top=1mm,
	bottom=1mm,
	fonttitle=\bfseries,
	arc=0mm,
	leftrule=1mm,
	rightrule=0mm,
	toprule=0mm,
	bottomrule=0mm,
	notitle,
	before=\par\smallskip\noindent,
	before upper={\textbf{#1: } },
}
\usepackage{caption}
\usepackage{array}

\usepackage{tikz}
\usetikzlibrary{shapes.geometric, arrows}
\tikzstyle{process} = [rectangle, rounded corners, minimum width=3.5cm, minimum height=1.5cm, text centered, draw=black, fill=gray!20]
\tikzstyle{arrow} = [thick,->,>=stealth]

\usepackage{tabularx}
\usepackage{booktabs}
\usepackage{makecell}

\usepackage[tableposition=top,labelfont=bf]{caption}
\newcolumntype{L}[1]{>{\raggedright\arraybackslash\hspace{0pt}}p{#1}}
\newcolumntype{C}[1]{>{\centering\arraybackslash\hspace{0pt}}p{#1}}
\newcolumntype{R}[1]{>{\raggedleft\arraybackslash\hspace{0pt}}p{#1}}

\usepackage{comment}
\usepackage[export]{adjustbox}
\usepackage{multicol}

\begin{document}


\title[Beyond the Hype]{Beyond the Hype: A Comprehensive Review of Current Trends in Generative AI Research, Teaching Practices, and Tools}




%
%

\author[Prather]{James Prather}
\authornote{Co-leader}
\orcid{0000-0003-2807-6042}
\affiliation{
  \institution{Abilene Christian University}
  \city{Abilene}
  \state{TX}
  \country{USA}
}
\email{james.prather@acu.edu}

\author[Leinonen]{Juho Leinonen}
\authornotemark[1]
\orcid{0000-0001-6829-9449}
\affiliation{
  \institution{Aalto University}
  \city{Espoo}
  \country{Finland}
}
\email{juho.2.leinonen@aalto.fi}

\author{Natalie Kiesler}
\authornotemark[1]
\orcid{0000-0002-6843-2729}
\affiliation{%
   \institution{Nuremberg Tech}
   \city{Nuremberg}
   \country{Germany}
}
\email{natalie.kiesler@th-nuernberg.de}

%
%

\author{Jamie Gorson Benario}
\orcid{0000-0002-0385-4357}
\affiliation{%
   \institution{Google}
   \city{Chicago}
   \state{IL}
   \country{USA}
}
\email{jamben@google.com}

\author{Sam Lau}
\orcid{0000-0002-3160-0151}
\affiliation{%
   \institution{UC San Diego}
   \city{La Jolla}
   \state{CA}
   \country{USA}
}
\email{lau@ucsd.edu}

\author{Stephen MacNeil}
\orcid{0000-0003-2781-6619}
\affiliation{%
   \institution{Temple University}
   \city{Philadelphia}
   \state{PA}
   \country{USA}
}
\email{stephen.macneil@temple.edu}

\author{Narges Norouzi}
\orcid{0000-0001-9861-7540}
\affiliation{%
   \institution{University of California Berkeley}
   \city{Berkeley}
   \state{CA}
   \country{USA}
}
\email{norouzi@berkeley.edu}

\author{Simone Opel}
\orcid{0000-0002-9697-9887}
\affiliation{%
   \institution{FernUniversität in Hagen}
   \city{Hagen}
   \country{Germany}
}
\email{simone.opel@fernuni-hagen.de}

\author{Vee Pettit}
\orcid{0009-0003-9021-3615}
\affiliation{%
   \institution{Virginia Tech}
   \city{Blacksburg}
   \state{VA}
   \country{USA}
}
\email{vpettit@vt.edu}

\author{Leo Porter}
\orcid{0000-0003-1435-8401}
\affiliation{%
   \institution{UC San Diego}
   \city{La Jolla}
    \state{CA}
   \country{USA}
}
\email{leporter@ucsd.edu}

\author{Brent N. Reeves}
\orcid{0000-0001-5781-1136}
\affiliation{%
   \institution{Abilene Christian University}
   \city{Abilene}
   \state{TX}
   \country{USA}
}
\email{brent.reeves@acu.edu}

\author{Jaromir Savelka}
\orcid{0000-0002-3674-5456}
\affiliation{%
   \institution{Carnegie Mellon University}
   \city{Pittsburgh}
   \state{PA}
   \country{USA}
}
\email{jsavelka@cs.cmu.edu}

\author{David H. Smith IV}
\orcid{0000-0002-6572-4347}
\affiliation{%
   \institution{University of Illinois}
   \city{Urbana}
   \state{IL}
   \country{USA}
}
\email{dhsmith2@illinois.edu}

\author{Sven Strickroth}
\orcid{0000-0002-9647-300X}
\affiliation{%
	\institution{LMU Munich}
	\department{}
	\city{Munich}
	\country{Germany}
}
\email{sven.strickroth@ifi.lmu.de}

\author[Zingaro]{Daniel Zingaro}
\orcid{0000-0002-1568-4826}
\affiliation{
  \institution{University of Toronto Mississauga}
  \city{Mississauga}
  \state{ON}
  \country{Canada}
}
\email{daniel.zingaro@utoronto.ca}

\renewcommand{\shortauthors}{James Prather et al.}


\begin{abstract}

Generative AI (GenAI) is advancing rapidly, and the literature in computing education is expanding almost as quickly. Initial responses to GenAI tools were mixed between panic and utopian optimism. Many were fast to point out the opportunities and challenges of GenAI. Researchers reported that these new tools are capable of solving most introductory programming tasks and are causing disruptions throughout the curriculum. These tools can write and explain code, enhance error messages, create resources for instructors, and even provide feedback and help for students like a traditional teaching assistant.  
In 2024, new research started to emerge on the effects of GenAI usage in the computing classroom. These new data involve the use of GenAI to support classroom instruction at scale and to teach students how to code with GenAI. In support of the former, a new class of tools is emerging that can provide personalized feedback to students on their programming assignments or teach both programming and prompting skills at the same time. With the literature expanding so rapidly, this report aims to summarize and explain what is happening on the ground in computing classrooms. We provide a systematic literature review; a survey of educators and industry professionals; and interviews with educators using GenAI in their courses, educators studying GenAI, and researchers who create GenAI tools to support computing education. The triangulation of these methods and data sources expands the understanding of GenAI usage and perceptions at this critical moment for our community.

\end{abstract}







\maketitle

%
%
%
%

\section{Introduction}
\label{sec:intro}

Computing education is undergoing a seismic shift due to the advances in generative AI (GenAI) ~\cite{denny2024computing,prather2023wgfullreport,cambaz2024use}. 
Beginning in early 2022, computing education researchers showed that these models had incredible accuracy solving programming problems and exam questions in multiple courses and contexts~\cite{finnieansley2022robots,finnie-ansley2023my,savelka2023thrilled,kiesler2023large,savelka2023cangenerative,reeves2023evaluating,hou2024more,ameryahia2023fromlarge}. 
Other early work focused on the ways in which GenAI can provide support to computing educators ~\cite{sarsa2022automatic,denny2022robosourcing,macneil2023experiences,leinonen2023comparing,leinonen2023using, becker2023generative,savelka2023efficient}. 
Others were quick to raise concerns about potential threats to education, such as over-reliance, bias in the models, and educational misconduct~\cite{becker2023programming,prather2023weird,prather2023wgfullreport,lau2023from}. Computing instructors at the K-12 level are also struggling with integrating GenAI into their curricula~\cite{grover2024enduring,tran2023prompt,barendsen2024wgabstract}. K-12 teachers outside of computing education are having similar conversations to those occurring in higher education~\cite{philbin2023impact,belghith2024testing,ruiz2024using,rahman2023chatgpt,murgia2023children,barendsen2024wgabstract}. Public perception of GenAI is mixed, partially because it is a ``black box'' and that lack of transparency often increases fear~\cite{brauner2023does}, which is one reason why GenAI should be designed to increase transparency to end users~\cite{tankelevitch2024metacognitive}.

However, the discussion has largely moved from threats, challenges, and opportunities~\cite{becker2023programming,kenthapadi2023generative,tolk2023chances,oestreicher2023new,silva2024chatgpt} to questions of practical adoption~\cite{macneil2024discussing,prather2024wgabstract}. The 2023 ITiCSE working group on GenAI summarized the activity within the computing education community and suggested that the next step is to determine reliable and safe ways to implement it into computing curricula \cite{prather2023wgfullreport}. An essay released along with the 2023 ACM/IEEE Computing Curricula suggested ways in which this could be accomplished based on preliminary data from a few studies and painted an optimistic picture of the future \cite{becker2023generative}. Indeed, many within the community are calling AI integration just another step in the advancement of educational technology~\cite{adair2023teaching}. Others are discussing how GenAI will change programming competencies and skills in the future~\cite{kiesler2023beyond,sheard2024instructor}. New assessment methods~\cite{kendon2023not-harmful} and ways to measure user interaction with GenAI~\cite{mozannar2024reading} are required, and although some are arguing for making assignments ``LLM-proof'' \cite{bopp2024case}, which seems to be an impractical goal given the rapid improvement in these models. 

Although some have advocated for banning GenAI entirely, that also seems impractical given its free availability to students outside the classroom~\cite{lau2023from}. Furthermore, professional developers are also discussing the role that GenAI will play within their work~\cite{guo2023six,tanimoto2023fivefutures,anewalt2023industry,weisz2022better,kuhail2024will,liang2024largescale}, lending credibility to the idea that students must be prepared for using it after university. Therefore, thoughtful integration and scaffolding appears to be the way forward~\cite{denny2024computing}. However, there is a lack of helpful and clear terminology to discuss the kinds of classroom interventions conducted so far. To this end, our report attempts to distinguish the following use cases:
(1) instructors teaching students about using GenAI tools in order to write code, and (2) instructors using GenAI tools to support the teaching of their course via help-seeking bots, code feedback, assignment creation, etc.

The first category is by far the most extensive. Researchers were quick to show that GenAI could automatically enhance programming error messages~\cite{leinonen2023using}, which was followed with two large-scale replications showing a direct benefit to students~\cite{wang2024large,taylor2024dcc}. Indeed, GenAI can provide other kinds of advanced and customized help and qualitative feedback to students working on computing assignments~\cite{macneil2023experiences,leinonen2023comparing,kiesler2023exploring,Azaiz2023feedback,azaiz2024feedback,lohr2024youre}. Other more recent work, such as the Harvard CS50 course, has focused on providing programming tutoring and help to students at scale through TA chatbots~\cite{liu2024teaching}. Researchers are only beginning to define how these should be designed, implemented, and evaluated~\cite{denny2024desirable}. The same applies to understanding how students actually use GenAI in authentic course settings, for example, in introductory programming courses~\cite{scholl2024analyzing,scholl2024noviceprogrammersuseexperience,kiesler2024novice}, and advanced computing courses~\cite{grande2024studentperspectives}.

Instructors can also use GenAI to create customized and unique assignments tailored to student interests~\cite{sarsa2022automatic,logacheva2024evaluating} as well as generate other educational content~\cite{doughty2024comparative,tran2023generating,sridhar2023harnessing}. 
Yet, this second category of GenAI usage in instruction is the least studied to date, possibly because many instructors have not yet thoughtfully integrated GenAI directly into their programming instruction~\cite{fernandez2024cs1-ai,walter2024embracing,haikal2024enhancing}. However, studies thus far are showing mixed results. Some are pointing to its introduction as a way to equip students to move faster through the curriculum than was ever possible before~\cite{vadaparty2024CS1-LLM}. Others are claiming that using GenAI has no negative effects on student learning outcomes~\cite{xue2024does}. However, some early work when GitHub Copilot was first released showed that students would flail and wander during programming tasks due to that tool's constant interruptions~\cite{prather2023weird}. Other more recent work has now shown that GenAI can significantly undercut student critical thinking during code writing and debugging tasks for those students who over-rely on it, decreasing their grades overall~\cite{jovst2024impact}. 
Similarly, in an observational study, researchers found that although a larger percentage of students are able to complete programming tasks with the aid of GenAI tools, students faced new metacognitive challenges~\cite{prather2024widening}.

As evidenced by the discussion above and our systematic literature review below, the literature on GenAI in computing education is expanding rapidly. With so much happening so quickly, it is difficult to know what has been done, why it is being done, what works, and where this is all headed. We therefore attempt to capture the zeitgeist of the present moment in computing history by defining terms, ordering and summarizing all of this for the reader.

\subsection{Goals}
With all of the above in mind, this report addresses the following overarching research goals:
\begin{enumerate}
\item[\textbf{(1a)}] How are instructors incorporating GenAI into teaching computing? 
\item[\textbf{(1b)}] And why are they making these choices?
\item[\textbf{(2a)}] How have the expectations towards skills in software development changed with the advent of GenAI?
\item[\textbf{(2b)}] Which computing competencies are required in the future, according to teachers and industry professionals?
\end{enumerate}

We address these overarching goals via several fine-grained research questions, and various methods: a systematic literature review, a survey study with educators and software developers in industry, and qualitative interviews with computing educators, computing researchers, and GenAI tool creators.

\subsection{Contributions} 
This working group report describes how and why computing instructors have chosen to integrate (or not integrate) GenAI into their courses, and what they expect with regard to future developments in curricula and industry usage of GenAI tools. We identify the current state-of-the-art by presenting the following deliverables: 

\begin{enumerate}[leftmargin=*,align=left]
    \item \textbf{Systematic Literature Review (\autoref{sec:litreview} and \ref{sec:litreview-results}):} 
    We review the existing literature on GenAI tools in computing education (through May 23, 2024) and present the studies in which educators report on (a) evidence of GenAI in computing education research, (b) educators using GenAI tools in their teaching practices, and (c) the rationale of educators to incorporate GenAI tools in a certain way.
    
    \item \textbf{Evaluating Instructional Practices to Teach Students How to Use GenAI Tools (\autoref{sec:educator-and-developer-views} and \ref{sec:educator-and-developer-views_results}):} We gather current integration practices through an international survey of computing instructors. The survey also focused on the tools, policies, motivational aspects, and the impact of GenAI on the competencies students require to succeed -- from an educator's perspective. 

    \item \textbf{Evaluating Instructors' Use of GenAI Tools (\autoref{sec:educator-and-developer-views} and \ref{sec:educator-and-developer-views_results}):} The survey of educators also revealed the use of GenAI tools by these educators, for example, to create tools that would support students.

    \item \textbf{Evaluating Instructors' Perspectives on Learning Outcomes and Future Developments (\autoref{sec:educator-and-developer-views} and \ref{sec:educator-and-developer-views_results}):} To capture the impact of GenAI tools on actual student outcomes, we conducted semi-structured interviews with instructors. The results reveal the disruptive character of GenAI tools in terms of the potential benefits and limitations of GenAI in computing education.
    
    \item \textbf{Exploring the Industry's Experience (\autoref{sec:educator-and-developer-views} and \ref{sec:educator-and-developer-views_results}):} 
    Another contribution of this report is the integration of the industry perspective, their experiences and usage patterns of GenAI tools. This encompasses policies, motivations, and expectations regarding the competencies future developers will need. 
\end{enumerate}

\subsection{Structure of the Report}

The structure of this working group report is as follows. To address related work, the authors present the systematic search and review of prior studies on how and why computing educators have integrated GenAI tools into their teaching practices. The methods of the literature review are presented in \autoref{sec:litreview} and the respective results of the systematic literature review are presented in \autoref{sec:litreview-results}.

The second major component of this working group report captures the perspective of both educators and software developers. We do so by presenting our methodology (\autoref{sec:educator-and-developer-views}), consisting of an international survey of computing educators, a series of qualitative interviews with computing educators, and an aligned survey of software developers to gather current industry practices and perspectives. As a part of the methodology, we briefly introduce prior work, fine-grained research questions, the process of developing the survey and interview questions, and the data analysis approaches we applied.

We present the results of our mixed-methods approach (i.\,e., surveys with educators and developers, and interviews with educators) in \autoref{sec:educator-and-developer-views_results} by triangulating the different data sources for every research question. In \autoref{sec:discussion}, we discuss the most interesting results, and how they build and extend prior work. 

In \autoref{sec:threats}, we summarize the threats to validity of the applied methodology, before concluding our work in \autoref{sec:conclusions}, and presenting pathways for future work (\autoref{sec:futurework}).

%
%
%
%
\section{Systematic Literature Review: Methods}
\label{sec:litreview}

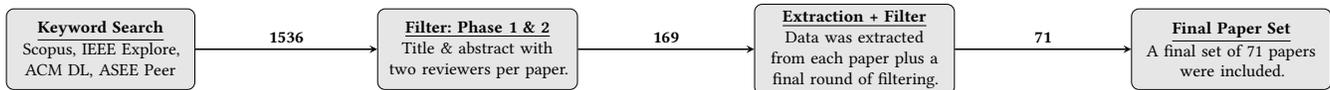
\begin{figure*}
    \centering
    \resizebox{\textwidth}{!}{
    \begin{tikzpicture}[node distance=3.5cm]

\node (search) [process, text width=3cm] {\underline{\textbf{Keyword Search}}\\Scopus, IEEE Explore,\\ACM DL, ASEE Peer};
\node (filter) [process, right of=search, xshift=3.5cm, text width=3.5cm] {\underline{\textbf{Filter: Phase 1 \& 2}}\\Title \& abstract with two reviewers per paper.};
\node (extract) [process, right of=filter, xshift=3.5cm, text width=3.5cm] {\underline{\textbf{Extraction + Filter}}\\Data was extracted from each paper plus a final round of filtering.};
\node (final) [process, right of=extract, xshift=3.5cm, text width=3.5cm] {\underline{\textbf{Final Paper Set}}\\A final set of 71 papers\\were included.};

\draw [arrow] (search) -- node[anchor=south] {\textbf{1536}} (filter);
\draw [arrow] (filter) -- node[anchor=south] {\textbf{169}} (extract);
\draw [arrow] (extract) -- node[anchor=south] {\textbf{71}} (final);

\end{tikzpicture}
    }
    \caption{The literature review collection and analysis pipeline.}
    \label{fig:lit-review-pipeline}
\end{figure*}

\todo[inline]{Add example papers found in the lit review to claims and results}

The systematic literature review (SLR) aims to identify how instructors are integrating generative AI into computing classrooms. The goal is to extend the many prior interview and survey studies that have looked at students' perceptions of how generative AI tools are used~\cite{lau2023from, zastudil2023generative, hou2024effects, amoozadeh2024trust} to focus instead on what is actually being done within classroom settings. Therefore the goal is to focus on pedagogies, tools, and classroom interventions that feature empirical data about students' experiences with those classroom interventions. Specifically, we investigate the following three research questions: 

\begin{itemize}[leftmargin=1em]
    \item[] \textbf{SLR-RQ1}. How can the reported evidence of Generative AI in CER be summarized?
    \item[] \textbf{SLR-RQ2}. How is generative AI being incorporated into teaching?
    \item[] \textbf{SLR-RQ3}. What are the motivations behind incorporating GenAI tools into teaching?
\end{itemize}

SLR-RQ1 and SLR-RQ2 relate to the overarching research goal (1a), while SLR-RQ3 relates to the overarching research goal (1b).

\autoref{fig:lit-review-pipeline} illustrates the overall process of the whole systematic literature review. We follow the literature review best practices by Kitchenham and Brereton~\cite{kitchenham2013systematic}. We first searched through various databases using a search string. Then, we checked for the inclusion of reference papers that should be found to confirm the quality of the search string. This was followed up by doing a title/\allowbreak{}abstract scan for relevance. After the title/\allowbreak{}abstract scan, we read the full papers and started extracting relevant information from them. In this stage, some papers were still rejected if they did not pass the inclusion/exclusion criteria. In the end, we had a set of 71 papers that passed the criteria and for which we extracted data. The included papers are listed in Table~\ref{tab:lit-included}.

\begin{table*}

\scriptsize

	\begin{center}
 	\caption{Papers included in the literature review (listed alphabetically by author, grouped by publication year). \\ Venue shows the status at the time when the article was included.  
  }
	\label{tab:lit-included}
		\begin{tabular}{p{2cm}p{12cm}p{2cm}cc}
			\toprule
			\textbf{AUTHOR} & \textbf{TITLE} & \textbf{VENUE} & \textbf{YEAR} & 
            \textbf{CITE} \\
   \toprule

\setlength{\arrayrulewidth}{0.4pt}

Jonsson and Tholander &	Cracking the code: Co-coding with AI in creative programming education &	C\&C	& 2022 & \cite{10.1145/3527927.3532801} \\

\specialrule{.2em}{.1em}{.1em} 

Crandall et al. &	Generative Pre-Trained Transformer (GPT) Models as a Code Review Feedback Tool in Computer Science Programs &	JCSC	& 2023 & \cite{crandall2023generative} \\
Denny et al. &	Can We Trust AI-Generated Educational Content? Comparative Analysis of Human and AI-Generated Learning Resources &	arXiv	& 2023 & \cite{denny2023trustaigeneratededucationalcontent} \\
Denny et al. &	Promptly: Using Prompt Problems to Teach Learners How to Effectively Utilize AI Code Generator &	arXiv	& 2023 & \cite{denny2023promptly} \\
Dos Santos and Cury &	Challenging the Confirmation Bias: Using ChatGPT as a Virtual Peer for Peer Instruction in Computer Programming Education &	FIE	& 2023 & \cite{Santos2023Challenging} \\
Hajj and Sah &	Assessing the Impact of ChatGPT in a PHP Programming Course &	ISAS	& 2023 & \cite{10391549} \\
Hu et al. &	Explicitly Introducing ChatGPT into First-year Programming Practice: Challenges and Impact &	TALE	& 2023 & \cite{hu2023explicitly} \\
Karnalim et al. &	Plagiarism and AI Assistance Misuse in Web Programming: Unfair Benefits and Characteristics &	TALE	& 2023 & \cite{karnalim2023plagiarism} \\
Kazemitabaar et al. &	Studying the effect of AI Code Generators on Supporting Novice Learners in Introductory Programming &	CHI	& 2023 & \cite{kazemitabaar2023studying} \\
Kumar et al. &	Bridging the Gap in AI-Driven Workflows: The Case for Domain-Specific Generative Bots &	BigData	& 2023 & \cite{10386894} \\
Kuramitsu et al. &	KOGI: A Seamless Integration of ChatGPT into Jupyter Environments for Programming Education &	SPLASH-E	& 2023 & \cite{kuramitsu2023kogi} \\
Liffiton et al. &	CodeHelp: Using Large Language Models with Guardrails for Scalable Support in Programming Classes &	Koli Calling	& 2023 & \cite{liffiton2023codehelp} \\
MacNeil et al. &	Experiences from Using Code Explanations Generated by Large Language Models in a Web Software Development E-Book &	SIGCSE TS	& 2023 & \cite{macneil2023experiences} \\
Markel et al. &	GPTeach: Interactive TA Training with GPT-based Students &	L@S	& 2023 & \cite{markel2023gpteach} \\
Perry et al. &	Do Users Write More Insecure Code with AI Assistants? &	CCS	& 2023 & \cite{perry2023do} \\
Prasad et al. &	Generating Programs Trivially: Student Use of Large Language Models &	CompEd	& 2023 & \cite{prasad2023generating} \\
Prather et al. &	``It's Weird That it Knows What I Want'': Usability and Interactions with Copilot for Novice Programmers &	TOCHI	& 2023 & \cite{prather2023weird} \\
Qureshi &	ChatGPT in Computer Science Curriculum Assessment: An analysis of Its Successes and Shortcomings &	ICSLT	& 2023 & \cite{Qureshi2023ChatGPTAssessment} \\
Sandoval et al. &	Lost at C: A User Study on the Security Implications of Large Language Model Code Assistants &	USENIX Security	& 2023 & \cite{sandoval2023lost} \\
Shanshan and Sen &	Empowering learners with AI-generated content for programming learning and computational thinking: The lens of extended effective use theory &	JCAL	& 2023 & \cite{shanshanempowering} \\
Shoufan &	Can Students without Prior Knowledge Use ChatGPT to Answer Test Questions? An Empirical Study &	ACM TOCE	& 2023 & \cite{Shoufan2023CanStudents} \\
Speth et al. &	Investigating the Use of AI-Generated Exercises for Beginner and Intermediate Programming Courses: A ChatGPT Case Study &	CSEE\&T	& 2023 & \cite{speth2023investigating} \\
Valový and Buchalcevova &	The Psychological Effects of AI-Assisted Programming on Students and Professionals &	ICSME	& 2023 & \cite{10336271} \\
Wang et al. &	Unleashing ChatGPT's Power: A Case Study on Optimizing Information Retrieval in Flipped Classrooms via Prompt Engineering &	IEEE TLT	& 2023 & \cite{10285884} \\
Wu et al. &	Research on the Construction of Intelligent Programming Platform Based on AI-generated Content &	ICETC	& 2023 & \cite{wu2023research} \\
Yilmaz and Yilmaz &	The effect of generative artificial intelligence (AI)-based tool use on students' computational thinking skills, programming self-efficacy and motivation &	Computers and Education: Artificial Intelligence	& 2023 & \cite{yilmaz2023effect} \\

\specialrule{.2em}{.1em}{.1em} 

Agarwal et al. &	Which LLM should I use?: Evaluating LLMs for tasks performed by Undergraduate Computer Science Students" &	arXiv	& 2024 & \cite{agarwal2024llm} \\
Arora et al. &	Analyzing LLM Usage in an Advanced Computing Class in India &	arXiv	& 2024 & \cite{arora2024analyzingllmusageadvanced} \\
Balse et al. &	Evaluating the Quality of LLM-Generated Explanations for Logical Errors in CS1 Student Programs &	arXiv	& 2024 & \cite{10.1145/3627217.3627233} \\
Barambones et al. &	ChatGPT for Learning HCI Techniques: A Case Study on Interviews for Personas &	IEEE TLT	& 2024 & \cite{Barambones2024ChatGPTforLearningHCI} \\
Bernstein et al. &	``Like a Nesting Doll'': Analyzing Recursion Analogies Generated by CS Students using Large Language Models" &	arXiv	& 2024 & \cite{bernstein2024like} \\
Cámara et al. &	Generative AI in the Software Modeling Classroom: An Experience Report with ChatGPT and UML &	IEEE Software	& 2024 & \cite{camara2024generative} \\
Chen et al. &	Learning Agent-based Modeling with LLM Companions: Experiences of Novices and Experts Using ChatGPT \& NetLogo Chat &	CHI	& 2024 & \cite{chen2024learning} \\
Choudhuri et al. &	How Far Are We? The Triumphs and Trials of Generative AI in Learning Software Engineering &	ICSE	& 2024 & \cite{choudhuri2024far} \\
Cipriano and Alaves &	``ChatGPT Is Here to Help, Not to Replace Anybody'' -- An Evaluation of Students' Opinions On Integrating ChatGPT In CS Courses &	arXiv	& 2024 & \cite{cipriano2024chatgpt} \\
Cipriano et al. &	A Picture Is Worth a Thousand Words: Exploring Diagram and Video-Based OOP Exercises to Counter LLM Over-Reliance &	arXiv	& 2024 & \cite{cipriano2024pictureworththousandwords} \\
Denny et al. &	Explaining Code with a Purpose: An Integrated Approach for Developing Code Comprehension and Prompting Skills &	arXiv	& 2024 & \cite{denny2024explaining} \\
Denny et al. &	Prompt Problems: A New Programming Exercise for the Generative AI Era &	SIGCSE TS	& 2024 & \cite{denny2024prompt} \\
Haindl and Weinberger &	Students' Experiences of Using ChatGPT in an Undergraduate Programming Course &	IEEE Access	& 2024 & \cite{10478015} \\
Hou et al. &	CodeTailor: Personalized Parsons Puzzles are Preferred Over AI-Generated Solutions to Support Learning &	arXiv	& 2024 & \cite{Hou2024CodeTailorLP} \\
Jacobs and Jaschke &	Evaluating the Application of Large Language Models to Generate Feedback in Programming Education &	arXiv	& 2024 & \cite{jacobs2024evaluating} \\
Jeuring et al. &	What Skills Do You Need When Developing Software Using ChatGPT? (Discussion Paper) &	Koli Calling	& 2024 & \cite{10.1145/3631802.3631807} \\
Jin et al. &	Teach AI How to Code: Using Large Language Models as Teachable Agents for Programming Education &	CHI	& 2024 & \cite{10.1145/3613904.3642349} \\
Jury et al. &	Evaluating LLM-generated Worked Examples in an Introductory Programming Course &	ACE	& 2024 & \cite{jury2024evaluating} \\
Kazemitabaar et al. &	CodeAid: Evaluating a Classroom Deployment of an LLM-based Programming Assistant that Balances Student and Educator Needs &	CHI	& 2024 & \cite{Kazemitabaar2024CodeAid} \\
Kimmel et al. &	Enhancing Programming Error Messages in Real Time with Generative AI &	CHI EA	& 2024 & \cite{kimmel2024enhancing} \\
Kosar et al. &	Computer Science Education in ChatGPT Era: Experiences from an Experiment in a Programming Course for Novice Programmers &	Mathematics	& 2024 & \cite{math12050629} \\
Kuramitsu et al. &	Training AI Model that Suggests Python Code from Student Requests in Natural Language &	Journal of Information Processing	& 2024 & \cite{kuramitsu2024training} \\
Liao et al. &	Scaffolding Computational Thinking with ChatGPT &	IEEE TLT	& 2024 & \cite{10508087} \\
Liu et al. &	Teaching CS50 with AI: Leveraging Generative Artificial Intelligence in Computer Science Education &	SIGCSE TS	& 2024 & \cite{liu2024teaching} \\
Lyu et al. &	Evaluating the Effectiveness of LLMs in Introductory Computer Science Education: A Semester-Long Field Study &	arXiv	& 2024 & \cite{lyu2024evaluating} \\
Ma et al. &	Enhancing Programming Education with ChatGPT: A Case Study on Student Perceptions and Interactions in a Python Course &	arXiv	& 2024 & \cite{ma2024enhancing} \\
Manley et al. &	Examining Student Use of AI in CS1 and CS2 &	JCSC	& 2024 & \cite{manley2024examining} \\
Moore et al. &	Teaching artificial intelligence in extracurricular contexts through narrative-based learnersourcing &	CHI	& 2024 & \cite{10.1145/3613904.3642198} \\
Nam et al. &	Using an LLM to Help With Code Understanding &	ICSE	& 2024 & \cite{10.1145/3597503.3639187} \\
Neyem et al. &	Exploring the Impact of Generative AI for StandUp Report Recommendations in Software Capstone Project Development &	SIGCSE TS	& 2024 & \cite{neyem2024exploring} \\
Neyem et al. &	Toward an AI Knowledge Assistant for Context-aware Learning Experiences in Software Capstone Project Development &	IEEE TLT	& 2024 & \cite{10518103} \\
Nguyen et al. &	How Beginning Programmers and Code LLMs (Mis)read Each Other &	CHI	& 2024 & \cite{nguyen2024beginning} \\
Pankiewicz and Baker &	Navigating Compiler Errors with AI Assistance -- A Study of GPT Hints in an Introductory Programming Course &	arXiv	& 2024 & \cite{pankiewicz2024navigating} \\
Pesovski et al. &	Generative AI for Customizable Learning Experiences &	Sustainability	& 2024 & \cite{pesovski2024generative} \\
Roest et al. &	Next-Step Hint Generation for Introductory Programming Using Large Language Models &	ACE	& 2024 & \cite{roest2024next} \\
Shah et al. &	Working with Large Code Bases: A Cognitive Apprenticeship Approach to Teaching Software Engineering &	SIGCSE TS	& 2024 & \cite{10.1145/3626252.3630755} \\
Sheese et al. &	Patterns of Student Help-Seeking When Using a Large Language Model-Powered Programming Assistant &	ACE	& 2024 & \cite{sheese2024patterns} \\
Singh et al. &	Bridging Learnersourcing and AI: Exploring the Dynamics of Student-AI Collaborative Feedback Generation &	LAK	& 2024 & \cite{Singh2024Bridging} \\
Sun et al. &	Would ChatGPT-facilitated programming mode impact college students’ programming behaviors, performances, and perceptions? An empirical study &	International Journal of Educational Technology in Higher Education	& 2024 & \cite{sun2024would} \\ 
Tanay et al. &	An Exploratory Study on Upper-Level Computing Students' Use of Large Language Models as Tools in a Semester-Long Project &	arXiv	& 2024 & \cite{tanay2024exploratorystudyupperlevelcomputing} \\
Taylor et al. &	dcc -{}-help: Transforming the Role of the Compiler by Generating Context-Aware Error Explanations with Large Language Models &	SIGCSE TS	& 2024 & \cite{taylor2024dcc} \\
Wang et al. &	A Large Scale RCT on Effective Error Messages in CS1 &	SIGCSE TS	& 2024 & \cite{wang2024large} \\
Woodrow et al. &	AI Teaches the Art of Elegant Coding: Timely, Fair, and Helpful Style Feedback in a Global Course &	SIGCSE TS	& 2024 & \cite{10.1145/3626252.3630773} \\
Xiao et al. &	Exploring How Multiple Levels of GPT-Generated Programming Hints Support or Disappoint Novices &	CHI EA	& 2024 & \cite{xiao2024exploring} \\
Zhang et al. &	Students' Perceptions and Preferences of Generative Artificial Intelligence Feedback for Programming &	AAAI	& 2024 & \cite{zhang2024students} \\







\bottomrule
		\end{tabular}
	\end{center}
\end{table*}

\subsection{Search String Construction}
\label{sec:search_string}

The research team iteratively constructed a search string aimed at including papers that matched our research interests according to four categories: 1) the \textit{domain} of the work was computer science or computer engineering education, 2) the \textit{topic} was related to the use of Generative AI (GenAI), 3) it aligned with the \textit{working group focus} on the impact of GenAI on pedagogy or teaching tools, and 4) it included the use of empirical \textit{methods}. The literature review team met several times to discuss various keyword permutations and ultimately arrived at the following sub-search string for each of the four categories:
\begin{itemize}[leftmargin=1em, rightmargin=1em, label={}, itemsep=0.2em, parsep=0.5em]
    \item \texttt{\textbf{Domain:}} ``Computer science education'' OR ``Computing education'' OR ``CS education'' OR ``CSEd'' OR ``CER'' OR ``Computing students'' OR ``Computing instructors'' OR ``CS students'' OR ``CS instructors'' OR ``Computer engineering education'' OR ``Programming education'' OR ``Introductory programming''
    \item \texttt{\textbf{Topic:}} ``Large language model'' OR ``Large language models'' OR ``LLM'' OR ``LLMs'' OR ``Generative AI'' OR ``ChatGPT'' OR ``GPT3'' OR ``GPT4'' OR ``Multimodal model'' OR ``Multimodal models'' OR ``Gemini'' OR ``Claude'' OR ``GenAI'' OR ``GPT-4'' OR ``GPT-3.5'' OR ``GPT-3'' OR ``Copilot'' OR ``Language model'' OR ``Language models'' OR ``Generative pre-trained transformer''
    \item \texttt{\textbf{Working Group Focus:}} ``Pedagogy'' OR ``Pedagogies'' OR ``Classroom'' OR ``Student'' OR ``Students'' OR ``Teaching approach'' OR ``Teaching tools''
    \item \texttt{\textbf{Method:}} ``Qualitative'' OR ``Quantitative'' OR ``Perceptions'' OR ``Investigating'' OR ``Exploratory'' OR ``Survey'' OR ``Interview'' OR ``Experiment'' OR ``Focus group''
\end{itemize}

The full search string was constructed by combining each of these categories to ensure a paper contained at least one search term per category. As such, the full search string is as follows:

\begin{quote}    
    \texttt{\textbf{Search String = }}
    \texttt{\textbf{Domain}} AND \texttt{\textbf{Topic}} AND \texttt{\textbf{Working Group Focus}} AND \texttt{\textbf{Method}}
\end{quote}

We decided to look for papers in a total of five different databases to ensure comprehensive coverage of work. The chosen databases were ASEE Peer, arXiv, Scopus, ACM Digital Library, and IEEE Xplore.

The final search resulted in 1536 papers after duplicate removals. The search was done on May 23rd, 2024, which is the cut-off date for included articles. The breakdown of the number of papers from each database prior to the removals is included in \autoref{tab:db_count}.

\begin{table}[htb]
\caption{The number of papers extracted from each database prior to duplicate removal.}
\label{tab:db_count}
\begin{tabular}{p{0.75\linewidth}r}
\toprule
\textbf{Source}     & \textbf{\#}  \\ \midrule
ASEE Peer           & 23  \\
arXiv               & 47  \\
Scopus              & 665 \\
ACM Digital Library & 721 \\
IEEE Xplore         & 258 \\ \bottomrule
\end{tabular}
\end{table}

\begin{figure*}
    \centering
    \begin{subfigure}[b]{0.49\textwidth}
        \centering
        \includegraphics[width=\textwidth]{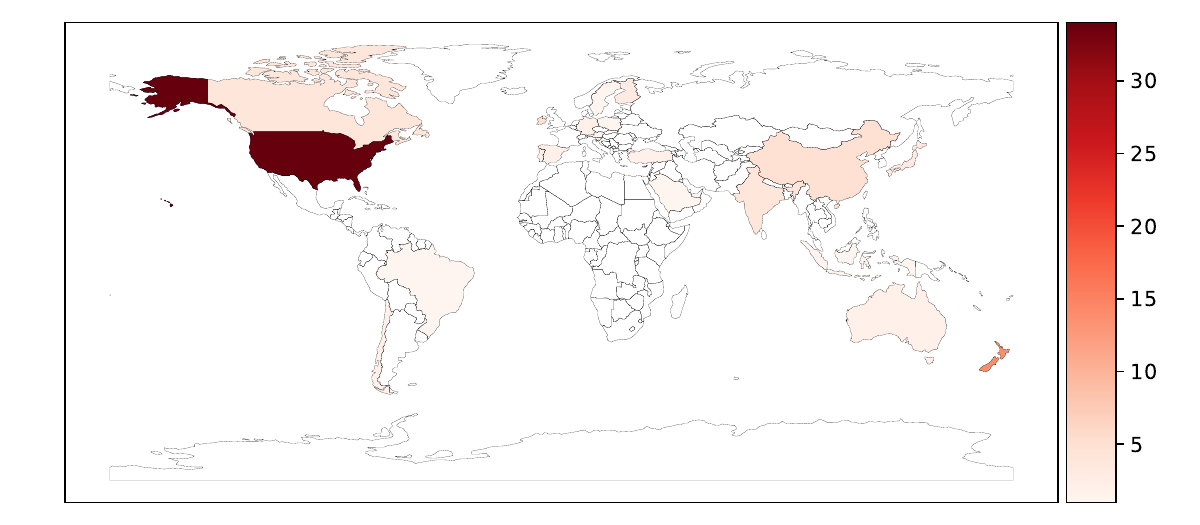}\\
        \hspace{-.55cm}\includegraphics[width=0.96\textwidth]{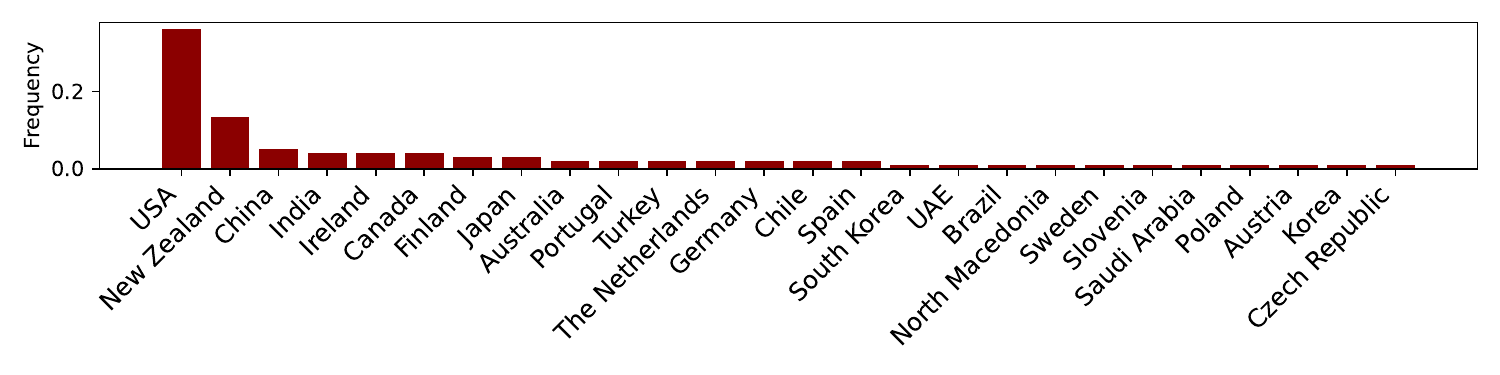}
        \caption{Authors locations}
        \label{subfig:author_map}
    \end{subfigure}
    \hfill
    \begin{subfigure}[b]{0.49\textwidth}
        \centering
        \includegraphics[width=\textwidth]{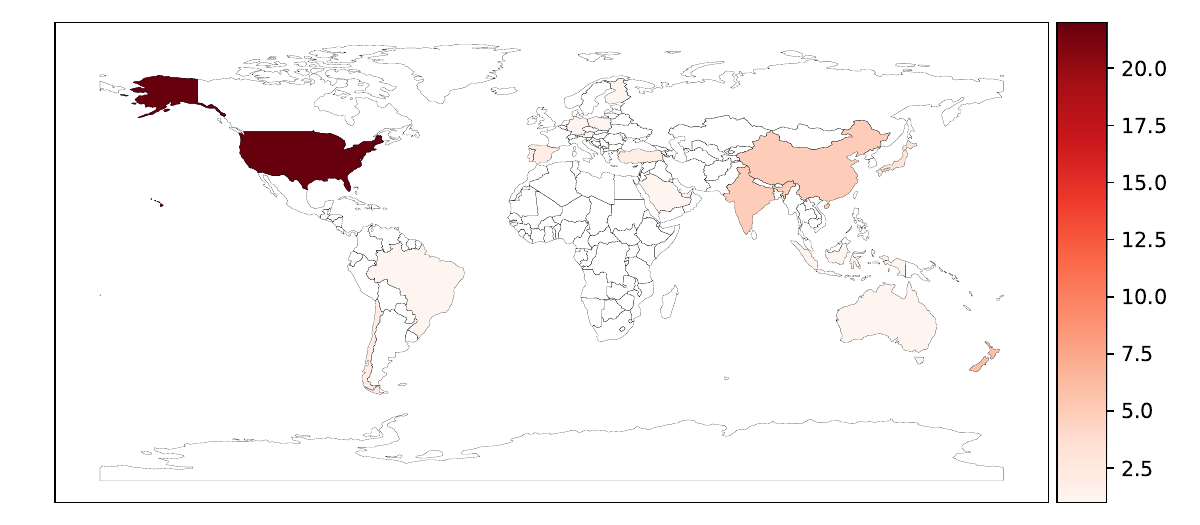}\\
        \hspace{-.68cm}\includegraphics[width=0.96\textwidth]{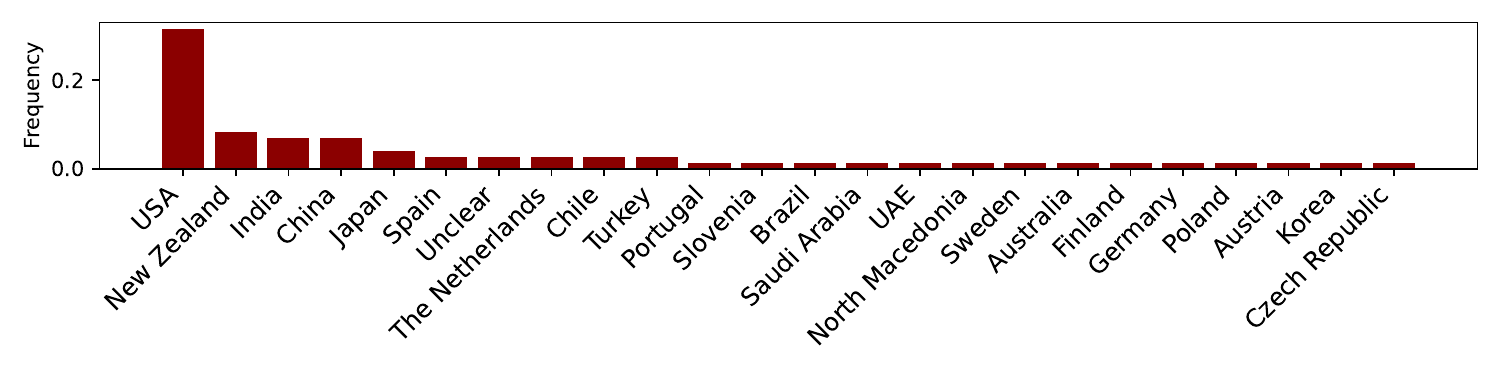}
        \caption{Students population locations}
        \label{subfig:student_map}
    \end{subfigure}
    \caption{Heatmaps of where authors publishing are located (Figure~\ref{subfig:author_map}) and where the student populations they are investigating are located (\autoref{subfig:student_map})}
\end{figure*}

To ensure that no relevant work was missed in our SLR, we broadly sampled the literature. To evaluate our search string for such comprehensive coverage, we selected ten papers that we would expect to find with our search string, shown in \autoref{tab:reference_papers}. All ten papers were found with the final search string, which provides evidence that the search string is adequate for finding relevant papers.

\begin{table*}[htb]
\caption{The reference papers used to evaluate the quality of the search string.}
\label{tab:reference_papers}
\begin{tabular}{p{15cm}l}
\toprule
Paper title & Citation \\ \midrule
Teaching CS50 with AI: Leveraging Generative Artificial Intelligence in Computer Science Education & \cite{liu2024teaching} \\
Prompt Problems: A New Programming Exercise for the Generative AI Era & \cite{denny2024prompt} \\
CodeHelp: Using Large Language Models with Guardrails for Scalable Support in Programming Classes & \cite{liffiton2023codehelp} \\
Experiences from Using Code Explanations Generated by Large Language Models in a Web Software Development E-Book & \cite{macneil2023experiences} \\
Evaluating LLM-generated Worked Examples in an Introductory Programming Course & \cite{jury2024evaluating} \\
Next-Step Hint Generation for Introductory Programming Using Large Language Models & \cite{roest2024next} \\
ChatGPT for Learning HCI Techniques: A Case Study on Interviews for Personas & \cite{Barambones2024ChatGPTforLearningHCI} \\
CodeAid: Evaluating a Classroom Deployment of an LLM-based Programming Assistant that Balances Student and Educator Needs & \cite{Kazemitabaar2024CodeAid} \\
``ChatGPT Is Here to Help, Not to Replace Anybody'' -- An Evaluation of Students' Opinions On Integrating ChatGPT In CS Courses & \cite{cipriano2024chatgpt} \\
Explaining Code with a Purpose: An Integrated Approach for Developing Code Comprehension and Prompting Skills & \cite{denny2024explaining} \\
\bottomrule
\end{tabular}
\end{table*}

\subsection{Paper Filtering Process}

Balancing the need to broadly include relevant papers while ensuring the central goals of the research remain in focus is a common challenge in SLRs. To address this, we developed exclusion criteria that were applied repeatedly throughout the filtering process, ensuring that only high-quality papers focusing on the use of generative AI in pedagogical practice were included in our final analysis.

A primary criterion for inclusion in the SLR was that the papers prominently featured some form of generative AI. This included studies where students used existing generative AI tools or where such tools were integrated into pedagogical practices. Papers that focused on teaching students about generative AI or its associated ethical aspects were also included.

We included only papers that focused on computing education at the tertiary level. Papers must have featured student participants of some kind and could not focus solely on K-12 or professional developers. However, papers that featured both tertiary students and K-12 or professional developers were included. The reason to focus on tertiary students was that these students are being trained to become computing professionals in the near future. 

In keeping with the goal of evaluating robust interventions within computing education, we excluded shorter papers, such as posters, demos, and other shorter formats. Since the typical length of papers in the field of computing education is six pages or more in a double-column format, we used this lower limit to exclude less comprehensive studies.

Based on these goals, we applied the following exclusion criteria at each stage of the filtering process:
\begin{enumerate}
    \item \textbf{Not GenAI}: Papers that did not include a generative AI component were excluded. Papers that simply provide implications for generative AI were also not included.  
    \item \textbf{Not Computing Education}: Papers that were not related to computing education were excluded. For example, papers primarily focused on professional developers were excluded.
    \item \textbf{No Human Evidence}: The paper did not contain empirical data or included only an expert evaluation.
    \item \textbf{No Intervention}: Papers that did not feature a classroom intervention were excluded. However, user studies with students were included. 
    \item \textbf{Exclusively K-12}: Papers where the participants were exclusively K-12 were excluded.     
    \item \textbf{Too Short}: Papers were excluded if they were under 5 pages for double-column articles and under 8 pages for single-column articles. 
    \item \textbf{Not an Article}: We excluded conference proceedings' front matters, white papers, or other non-research content.
\end{enumerate}

In the first phase, we had two reviewers evaluate the title and abstract of the paper for inclusion. If either of the reviewers thought that the paper was relevant, it was chosen for the next stage -- extraction.

\subsection{Extraction}

For all the papers that were not excluded in the title/abstract scanning phase, we thoroughly read the full paper. A paper could still be rejected at this stage if it did not pass the inclusion/exclusion criteria. 
To consistently systematically extract the content from each paper, we developed a structured form shown in Appendix~\ref{app:appendix-extraction} to extract information for all papers that were included in the review. Four researchers extracted the content from the final set of 71 papers.

\section{Systematic Literature Review: Results}
\label{sec:litreview-results}

In this section, we present the results of the systematic literature review. They are based on the information extracted from the final set of 71 papers that resulted from the process described in the previous section.

\subsection{Descriptive Statistics}

From the evaluation form we collected a variety of descriptive data that relates to: 1) the authors and the students they evaluated, 2) the characteristics of the studies they conducted, 3) the types of courses in which these studies took place, and 4) the custom tools that have been developed.

\paragraph{Course Information}
A variety of courses were used to study AI tools.  We saw 26 upper division courses including three masters' level courses. CS1 was studied in 20 papers. Eight papers included more than one course. Examples of other courses were Human Computer Interaction, Software Modeling, and Embedded Systems.

\paragraph{Author and Student Information}

As shown in \autoref{subfig:author_map}, the locations of the authors' institutions varied widely with the largest proportion of articles having authors from the United States (34\,\%) and New Zealand (14\,\%). Additionally, they were primarily from academic institutions (n=67) with very few coming from industry or involving industry-academic collaborations (n=2), or just from industry (n=1). As might be expected, the majority of studies took place at one or more of the authors' institutions which results in the distribution of student populations that were investigated looking quite similar to the one for authors (\autoref{subfig:student_map}).

\begin{table}[t]
    \centering
    \caption{Characteristics of the studies and their methods.}
     \begin{subtable}[t]{\linewidth}
        \centering
        \begin{tabular}{p{0.75\linewidth}r}
    \toprule
    \textbf{Methodology} & \textbf{\#} \\
    \midrule
    Both         & 36 \\
    Quantitative & 25 \\
    Qualitative  & 9  \\
    \bottomrule
\end{tabular}
        \caption{Methodologies Used}
        \label{stab:methodology}
    \end{subtable}
    \hfill
        \begin{subtable}[t]{\linewidth}
        \centering
        \begin{tabular}{p{0.75\linewidth}r}
    \toprule
    \textbf{Study Type} & \textbf{\#} \\
    \midrule
    Unsupervised study & 41 \\
    Supervised study & 24 \\
    Both & 2 \\
    Unclear & 3 \\
    \bottomrule
\end{tabular}
        \caption{Level of Supervision}
        \label{stab:supervision}
    \end{subtable}
    \label{tab:study_types}
\end{table}

\begin{figure}[htb]
    \centering
    \includegraphics[width=0.95\columnwidth]{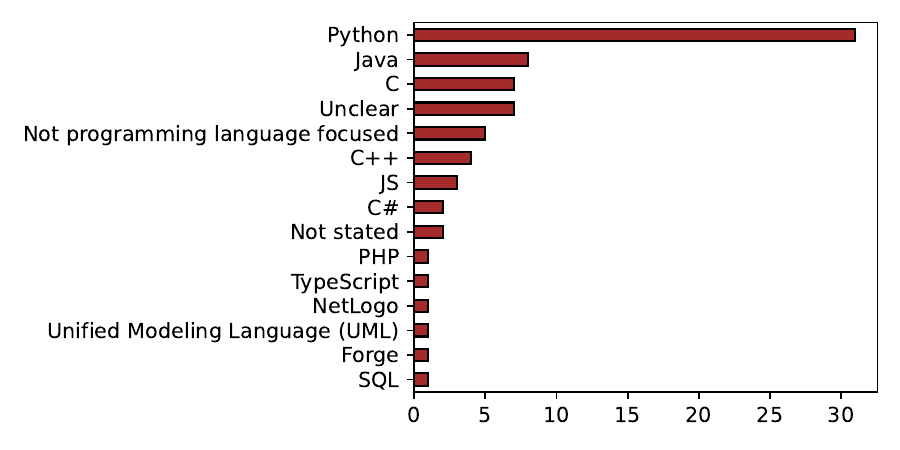}
    \caption{The programming languages reported on by the studies.}
    \label{fig:programming_languages}
\end{figure}

\begin{figure}[htb]
    \centering
    \includegraphics[width=0.95\columnwidth]{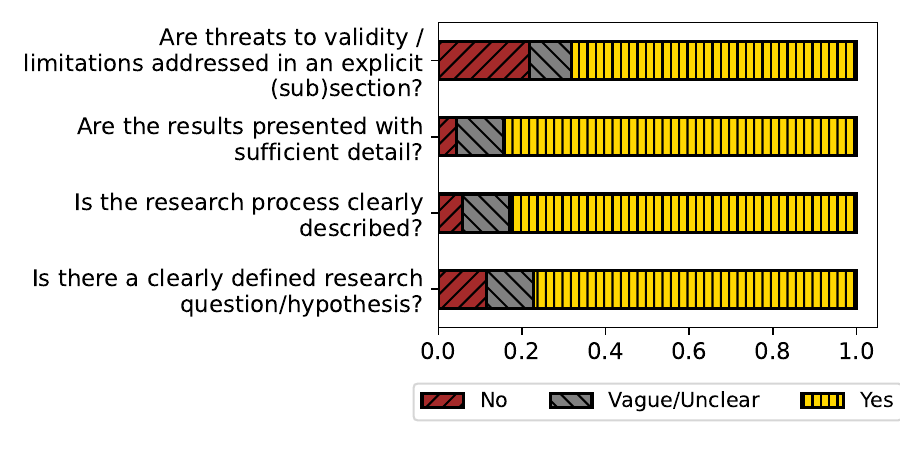}
    \caption{Quality of work as evaluated using the paper quality metrics of \citet{hellas2018predicting}.}
    \label{fig:paper_quality}
\end{figure}

\paragraph{Study Context and Quality} The majority of the evaluations used a
mixed methods approach, followed by quantitative methods (\autoref{stab:methodology}). 
With
respect to the student populations under investigation both mixed-methods
(median=52.0 IQR=(24.0-105.0)) and quantitative (median=56.0 IQR=(49.0-160.0)) studies
investigated student populations of similar sizes. Purely qualitative studies,
as might be expected, had the smallest number of participants (median=21.0
IQR=(12.0-72.0)). Additionally, the majority of studies conducted were done in
an unsupervised (i.\,e., uninvigilated) manner (\autoref{stab:supervision}) and Python was by far the
most common language being used (\autoref{fig:programming_languages}).

Using the quality metrics presented by \citet{hellas2018predicting}, we
evaluated each paper according to its presentation of: 1) addressed
limitations, 2) descriptions of the presented results, 3) description of the
research process, and 4) presentation of the research questions. Overall, we
find the majority of the empirical work we evaluated were sufficient across all
four of these dimensions (\autoref{fig:paper_quality}). The dimension that saw
the largest degree of vagueness (20\,\%) or exclusion (11\,\%) was that of
clearly defined limitations.

\paragraph{Custom Tools} In the course of our literature review we uncovered a wide variety of custom tooling that has been developed for the support of instructors and students alike. These included:  CodeAid~\cite{Kazemitabaar2024CodeAid}, CodeHelp~\cite{liffiton2023codehelp, sheese2024patterns}, WorkedGen~\cite{jury2024evaluating}, LLM Hint Factory~\cite{xiao2024exploring}, Tutor Kai~\cite{jacobs2024evaluating}, CodeTutor~\cite{lyu2024evaluating}, GPTeach~\cite{markel2023gpteach}, Charlie the Coding Cow~\cite{nguyen2024beginning}, KOGI~\cite{kuramitsu2023kogi}, NetLogo Chat~\cite{chen2024learning}, Promptly~\cite{denny2023promptly, denny2024prompt}, IPSSC~\cite{10508087}, CS50 Duck~\cite{liu2024teaching}, Gilt~\cite{10.1145/3597503.3639187}, CodeTailor~\cite{Hou2024CodeTailorLP}, and a variety of others that lacked names~\cite{10386894, 10.1145/3613904.3642198, 10.1145/3626252.3630773, 10518103}.

\paragraph{Using Generative AI for Teaching vs Teaching Students to Use Generative AI} In relation to the research goals of the group, we categorized the literature into two broader categories: studies where generative AI was \textit{used} for some pedagogical purpose and studies where the main goal was to \textit{teach} students about generative AI specifically or to use a specific generative AI tool.
Of the papers reviewed, 57 fell into the \textit{use} category and the remaining 14 were categorized as \textit{teach}.

\subsection{How Is Generative AI Being Incorporated into Teaching?}
\label{sec:litreview_HOW}

In the extraction form, we had an open-ended question on ``How Instructors Incorporate Generative AI into Teaching Computing?'', which directly maps to our SLR-RQ2.

To analyze the results for this question, three authors thematically analyzed the open-ended responses and came up with initial tags. They discussed the initial tags and then combined them into definitive tags along three axes: tool type, purpose from the student's point of view, and whether students received guidance on how to use generative AI.

Our tags for each axis are below.

\begin{itemize}
    \item Tool type: general purpose (e.\,g., ChatGPT), task-specific (e.\,g., GitHub Copilot), instructor-provided guardrails (e.\,g., CodeHelp~\cite{liffiton2023codehelp}, Promptly~\cite{denny2023promptly}, CodeAid~\cite{Kazemitabaar2024CodeAid}, CodeTailor~\cite{Hou2024CodeTailorLP}).
    \item Purpose: hints, debug, learning resources, writing code, teacher training, code comprehension, code review, teach GenAI, motivation, UML, multiple, not specified.
    \item Guidance on GenAI use: yes, no, unclear.
\end{itemize}

For tool type, we categorized tools as general purpose (such as ChatGPT), whether the tool is task-specific (i.\,e., the tool is meant for a specific task, but not education-focused, such as GitHub Copilot), and whether there are instructor-provided guardrails (i.\,e., there were pedagogical guardrails or other constraints in the tool).

For purpose from the student's point of view, we looked at the tasks that students worked on with the support of GenAI. During the thematic analysis, we came up with the following categories.

\begin{itemize}

    \item \textbf{Writing code} -- GenAI was used for code writing support. For example, CodeTailor helps students write code and interact with Parsons problems~\cite{Hou2024CodeTailorLP}.

    \item \textbf{Code comprehension} -- GenAI was used to teach code comprehension. For example, Gilt provides scaffolding contextualized to select sections of students' code~\cite{10.1145/3597503.3639187}.
    
    \item \textbf{Hints} -- GenAI was used to produce hints for students. For example, next-step hints~\cite{roest2024next}.

    \item \textbf{Learning resources} -- GenAI was used to create or improve learning resources that could be used by other students too. For example, students were instructed to generate analogies using LLMs~\cite{bernstein2024like}.

    \item \textbf{Teach generative AI} -- the article describes an approach or tool to teach students how to use GenAI. For example, Promptly provides scaffolding for students to learn how to prompt GenAI~\cite{denny2023promptly}.

    \item \textbf{Multiple tags} -- GenAI was used for multiple purposes. For example, Choudhuri et al. investigated students' use of GenAI for multiple different software engineering tasks such as debugging and refactoring~\cite{choudhuri2024far}.

    \item \textbf{Debug} -- GenAI was used for debugging help, such as by explaining error messages~\cite{shanshanempowering}.
    
    \item \textbf{Code review} -- GenAI was used for code review. For example, by integrating it into an assignment submission system~\cite{crandall2023generative}.

    \item \textbf{UML} -- GenAI was used to assist in generating UML diagrams. This was done by C{\'a}mara et al.~\cite{camara2024generative}.

    \item \textbf{Teacher training} -- GenAI was used for teacher or teaching assistant training. GPTeach creates LLM agents that act as students to train new TAs~\cite{markel2023gpteach}.
    
    \item \textbf{Motivation} -- GenAI was used to increase student motivation. This was done by Moore et al. in a narrative-based learnersourcing platform~\cite{10.1145/3613904.3642198}.
    
    \item \textbf{Not specified} -- the purpose of GenAI was not specified in the paper or was unclear.

\end{itemize}

For guidance on GenAI use, we categorized each paper as a ``yes'', `no'', or ``unclear'' (i.\,e., whether students received guidance or instructions on how to use generative AI).

The results of this analysis are presented in \autoref{tab:study_characteristics}. Based on the findings, most studies focused on using generative AI for writing code, code comprehension, hints, and generating learning resources. Slightly under half (32/71) of the studies used tools with instructor-provided guardrails, for example, a custom tool with pedagogical guardrails. However, most commonly (34/71), studies used a general purpose generative AI tool such as ChatGPT. In the majority of studies (49/70), it was not reported that students would have been instructed on how to use generative AI.

\begin{table}[h]
    \caption{The characteristics of studies in terms of (a) their intended purpose, (b) the type of the tool, and (c) guidance provided on how to use GenAI. }
    \vspace{0.1cm} 
    \begin{subtable}[t]{0.45\linewidth}
        \centering
        \begin{tabular}{p{0.7\linewidth}r}
\toprule
\textbf{Task} & \textbf{\#} \\ 
\midrule
writing code & 26 \\ 
code comprehension & 10 \\ 
hints & 8 \\
learning resources & 7 \\ 
teach GenAI & 6 \\ 
multiple & 5 \\
debug & 2 \\ 
code review & 2 \\ 
UML & 1 \\ 
teacher training & 1 \\ 
motivation & 1 \\ 
not specified & 2 \\ 
\bottomrule
\end{tabular}
        \vspace{0.1cm} 
        \caption{Intended Purpose}
        \label{stab:intpurpose}
    \end{subtable}
    \hfill
    \begin{subtable}[t]{0.45\linewidth}
    
        \centering    
        
        \vspace{-2.75cm}
        \begin{tabular}{p{0.6\linewidth}r}
\toprule
\textbf{Response} & \textbf{Count} \\ 
\midrule
general purpose & 34 \\
instructor guardrails & 32 \\
task-specific & 4 \\
unclear & 1 \\
\bottomrule
\end{tabular}

        \vspace{0.1cm} 
        \caption{Tool Type}
        \label{stab:toolscaffolding}
        
        \vfill 

        \vspace{0.5cm}
        \begin{tabular}{p{0.6\linewidth}r}
\toprule
\textbf{Response} & \textbf{Count} \\ 
\midrule
no & 49 \\ 
yes & 21 \\  
unclear & 1 \\ 
\bottomrule
\end{tabular}
        \caption{Guidance on GenAI Use}
        \label{stab:pedagogicalscaffolding}
        
    \end{subtable}
    
    \label{tab:study_characteristics}
\end{table}

We also looked at the type of evidence used, categorizing it into perceptual (e.\,g., opinions of activity) or behavioral (e.\,g., correctness of produced code). This was contrasted with the nature of the findings, which were categorized as positive, negative, mixed, or neutral. The results of this analysis are presented in \autoref{tab:evidencevsfindings}. The results suggest that the majority of studies found positive results, with studies using behavioral evidence slightly more likely to report positive findings. 54\,\% of studies with perceptual evidence reported positive findings whereas 69\,\% of studies with behavioral and 70\,\% of studies with both types of evidence reported positive findings.

\begin{table}[h]
\centering
\caption{Type of evidence versus nature of findings. Note that one study had positive findings, but unclear type of evidence, leading to the numbers only summing to 70.}
\begin{tabular}{lcccc|c}
\toprule
 & Positive & Negative & Mixed & Neutral & Total \\
\midrule
Perceptual & 13 & 1 & 9 & 1 & 24 \\
Behavioral & 11 & 1 & 1 & 3 & 16 \\
Both & 21 & 3 & 5 & 1 & 30 \\
\midrule
Total & 45 & 5 & 15 & 5 & 70 \\
\bottomrule
\end{tabular}

\label{tab:evidencevsfindings}
\end{table}

\begin{table*}[h!]
\centering
\caption{Comparing the types of tools and level of guidance provided.}
\begin{tabular}{cl|cccc|c}
\toprule
\textbf{Guidance} & \textbf{Tool type} &  \textbf{Positive} & \textbf{Negative} & \textbf{Mixed} & \textbf{Neutral} & \textbf{Total} \\
\midrule
no & general                  & 12   & 3    & 4    & 3       & 22  \\ 
no & task-specific                & 1    &      &      &         & 1   \\ 
no & guardrailed          & 19   &      & 5    & 2       & 26  \\ 
\midrule
yes & general                 & 7    & 1    & 4    &         & 12  \\ 
yes &task-specific                 & 2    &      & 1    &         & 3   \\ 
yes &guardrailed          & 4    & 1    & 1    &         & 6   \\
\bottomrule
\end{tabular}
\label{tab:scaffolding_results}
\end{table*}

\begin{table*}[h!]
\caption{Comparing outcomes based on the task type.}
\centering
\begin{tabular}{lcccc|c}
\toprule
\textbf{} & \textbf{Positive} & \textbf{Negative} & \textbf{Mixed} & \textbf{Neutral} & \textbf{Total} \\ \midrule
\textbf{Writing code}          & 15 & 2 & 8 & 1 & 26 \\
\textbf{Code comprehension}    & 8  & 1 & 1 &   & 10 \\
\textbf{Hints}                 & 4  &   & 2 & 2 & 8  \\
\textbf{Learning resources}    & 6  &   & 1 &   & 7  \\
\textbf{Teach GenAI}           & 4  & 1 & 1 &   & 6  \\
\textbf{Multiple}    & 2  & 1 & 2 &   & 5  \\
\textbf{Debug}                 & 1  &   &   & 1 & 2  \\
\textbf{Code review}           & 2  &   &   &   & 2  \\
\textbf{UML}            & 1  &   &   &   & 1  \\
\textbf{Teacher training}      & 1  &   &   &   & 1  \\
\textbf{Motivation}            & 1  &   &   &   & 1  \\
\textbf{Not specified}         & 1  &   &   & 1 & 2  \\ \midrule
\textbf{Total}                 & 46 & 5 & 15 & 5 & 71 \\ \bottomrule
\end{tabular}
\label{tab:task_outcomes}
\end{table*}

\subsection{What Are the Motivations Behind Incorporating GenAI Tools into Teaching?}
\label{sec:litreview_WHY}

In the extraction form, we had an open-ended question on ``And why do they incorporate GenAI tools that way?'', which directly maps to our SLR-RQ3. In the previous section, we outlined \textit{how} instructors are incorporating generative AI into their teaching. Part of the \textit{why} is overlapping -- the ways it is incorporated often tell about the \textit{why}. Thus, we here focus on \textit{in benefit of whom} it is being integrated. We tagged each of the included papers with ``students'', ``instructors'' or ``both''. Out of the 71 papers, in 59 
GenAI was incorporated in benefit of students. In 5 papers it was in benefit of teachers. In 7 papers it was in benefit of both students and teachers.

\todo[inline]{Add examples of each category}

\subsection{Recommendations for Incorporating GenAI}

To analyze the effectiveness of incorporating generative AI into computing classrooms, we cross-tabulated the type of tool and guidance on GenAI use provided to students with the nature of the findings (positive, negative, mixed or neutral). The results of this analysis are reported in \autoref{tab:scaffolding_results}. The main result that can be seen from the table is that when there is no guidance on using GenAI from the instructor, it is recommended to use a tool that includes instructor-provided guardrails, e.\,g., pedagogical guardrails. When students are not provided guidance and use a general purpose generative AI model (e.\,g., ChatGPT), only 55\,\% (12/22) of studies found positive results. However, even without guidance, if the tool included instructor-provided guardrails, positive results were found in 73\,\% (19/26) of studies. When students are given instructions on how to use generative AI, whether the tool has built-in pedagogical guardrails does not seem to matter as much -- studies that used general purpose tools found positive results in 58\,\% (7/12) of cases and studies that used instructor-guardrailed tools found positive results in 67\,\% (4/6) of cases.

In a similar vein, we cross-tabulated the nature of the findings with the task that GenAI was used for to examine what computing education tasks might benefit most from generative AI. The results of this analysis are reported in \autoref{tab:task_outcomes}. There are differences between tasks in whether findings of the studies have been positive or not. For code writing, 58 \% (15/26) studies reported positive results, whereas for code comprehension, 80 \% (8/10) of studies reported positive results. According to our results, hint generation is an area where generative AI could still improve, since only half of the studies (4/8) reported positive results. Better results have been observed for generating learning resources, where 86 \% (6/7) studies found positive results.

%
%
%
%

\section{Educator and Developer Views: A Mixed-Methods Approach}
\label{sec:educator-and-developer-views}

%
%


Computing educators want to prepare students for a successful career in software development. It is thus important to understand the experiences of software developers in addition to educators' perspectives to evaluate how different the two perspectives are. For this reason, we conducted a survey study with both target groups, and an interview study with educators. 

Through the educator survey study, we aimed to gather a large sample of educator perspectives on whether (and how) they incorporate GenAI in their classroom, their motivations for this decision, and how they see competencies changing for programming education. We also surveyed developers to build a landscape of how GenAI tools are currently being used in industry settings and how developers view the changing competencies required for programming. The larger sample size of the survey also allows us to begin to explore equity-related questions related to student access and exposure to GenAI tools. 

Through the interview study, we aimed to gather first-hand accounts, and deeper insights from educators on changes to their classroom as a result of emerging GenAI tools. We interviewed educators who teach students how to use GenAI tools to engage with class materials as well as how to use such tools in their professional careers. We also included educators who took a deliberate (explicit) stand to disallow the use of GenAI tools in their classes. Further, we interviewed developers of LLM-powered educational tools as well as researchers focused on the use of GenAI in computing education.

In this section, we first summarize prior studies that similarly gathered perspectives of educators and developers. Then we present the ten research questions that we answer through both the surveys and interviews. Next, we describe the applied methodology. 



\subsection{Prior Studies of Educators' and Developers' Perceptions}

Computing educators' perspectives on GenAI tools have been reported in several prior studies. For example, Chan and Lee ~\cite{chan2023ai} surveyed 184 educators and 399 students, primarily from Hong Kong but across different disciplines. They focused on the perceptions, experience, and knowledge about GenAI, and compared educator and student responses. They found that the educators seemed more concerned about students' over-reliance and ethical issues and were skeptical towards GenAI tools and their capabilities. The need for policies and guidelines ensuring academic integrity and equitable learning conditions was yet another outcome~\cite{chan2023ai}.  

Amani et al.~\cite{amani2023generative} also investigated students' and instructors' perceptions towards GenAI through two surveys at Texas A\&M university. The goal of the instructor survey was to capture how GenAI affected their recent courses and how they think students should use respective tools. One important finding is that the responses from 243 staff members emphasize the need for teaching practices to adapt~\cite{amani2023generative}. Another survey of educators was conducted by Prather et al.~\cite{prather2023wgfullreport}. The 57 respondents elaborated on their perceptions, experience, usage, course policies, expectations, and beliefs, indicating that educators should stay abreast of the technological developments. The results also highlight the need to provide guidance for students regarding the ethical use of GenAI.

In addition to these surveys, a number of recent research studies used interviews to gather the instructor perspective on the impact of GenAI on Computing education research and practice~\cite{lau2023from, zastudil2023generative, wang2023towards, rajabi2023exploring,prather2023wgfullreport,sheard2024instructor}. For example, Lau and Guo conducted interviews with 20 instructors about their intentions to adapt their teaching to emerging GenAI tools (e.\,g., ChatGPT and Copilot)~\cite{lau2023from}. Wang et al.~\cite{wang2023towards} also interviewed 11 instructors about their perceptions. Despite sharing concerns about over-reliance and misuse of GenAI tools, the instructors did not have plans to adapt their courses due to the lack of effective strategies at the time. 
Another example is the interview study with 40 instructors by Rajabi et al.~\cite{rajabi2023exploring}, which showed that educators were cautious about banning AI tools, because students would find ways of using them regardless. At the same time, instructors seemed concerned about increasing anxiety among students by focusing too heavily on exams, which is a well-known issue~\cite{latulipe2015structuring, macneil2016exploring, khan2018active}. A recent interview study with educators was conducted by Sheard et al.~\cite{sheard2024instructor}. They focused on educators' current practices, concerns, and planned adaptations relating to these tools. They found, for example, that educators appreciate that the tools can be a source of support for students, but that these tools may lead to students missing out on learning. Finally, Zastudil et al.~\cite{zastudil2023generative} conducted an interview study comparing the perspectives of 6 instructors and 12 students, finding alignment in their concerns about over-reliance and misalignment in their motivations and relative knowledge about generative AI.

While there has been some work in studying how developers can use GenAI tools to enhance their workflow, little has been done to understand the perceptions of how software developers see the practice changing. Specifically, one study found that software developers are using GenAI tools at many points in the programming process, including creating, modifying, debugging, and explaining code~\cite{Barke2023-sw}. Recent work reports that GenAI automated code generation increases developer productivity~\cite{li2024sheetcopilot}, and GenAI tools are particularly useful for assisting in resolving technical issues~\cite{Coutinho2024-up}. Typically, productivity is measured in terms of the acceptance rates of coding suggestions~\cite{Ziegler2024-fi}. 

In summary, we identified several survey and interview studies about the perceptions of educators regarding the \emph{prospective} adoption of courses, learning objectives, assessments, or institutional policies, but not the \emph{actual} adoption. Moreover, we found fewer studies focused on developers. In our survey study, we focus on both instructors and developers, to understand the perspectives of experts both within the learning environment and within the industry.

\subsection{Research Questions}
\label{sec:research-questions}

Based on the overarching research goals (see \autoref{sec:intro}), the systematic literature review, and the identified gap on actual integration practices in education and industry, we constructed the following research questions guiding our work: 
\begin{enumerate}
\item[\textbf{RQ1}] \textbf{Policies}: What are the existing policies and practices around using GenAI in Computing courses? (1b)

\item[\textbf{RQ2}] \textbf{Instruction-use}: How are instructors teaching students about using GenAI tools in order to program? (1a)
\item[\textbf{RQ3}] \textbf{Instruction-support}: How are instructors using generative AI tools to support the teaching of their courses? (1a)
\item[\textbf{RQ4}] \textbf{Motivations}: Why are instructors making these choices around using GenAI? (1b)

\item[\textbf{RQ5}] \textbf{Tools}: What kinds of GenAI tools are emerging in Computing education? (1a)
\item[\textbf{RQ6}]\textbf{Perceived Outcomes}: How has GenAI impacted instructor perceptions of student outcomes (compared to before)? (2a)
\item[\textbf{RQ7}] \textbf{Industry usage}: How are industry developers using GenAI tools? (2a)
\item[\textbf{RQ8}] \textbf{Competencies}: How has GenAI impacted desired student competencies? (2a, 2b)
\item[\textbf{RQ9}]\textbf{Equity}: How has GenAI impacted student equity? (1a, 1b, 2b)
\item[\textbf{RQ10}] \textbf{Future}: How is GenAI shaping the future of Computing education? (2b)

\end{enumerate}

We intentionally divided RQ2 and RQ3 into two distinct research questions as the Working Group believes it is valuable to emphasize that GenAI tools are used in the classroom and also impacting course learning outcomes. Specifically, RQ2 focuses on investigating how educators prepare students for the future of programming where GenAI tools are available. This may include instructing them about GenAI capabilities for programming or allowing GenAI tools to be used by students as a method to teach students how to effectively incorporate tools into their workflow. RQ2 is in contrast to RQ3 which explores how educators incorporate GenAI tools into the classroom and their workflow. This could include using GenAI tools for grading, assignment creation, student feedback, or providing help (like a TA-bot). 

\subsection{Mixed-Methods} 

To address the research questions regarding educators' and software developers' viewpoints, we employed a mixed-methods approach. To capture the breadth of GenAI usage and perspectives in computing education, we conducted two surveys, one targeting educators and the other targeting developers. To more deeply understand current views and potential future uses for GenAI, we carried out an interview study involving tool creators, educators studying GenAI, and educators using GenAI. 

The subsequent sections describe the design of both the surveys and interview study. Since our studies were structured around our research questions, we have organized the results section accordingly. Rather than presenting survey and interview results separately, we integrate the findings from both methods for each research question. For clarity, we present the mapping of research questions and survey questions in~\autoref{tab:RQs-to-survey}. 

\begin{table*}
  \centering
  \caption{Mapping of the research questions (1st column) and respective survey questions (2nd column) in the Educator Survey (ES) and Developer Survey (DS).}
  \smaller
  \begin{tabularx}{\linewidth}{@{} L{0.14\linewidth} X @{}}
    \toprule
    \textbf{Research Questions} & \textbf{Survey Questions}                                                                      \\
    \midrule
    RQ1: Policies
                               &
    ES-1: Do you explicitly disallow students to use GenAI tools for all of your computing courses (within the last 12 months)? \newline
    ES-2: Are you incorporating GenAI tools (e.g., actively integrating it into the curriculum or exercises) into your recent courses (within the last 12 months)? \newline
    ES-10: Are you doing anything to prevent GenAI tools' use in your course? \newline
    ES-11: What are you doing to prevent GenAI tools' use in your course? \newline
    ES-21: Please describe any changes you have made to your teaching approaches in courses you are teaching as a result of GenAI tools. \newline
    ES-22: Please describe any changes you have made to your assessment approaches in courses you are teaching as a result of GenAI tools. \newline
    DS-14: What is the main reason that you do not use GenAI tools for professional software development?                                                                                                 
                                                                           \\
        \midrule
    RQ2:~Instruction-use and
    RQ3:~Instruction-support
                               &
    ES-12: We ask you to think of a recent course (within the last 12
months) that you teach that is most influenced by GenAI
tools. \newline
    ES-13: Select the size of the recent course that you teach that is most
influenced by GenAI tools \newline
    ES-14: Who uses (or is expected to use) GenAI tools in your course(s)? \newline
    ES-15: If students are allowed to use GenAI tools, how do you expect students to access them? \newline
    ES-16: Which type of GenAI tools are you incorporating into your recent course that is most influenced by GenAI tools? \newline
    ES-17: In what ways have you incorporated GenAI tools into your recent course that is most influenced by GenAI tools?                                                                                 \\
   
    \midrule
    RQ4: Motivations
                               &
    ES-9: Why don't you allow GenAI tools in your courses? \newline
    ES-18: Why have you incorporated GenAI tools into your recent course? \newline
    ES-20: Why have you not incorporated GenAI tools (e.g., actively integrating it into the curriculum or exercises) into your recent courses (within the last 12 months)?                               \\
    \midrule
    RQ5: Tools    &
    ES-16: Which type of GenAI tools are you incorporating into your recent course that is most influenced by GenAI tools? \newline                                                                                \\
    \midrule
    RQ6: Perceived Outcomes
                               &
    Interviews only                                                                                                                                                                                       \\
    
    \midrule
    RQ7: Industry usage
    %
                               &
    ES-7: How often do you believe professional software engineers are using GenAI tools as part of their professional role? \newline
    ES-8: For which tasks or contexts are industry professionals using GenAI tools? \newline
    DS-1: Do you use GenAI? \newline
    DS-2: How often do you use GenAI? \newline
    DS-3: What types? \newline
    DS-4: Describe how you use them \newline
    DS-5: Select the tasks for which you use GenAI \newline
    DS-6: How not useful or useful have GenAI tools been to your software development? \newline
    DS-7: Have GenAI tools made your software development more or less efficient? \newline
    DS-8: How not harmful or harmful have GenAI tools been to your software Development? \newline
    DS-9: If you consider GenAI tools harmful, please describe a situation you have experienced, e.g., what were you doing, what did you expect, why was the use of the GenAI tools harmful and to whom? \newline
    DS-14, DS-15: What is the main reason that you do not use GenAI tools for professional software development?                                                                                          \\
   
    \midrule
    RQ8: Competencies
                               &
    ES-3: Do you believe the skills to create software have changed after the advent of GenAI tools? \newline
    ES-4: Please elaborate on your last response why skills have not changed. \newline
    ES-5: In what ways do you think the skills needed to create software have changed with the introduction of GenAI tools? \newline
    ES-6: When using GenAI tools to create (parts of) software, which skills become the most important (select up to 3)?  \newline
    ES-19: Have you changed any of the learning objectives of your recent course based on the capabilities of GenAI tools? \newline
    DS-10: Did the competencies (i.e., knowledge, skills, dispositions in context of a task) required to professionally develop software change with the availability of GenAI tools? \newline
    DS-11: If you have seen changes, from your experience with GenAI tools, what do you believe are new relevant competencies to professionally develop software with GenAI tools? \newline
    DS-12: If you have seen changes, from your experience with GenAI tools, what do you believe are competencies that are no longer or less relevant to professionally develop software with GenAI tools? \\

    \midrule
    RQ9: Equity
                               &
    ES-26: Do you teach at an institution that serves a minority population in your country?                                                                                                              \\
    \midrule
    RQ10: Future
                               &
    Interviews only                                                                                                                                                                                       \\

    \midrule
    All
                               &
    DS-13, DS-16: What advice would you give to novice programmers regarding the use of GenAI tools?                                                                                                      \\
    \bottomrule
  \end{tabularx}
  \label{tab:RQs-to-survey}
\end{table*}

\subsection{Survey Development} 

To address our research questions, we developed surveys to quantify the opinions and behaviors of a broad range of CS educators and professional programmers. Because CS educators have different responsibilities and may have different perceptions than professional programmers, we created two separate surveys: an educator survey and a developer survey.

To design the questionnaires, we collaborated within our group to draft survey questions for each overarching research question, including closed-ended, open-ended, and rating scale question types. As a part of this process, we reviewed the literature (e.\,g.,\,~\cite{prather2023wgfullreport}) and questions from existing surveys. However, few questions could be reused due to our focus on the rationale and actual integration of GenAI tools rather than expectations or anticipated use.

We then reviewed and edited these questions to increase clarity, relevance, neutrality, and specificity, following best practices from survey design~\cite{krosnick2018questionnaire}. When appropriate, questions were shared between the educator and developer surveys to allow us to compare responses between the two audiences. As part of our development process, we also piloted our survey with educators and developers to increase alignment with our target audiences.

\subsubsection{Educator Survey}
The educator survey was developed to address the two overall goals of the working group, and most of the research questions mentioned earlier (see Section \ref{sec:research-questions} and \autoref{tab:RQs-to-survey} for the precise mapping of survey questions and research questions).
As the previous year's working group had surveyed educators~\cite{prather2023wgfullreport}, we also drew from their survey questions where possible to facilitate potential comparisons. We expected educators to have a wide variety of views on the topic of GenAI and designed the survey to be multi-pathed based on their earlier responses. The full survey comprises 30 questions. It can be found in \autoref{app:appendix-edusurvey}. The precise mapping of survey questions to research questions is shown in~\autoref{tab:RQs-to-survey}.

Prior studies conducted in the very early days of LLMs emergence and general availability necessarily needed to focus on short- and long-term concerns of stakeholders, as many educators were unsure of how to proceed. Moreover, educators had broadly taken one of the two approaches to GenAI: either they do not use GenAI and disallow students from using it, or they actively include it in their teaching or at least allow students to use it~\cite{lau2023from}. There are still open questions, of course, but now the community has had more time to process the upheaval and more research to inform their teaching. As such, our survey shifts to focus on specific reasons, opportunities, and constraints for why educators and industry professionals use or do not use GenAI. To increase specificity and to capture a diverse range of perspectives, our educator survey also has separate branches for instructors that allowed or did not allow their students to use GenAI tools. We are also asking about concrete teaching approaches and assessments.

We are also particularly interested in the ways that learning to program, and programming in industry, are changing. To that end, we have asked several questions around whether educators believe that programming skills are shifting, how competencies (meaning knowledge, skills and dispositions in context of a task~\cite{raj2021professional}) have or have not changed, and which of them may become more or less relevant. The latter has been subject to discussion ever since GenAI tools have emerged~\cite{becker2023generative}. Another ongoing concern is around equitable access to these tools. As such, in our educator survey, we ask which tools students are expected to use and how they will access those tools (including whether there is a cost). 

\subsubsection{Developer Survey}
The developer survey was created for three purposes: (1) to capture an overarching view on how GenAI tools are being used in professional software development, (2) to capture developer viewpoints on how the tools are impacting their workflow, and (3) to capture developer viewpoints on the expected changes in competencies required to develop software. Another aspect we are addressing is whether and how developers have experienced issues regarding ethical concerns, or harm being done by the use of GenAI tools.

Gathering the developers' perspectives helps us develop a ground\-ed understanding of how software development is changing in the field. Moreover, we can compare their viewpoint with the educators' perspective. This enables a more realistic evaluation of the current state-of-the-art in industry, which may eventually contribute to adapting teaching practices and curricula. All 19 questions of the developer survey are available in \autoref{app:appendix-devsurvey}. It was designed to address both users and non-users. The mapping of survey questions to research questions is shown in~\autoref{tab:RQs-to-survey}.

\subsection{Distribution of the Surveys}

To recruit instructors, we sent emails to colleagues, public mailing lists, and private mailing lists. We also asked recipients to forward the survey links to peers to increase its distribution.
We sent emails that contained recruitment for the educators survey to the following groups and mailing lists:
\begin{itemize}
    \item SIG ``Computer Science Education'' (ACM)
    \item SIG ``Educational Technology'' of the German Computer Science society
    \item SIG ``Human-Computer-Interaction'' of the German Computer Science Society
    \item Participants at ITiCSE 2024 conference
    \item Contacts at Minority Serving Institutions (MSIs)
    \item A Slack group for teaching faculty who are predominantly in the United States
    \item Departmental mailing lists of the authors
    \item Colleagues of the authors
    \item LinkedIn networks of the authors
\end{itemize}

To recruit developers in industry, we sent emails to personal contacts, public and private mailing lists. We also asked recipients to forward the survey links to peers to increase its distribution.
We sent emails with the developer's survey link to the following mailing lists and social networks:
\begin{itemize}
    \item SIG ``Educational Technology'' of the German Computer Science society
    \item SIG ``Human-Computer-Interaction'' of the German Computer Science Society
    \item Personal contacts through online networks (e.\,g. LinkedIn)
    \item Miscellaneous company lists
    \item CS department alumni groups 
\end{itemize}
In addition, the German ``Fraunhofer Gesellschaft'', a non-profit organization for research and development, helped share the survey among their scientific staff and industrial partners. The same applies to the authors' network with the Leibniz Association---a connection of 96 research institutes in Germany.

\subsection{Survey Data Analysis}

The survey data were analyzed using a mixed-methods approach to accommodate both closed-ended and open-ended questions. Quantitative responses from closed-ended items were processed using descriptive statistics and statistical data visualizations. 

Qualitative data from open-ended questions underwent a thematic analysis~\cite{braun2012thematic}. Two members of the research team coded responses, identified recurring themes, and categorized them into broader conceptual groups. To improve reliability, the two re\-searchers independently coded a subset of responses, compared their coding schemes, and resolved discrepancies through discussion until reaching an agreement. 

For the educator survey, we received 209 responses (cutoff date July 29, 2024). However not all responses were used. We did not use the responses for which the participants did not agree to the consent form or which were incomplete (100 responses). In addition, we had some test data that were also not included (33 responses). Based on these criteria, we have \textit{N}\,=\,76 fully complete responses for the present analysis. 

For the developer survey, we received 94 responses (cutoff date 29 July 2024). Again, not all answers were used: we did not use the responses for which the participants did not agree to the consent form or which contained no data.
Based on these criteria, we analyze the resulting \textit{N}\,=\,39 response sequences, 29 of which contain responses to all survey questions (and are thus fully complete).



%
%
%
%

\subsection{Interviews Method}
Before conducting the interviews, we compiled a list of individuals who fit into one of the following three categories of interest:
\begin{itemize}
    \item \textbf{Tool creators:} This group comprises faculty, graduate students, software developers, and tech leads who design, build, and refine GenAI systems for educational use. These tools focus on enhancing coding efficiency, supporting educational needs, and integration into various computing environments. Insights from this group provide a critical understanding of the technical challenges and opportunities associated with the adoption of LLMs in computing education.
    
    \item \textbf{Educators studying GenAI:} Educators studying GenAI are primarily researchers and academic professionals who investigate these technologies' implications, efficacy, and educational potential. This group includes faculty members, educational researchers, and curriculum developers who are exploring how LLMs can be leveraged to enhance learning outcomes, transform teaching methodologies, and address the evolving needs of computing education. They examine the theoretical underpinnings, practical applications, and ethical considerations of using AI in educational settings.

    \item \textbf{Educators using GenAI:} Educators using GenAI are instructors actively incorporating LLMs into their teaching practices. This group spans faculty members from computer science and related disciplines, teaching both majors and non-majors. These educators use AI tools to facilitate learning, provide personalized support, and improve the overall educational experience. They bring practical perspectives on the benefits and challenges of integrating AI into the classroom, including its impact on student engagement, assessment, and skill development. Their experiences offer a practical understanding of how GenAI can be effectively implemented to enhance teaching and learning in computing education.
\end{itemize}

The interviewees have been active in the computing education community in at least one of the areas identified above. The research team completed 4, 5, and 7 interviews in each of the respective areas (in the order as listed above), and one interview with a thoughtful non-user. We additionally interviewed the thoughtful non-user to obtain broad perspectives.

We used the research questions outlined in \autoref{sec:educator-and-developer-views} to pursue our overall research goals. We used these research questions to guide our analysis of the transcripts.

\subsubsection{Interview Question Development} 
Developing interview questions was a collaborative and iterative process involving a team of four members. This team was responsible for designing a set of questions associated with each category of individuals being interviewed: tool creators, educators studying GenAI, and educators using GenAI. The primary aim was to ensure that the questions would elicit key insights from these individuals while addressing the main research questions of the study.

Initially, the team drafted a comprehensive list of questions for each category, focusing on capturing the unique perspectives and experiences of the interviewees. The questions were designed to explore various aspects of using LLMs and GenAI in computing education, including their development, implementation, and impact.

Once the initial set of questions was prepared, it was presented to the larger group of researchers for feedback. During this revision phase, several modifications were made to refine the questions further. Modifications include:
\begin{itemize}
    \item \textbf{Content Overlap:} Some questions were removed due to redundancy and overlap in content to ensure that each question addressed distinct aspects of the research.
    \item \textbf{Duration Management:} To keep the interviews within a manageable duration of one hour, certain questions were cut. This decision was made to respect the interviewees' time and maintain the focus and depth of the discussions.
    \item \textbf{Improving Quality and Directedness:} The wording of several questions was revised to enhance clarity, specificity, and directedness. This helped ensure the questions were straightforward and elicited detailed, relevant responses.
    \item \textbf{Adding Follow-Up Questions:} To extract more context and deeper insights, follow-up questions such as ``Why?'' and ``Can you elaborate more?'' were incorporated. These prompts encouraged interviewees to provide more detailed explanations and examples.
\end{itemize}

\subsection{Interview Process}
The interview process was designed and implemented to gain comprehensive insights. To ensure a diverse and representative sample from each category, we selected individuals for an interview based on the results of our literature review, i.\,e., who is writing about this topic from the perspectives of educators using GenAI in their classes or researching GenAI in computing or creating GenAI tools for computing classrooms. A total of \textbf{17} individuals were selected to interview, with 4, 5, and 7 individuals from each of the previously identified categories, respectively. The one remaining participant could be classified as a thoughtful non-user because we felt it important to hear from someone who was intentionally and thoughtfully trying to avoid using it in their computing classroom. These participants were invited to participate in one-on-one interviews, and they accepted and signed the consent form.

\subsubsection{Conducting Interviews}
A single team member conducted each interview via Zoom. The scheduling of these interviews ensured that each session lasted for no more than one hour to adhere strictly to the allocated time frame. Using Zoom facilitated a convenient and efficient way to conduct these interviews remotely.

\subsubsection{Recording and Transcription}
All Zoom sessions were recorded with participant permission to create accurate transcriptions. To comply with Institutional Review Board (IRB) protocol requirements, these recordings were deleted once the transcriptions were completed. We used secure transcription services to create the transcripts and cleaned them by removing typographic errors.

\subsubsection{Interview Structure}
The interview questions were posed to each interviewee sequentially. It is important to note that these questions were not shared with the participants prior to the interview. This approach aimed to elicit spontaneous and genuine responses to provide more authentic and valuable data for the study.

\subsection{Interview Data Analysis}
This section describes the process used to analyze the interview transcripts. The analysis was conducted systematically, involving multiple team members to ensure consistency and reliability.

\subsubsection{Initial Review and Consensus Building}
Three team members jointly reviewed one transcript to establish a consistent tagging methodology. During this review, they tagged sections (which could be one sentence, multiple sentences, one paragraph, or multiple paragraphs) with relevant categories. The predefined categories used for tagging were: Instruction-use, Instruction-teach, Tools, Policies, Motivations, Competencies, Outcomes, Future, Industry, and Equity, 
aligned with the research questions described in \autoref{sec:research-questions}. This collaborative step ensured that all team members were aligned on the tagging criteria and approach.

\subsubsection{Individual Tagging and Data Compilation}
After reaching a consensus on tagging methodology, the remainder of the transcripts were independently tagged by one researcher, 
who was not the interviewer. This approach helped maintain objectivity and consistency in the tagging process. The tagged segments from all transcripts were then compiled into an Excel sheet. This compilation step included all identified segments categorized by theme.

Each segment entry in the Excel sheet was summarized into one or a few sentences. This step was used to capture the essence of the responses without losing the context or the main points expressed by the interviewees.

\subsubsection{Thematic Analysis} With the summarized segments, we conducted a thematic analysis across all interviews. This analysis involved 1) grouping summaries by category, 2) identifying patterns and themes, and 3) synthesizing community thoughts.

Through this process, we aim to derive meaningful conclusions from the interview data and understand the community's views on the various aspects covered in the interviews.

%
%

\section{Educator and Developer Views: Results}
\label{sec:educator-and-developer-views_results}

We now present the results of our surveys and interviews.
We begin with the participants' demographics from both surveys and the interviews. Following this, we present the aggregated survey and interview results for each research question.

\subsection{Demographics of the Samples}

First, we present the characteristics of the surveyed educators and developers by providing details on their demographics. We also present the demographics of the interviewed educators.

\subsubsection{Educator Survey Demographics}

In the educator survey (N\,=\,76), educators from the USA~(28), Germany~(13), Canada~(11), UK~(5), and more countries participated, as shown in \autoref{tab:educator-country}. Overall, we gathered educators' perspectives from North America, Europe, and Australia. However, 6 respondents did not indicate the country of their institution. 

\begin{table}[htb]
  \centering
  \caption{Educator Survey -- Country Institution is Located}
  \label{tab:educator-country}
  \small
  \begin{tabular}{p{2in}r}
    \toprule
    Country & Percent Respondents \\
    \midrule
    USA & 36.8\,\% \\
    Germany & 17.1\,\% \\
    Canada & 14.5\,\% \\
    UK & 6.6\,\% \\
    Sweden & 5.3\,\% \\  
    Australia &	1.3\,\% \\
    Finland	 & 1.3\,\% \\
    France	 &	1.3\,\% \\
    Iceland	 &	1.3\,\% \\
    Ireland	 &	1.3\,\% \\
    Netherlands	 &1.3\,\% \\
    Poland	 &1.3\,\% \\
    Spain	 &	1.3\,\% \\
    Switzerland	 & 1.3\,\% \\
    Ukraine	 &	1.3\,\% \\
    No country given &	6.6\,\% \\
   
    \bottomrule
\end{tabular}
\end{table}

31.6\,\% (24 out of 76) of the educators identify themselves as females, 57.9\,\% (45 out of 76) as males, 2.6\,\% (2 out of 76) as non-binary or gender diverse, and 5.3\,\% (4 out of 76) prefer not to disclose (ES-29). 
The majority of the educator respondents described their institutions (ES-24) as a university that grants graduate degrees  (52 out of 76), and the others teach at colleges or other types of institutions, as shown in \autoref{tab:educator-institution}. 

\begin{table}[htb]
  \centering
  \caption{Educator Survey -- Institutional Characteristics (ES-24)}
  \label{tab:educator-institution}
  \small
  \begin{tabular}{p{2in}r}
    \toprule
    Institution & Percent Respondents \\
    \midrule
    Secondary  & 11.84\,\% \\
    2-year college (associates) & 1.3\,\% \\
    Vocational school & 2.6\,\% \\
    College (bachelor's degree granting) & 11.8\,\% \\
    University (graduate degree granting) & 68.4\,\% \\
    Other (research institutes) & 2.6\,\% \\
    No institution given & 1.3\,\%\\
    \bottomrule
\end{tabular}
\end{table}


Most of the educators teach \textit{CS1 -- Introduction to programming} (50.6\,\%, 39 out of 76; cf. \autoref{tab:educator-area}), which is one of the courses we assume would be most influenced by the abilities and potential of GenAI. At the same time, it should be noted that the respondents teach various other crucial computing courses, a total of 193 classes as displayed in  \autoref{tab:educator-area}. Among the courses listed as ``other'' are the following: Data Science (5 responses), Data Base Systems (3 responses), Computing Education, and Web Technologies and Development (2 responses each). The diversity of courses also reflects upon the diversity of the responding educators. 

\begin{table}[htb]
  \centering
  \caption{Educator Survey -- Course Characteristics (ES-12); 193 responses from 76 educators}
  \label{tab:educator-area}
  \small
  \begin{tabular}{p{2in}r}
    \toprule
    Primary area & Percent Responses \\
    \midrule
    Algorithms and Complexity & 7.3\,\%\\
    Architecture and Organization & 2.6\,\% \\
    Artificial Intelligence/ML & 5.2\,\% \\
    Computational Science & 2.6\,\% \\
    CS 1 -- Introduction to Programming & 20.2\,\% \\
    CS 2 -- Introduction to Data Structures & 10.4\,\% \\
    Discrete Structures & 1.6\,\% \\
    Human-Computer Interaction & 5.7\,\% \\
    Information Assurance and Security & 0.5\,\% \\
    Graphics and Visualization & 3.6\,\% \\
    Information Management & 2.1\.\% \\ 
    Networking and Communications & 1.6\,\% \\
    Operating Systems & 2.6\,\% \\
    Parallel and Distributed Computing & 1.6\,\% \\
    Platform-based Development & 0.5\,\% \\
    Programming Languages & 4.7\,\% \\
    Robotics & 0.5\,\% \\
    Social Issues and Professional Practice & 2.1\,\% \\
    Software Development Fundamentals & 5.7\,\% \\
    Software Engineering & 7.3\,\% \\
    Systems Fundamentals & 1.6\,\% \\
    Teacher Preparation (age 5--18) & 2.1\,\% \\
    Other & 8.3\,\% \\
    \bottomrule
\end{tabular}
\end{table}


We furthermore asked the educators how many years they have been teaching (ES-28). Most of the surveyed educators proved to have more than a decade of teaching experience, as the average number of years is 15.8, with a median of 14.5\,years (cf. \autoref{tab:educator-years}). The largest groups of educators have been teaching for 11 to 15 years (16 out of 76) and from 16 to 25 years (20 out of 76), so they can be described as experienced on average.

\begin{table}[h]
  \centering
  \caption{Educator Survey -- Years of Teaching (ES-28)}
  \label{tab:educator-years}
  \small
  \begin{tabular}{p{1.2in}r r}
    \toprule
    Years & Number & Percent Respondents \\
    \midrule
    0--2 & 7 & 9.2\,\% \\
    3--5 & 9 & 11.8\,\% \\
    6--10 & 14 & 18.4\,\% \\
    11--15 & 16 & 21.1\,\% \\
    16--25 & 21 & 27.6\,\% \\
    >\,25 & 8 & 10.5\,\% \\
    no answer & 1 & 1.3\,\% \\
    \bottomrule
\end{tabular}
\end{table}

\subsubsection{Developer Survey Demographics}
Based on our selection process, \textit{N}\,=\,39 developers provided reasonably relevant responses to our survey. As mentioned, 29 of them provided answers to all questions (i.\,e., finished the survey). In this section, we present the results of survey questions DS-17--19 (in the developer survey) to characterize our sample. 

18 developers provided the name of the country in which they are currently employed (DS-17). 
Most developers who participated are employed in the United States (66.7\,\%, 12 out of 18), or Germany (22.2\,\%, 4 out of 18). Others work in France (5.6\,\%, 1 out of 18), or the United Kingdom (5.6\,\%, 1 out of 18).

Moreover, we asked the developers about their job title (DS-18, n=23). 
Most developers identified their job title as \textit{software developer} (70\,\%, 16 out of 23) followed by \textit{research engineers} (13\,\%, 3 out of 23), \textit{(scientific) researchers} (13\,\%, 3 out of 23), and a \textit{software engineer} (4\,\%, 1 out of 23).

The companies of the developers (DS-19, n=27) can be characterized as summarized in \autoref{tab:developer-institution}. 
The largest percentage of developers (59.3\,\%, 16 out of 27) work in a large company with more than 500 employees. A smaller number of developers work at a research institute (11.1\,\%, 3 out of 27), a small or medium software company (7.4\,\% each, 2 out of 27), a non-software focused company (7.4\,\%, 2 out of 27), or for the government (3.7\,\%, 1 out of 27). One respondent (3.7\,\%) selected the option ``other'' without specifying the type of company any further. None of the participants are employed in a start-up or non-profit organization. 

\begin{table}[htb]
  \centering
  \caption{Developer Survey -- Company Characteristics}
  \label{tab:developer-institution}
  \small
  \begin{tabular}{p{2.4in}r}
    \toprule
    Corporation Type & Percent Resp. \\
    \midrule
    Start-up (10 engineers or less) & 0 \\
    Small Software Company (11--50 engineers)  & 7.4 \\ 
    Medium Software Company (51--500 engineers) & 7.4 \\ 
    Large Software Company (more than 500 engineers) & 59.3 \\ 
    Non-profit & 0 \\
    Non-software focused company & 7.4 \\ 
    Government & 3.7 \\ 
    Research Institute & 11.1\\ 
    Other (not specified) & 3.7 \\ 
    \bottomrule
\end{tabular}
\end{table}

\subsubsection{Interview Sample Demographics}
The authors initially built a list of potential interviewees based on personal knowledge of who was actively teaching with AI, publishing research on its use in computing education classrooms, or building tools for use in computing education. We then augmented that list based on the results of the literature review, personal networking at the ITiCSE 2024 conference, and an email to a large mailing list. Next, we selected people from the list to interview based on what would provide a diverse set of experiences and ideas in terms of interviewee position, seniority, nationality, gender, and race. Not everyone agreed to be interviewed, so we made some substitutions with others on the list that were not initially selected. Finally, we also intentionally sought out some interviews from people who are less likely to be selected, such as K-12 teachers, professors who teach computing outside of computer science, and thoughtful non-users who have intentionally decided to resist generative AI in all forms. 

\subsection{RQ1 -- Policies}
\label{sec:rq1-policies}

This research question was addressed by the survey for educators, as well as the interview study. Policies applicable for software developers were also gathered through the developer survey. 

\subsubsection{Educator Policies preventing, tolerating, or integrating GenAI}
In the survey study, we asked educators about their classroom policies regarding the use of GenAI tools by students (ES-1, n\,=\,76).
The majority of faculty do not explicitly disallow students from using GenAI tools (59 out of 76, 77.6\,\%). Roughly a third of faculty actively incorporate GenAI into their courses (ES-2; 27 out of 76, 35.5\,\%). This is particularly interesting as the majority of faculty (57 out of 76, 75\,\%) report believing the skills to create software have changed after the advent of GenAI (ES-3, see Section~\ref{sec:competencies} for more details on how the competencies have changed). Thus, incorporating GenAI seems to be lagging behind the percentage of faculty who think that skills have changed.

Among faculty who are disallowing students from using GenAI, we learned from ES-10 (``Are you doing anything to prevent GenAI tools' use in your course?'') that 58.8\,\% (10 out of 17) are working to prevent their use. 

Among those faculty trying to prevent the use of GenAI, they reported, in response to ES-11, (``What are you doing to prevent GenAI tools' use in your course?'') on different actions they had taken. For example, 36.0\,\% (9 out of 25) reported that, together with the students, they carry out code reviews in various forms on assignments and tests, to see whether the students have understood their own code or whether typical AI markers can be found.
Another 28.0\,\% (7 out of 25) of the responses describe changes they have made to assignments and exams. 
Another point mentioned several times is the appeal to the students' social responsibility (20.0\,\%; 5 out of 25). 
Technical measures and the introduction of workflows to be documented were rarely mentioned (8.0\,\%; 2 each).

The following quote in response to ES-11 is an example of a detailed explanation of an educator regarding their attempts to prevent students from using GenAI tools in their courses:

\begin{quote}
``\textit{We tell the students it's not allowed, and make sure they see that by putting a question on our course rules quiz. We, for now, have a few patterns we look for that are typical of AI code but which we don't teach in our class that we automatically scan for as students submit assignments. When we see these we do a human check of whether we think the code was probably written by an AI or not. Last semester, in 13/14 cases, accused students admitted they used AI (the other person learned some extra concepts from alternate resources and was able to demonstrate understanding of them). For next semester, we're also introducing a citation policy: If students use stuff they learned outside of our teaching resources, they are required to cite where they learned it. This is good practice in writing code anyway, but it also gives us an extra way to distinguish AI coding from people using third-party resources legitimately.}''
\end{quote}



Based on responses to question ES-2 (``Are you incorporating GenAI tools (e.\,g., actively integrating it into the curriculum or exercises) into your recent courses (within the last 12 months)?'') we found that 35.5\,\% (27 out of 76) of instructors are integrating GenAI into their courses. 
In the follow-up question ES-21, we asked those educators to elaborate on how their teaching approaches have changed in an open-question format.
Two members of the working group performed a thematic analysis of the 60 open-ended responses. The following five themes summarize how educators changed their \textbf{teaching approaches}:

\begin{enumerate}[leftmargin=*]
    \item \textbf{Focus on different programming skills}: This theme captured responses from the educators who have refocused the skills that they teach. Some respondents refer to existing skills that are nonetheless now more important than before, such as reading code (8 respondents), modifying code (1), testing code (1), and problem decomposition (2). Others refer to entirely new skills that did not exist prior to the use of GenAI, such as writing prompts (4) and incorporating code from GenAI into larger projects (1). Still others refer to a lessening importance of existing skills; for example, that syntax is now less important than before and that the focus should be on a higher level of abstraction (3).
    \item \textbf{Policy on when and how to use GenAI}. Respective respondents report setting expectations to students of tasks where they should use GenAI and when they should not (7 participants). In addition, if students are allowed to use these tools, these respondents require that students describe how they used the tools including clear attribution to GenAI (3), e.\,g., ``\textit{I actively encourage students to use GenAI tools and set expectations where they should be used and where they should not. I have created a policy that when in doubt students are allowed to use the tools but they have to clearly and fully describe the use.}''
    \item \textbf{Change lecture to adapt to GenAI}. Participants report demonstrating working with GenAI in class (12), incorporating GenAI into in-class activities (2), using more active learning in class (2), and showing non-deterministic responses from the GenAI in class (1).
    \item \textbf{Teach students to use GenAI effectively}. Some participants reported encouraging students to use GenAI (2) and others reported trying to teach students how to be more effective when working with GenAI beyond the demonstrations in class (2). The following quote illustrates: ``\textit{Focusing on teaching students prompt design techniques like tree of thoughts, etc. that utilize the underlying architecture of transformer models.}''
    \item \textbf{Not yet / need to at some point}. Some participants responded that they had not changed in class activities yet but recognized they may need to in the future (8 participants). ``\textit{None, so far.  I think their negative impact on a course is, at this point, less than other types of illicit support that students might seek}'', ``\textit{Not much at the moment, but I feel I need to do more here.}''
\end{enumerate}



Yet another follow-up question for educators integrating GenAI asked how assessment approaches have changed as a result of GenAI tools (ES-22). Again, two members of the working group performed a thematic analysis of the 63 open-ended responses to question ES-22 and any responses to question ES-21 that related to changes in assessments. We uncovered six themes representing \textbf{how assessment approaches have changed}:

\begin{enumerate}[leftmargin=*]
    \item \textbf{Process over product}. Six participants described assessing students on their process of creating software more than on the correctness of the final result. One participant is ``\textit{... shifting things to watching the process (ongoing use of repos, iteration, trackable code growth, etc.) vs final artifact evaluation}''.  Another focuses on ``\textit{More incremental checkpoints and questioning the process before completion... the end work is less weighted and the process towards the solution is more emphasized.}''
    \item \textbf{Invigilated exams}. Thirteen participants reported placing more emphasis on the proctoring of assessments. Six described more emphasis on proctoring, for example, ``\textit{programming assignments are now preparation for the proctored weekly quizzes.}'' Three participants spoke of the importance of giving exams on paper.  Seven spoke of adding or placing more emphasis on oral exams: ``\textit{Oral exams are now a vital part of student assessment.}'' 
    \item \textbf{Increased weighting on exams}. Ten participants decreased the weight of unproctored assessments and increased the weight of proctored exams in assigning grades. One participant wrote, ``\textit{Minimized unsupervised assessment weight (e.\,g., assignments and projects) ... Increased supervised assessment weight (e.\,g., tests and exams).}''  One participant described instituting ``\textit{a minimum cutoff for exams}''.
    \item \textbf{No change (yet)}. Nine participants said they had not made any changes with an additional six participants saying they had not made changes yet but had plans to in the future.
    \item \textbf{Confuse the LLM}.  Five participants chose to create assignments where the questions were designed to make the LLMs do poorly.  One participant wrote, ``\textit{Homework assignments have been customized so that LLMs can't solve them well.}''
    \item \textbf{Describe AI use}. Three participants are asking students to describe how they use GenAI on their assignments and provide chat logs with the LLM, as exemplified in the following quote: ``\textit{We require students to disclose if they used GenAI (as part of any outside help), and if they do, they are required to provide the prompts that they used to get their solution.}''
\end{enumerate}

In addition, interview results revealed concerns of educators and researchers around AI policy. As GenAI tools are appearing more and more in the classroom, both institutions and individual instructors are adding policies around GenAI use. One concern repeated throughout the data was one of user privacy. There is the concern that students' private data will be leaked through the use of publicly available AI models or that it may be stored in a country that has different data privacy laws than the institution's host country. Some also expressed concern that these models will learn off of private course assignments and teaching materials. Then there are academic concerns, which primarily take three forms. First, some institutions will attempt to ban or limit their use. This was particularly apparent at the K-12 level where many schools and school districts have set policies in place to ban AI tools like ChatGPT entirely. Second, some individual instructors are adding policies to their courses either banning or limiting the use of generative AI tools. Third, some instructors are allowing or even encouraging generative AI usage, but requiring students to use specific tools that will not reveal the answer or requiring that a student must be able to explain the code that they submit. Sometimes these tools are built by the university and housed locally, which also addresses the privacy concerns. 

Thinking about privacy policies and its impact on instruction, one instructor said:

\begin{quote}
``\textit{I know at our university we are concerned about [privacy], and are trying to get certain tools with basically a university wide license for our students. I think we were trying to get a university-wide license for [GitHub] Copilot.}''
\end{quote}

Speaking on their own course policies, another instructor said:
\begin{quote}
``\textit{The bottom line is you need to demonstrate that you understand the material and anything that you turn in. It needs to be something that you know you've created and can fully understand what's going on}''
\end{quote}

One K-12 teacher said:
\begin{quote}
``\textit{One of the major changes that's gonna happen is that [anonymous state] has a whole bunch of new policy regulation guidance and support around generative AI and schools and teachers are expected to follow that. Now, teachers are going to have to know about those policy pieces, and we're going to have to figure out how to include that in the learning that they're doing on top of all of the other policy work that they're doing. In addition to just the classroom work, [teachers] now have to also be aware of all of this extra policy and regulation about it, too.}''
\end{quote}

\subsubsection{Industry Policies}
Eight developers stated that they do not use GenAI tools (cf. \autoref{sec:rq7}). Regarding industry policies around this non-use (subset of DS-14 and DS-15),
one developer answered that ``\textit{it's currently not allowed to use GenAI tools but we expect it will be in the near future}''. Interestingly, none of the other replies concerned their company's policies.

\subsection{RQ2 and RQ3 -- Instruction-use and Instruction-support}
\label{sec:rq2rq3}

The second and third research question pertain to teaching students how to use GenAI tools effectively, and how to support them doing so. For example, students may be taught explicitly how to use GenAI to help them write software. Of the educators who participated in our survey, 27 were actively incorporating GenAI into their courses. This section reports on these instructors and their replies, before reporting on the educators who were interviewed.


Educators incorporating GenAI teach a variety of courses (ES-12).
Most of the surveyed instructors either taught introduction to programming courses (11 out of 27, 40.7\%) or software engineering courses (9 out of 27, 33.3\%). The remaining instructors (7 out of 27, 25.9\%) taught more advanced courses such as databases, data visualization, and artificial intelligence. 

\begin{figure}[htb]
    \centering
    \includegraphics[width=0.8\linewidth]{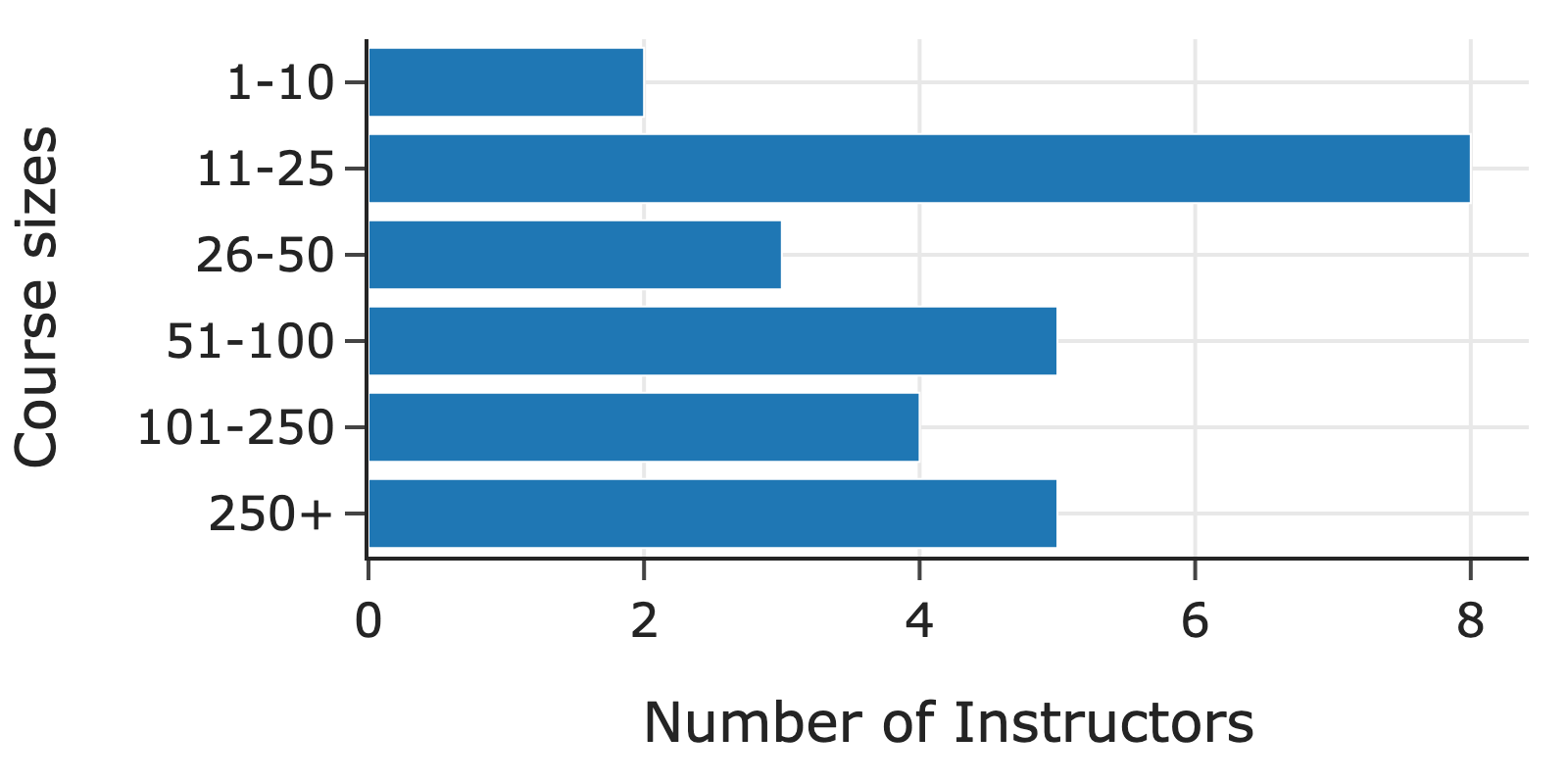}
    \caption{Instructors integrated GenAI into a wide variety of course sizes (ES-13).}
    \label{fig:q13}
\end{figure}

Course sizes varied widely (ES-13), as shown in \autoref{fig:q13}---half the courses were 50 students or smaller (13 out of 27, 48.1\%), while the other half were larger than 50 students (14 out of 27, 51.9\%). 

The large majority of instructors who incorporated GenAI into their courses reported that both instructors and students were expected to use GenAI for their courses (20 out of 27, 74.1\%), as summarized in \autoref{tab:educator-user} (ES-14).
However, students using the tools does not necessarily mean they are being trained in how to use them, as the instructor may have used the GenAI tools as a tutor.

\begin{table}[htb]
  \centering
  \caption{Most instructors who integrated GenAI expected that both instructors and students would use GenAI (ES-14).}
  \label{tab:educator-user}
  \small
  \begin{tabularx}{\linewidth}{X r}
    \toprule
    Who uses GenAI & Number of Respondents \\
    \midrule
    instructors only & 3\\
    students only & 4 \\
    instructors and students & 20 \\
    \bottomrule
\end{tabularx}
\end{table}

\begin{figure}[htb]
    \centering \includegraphics[width=0.9\linewidth,trim={0 0 0 2.7cm},clip]{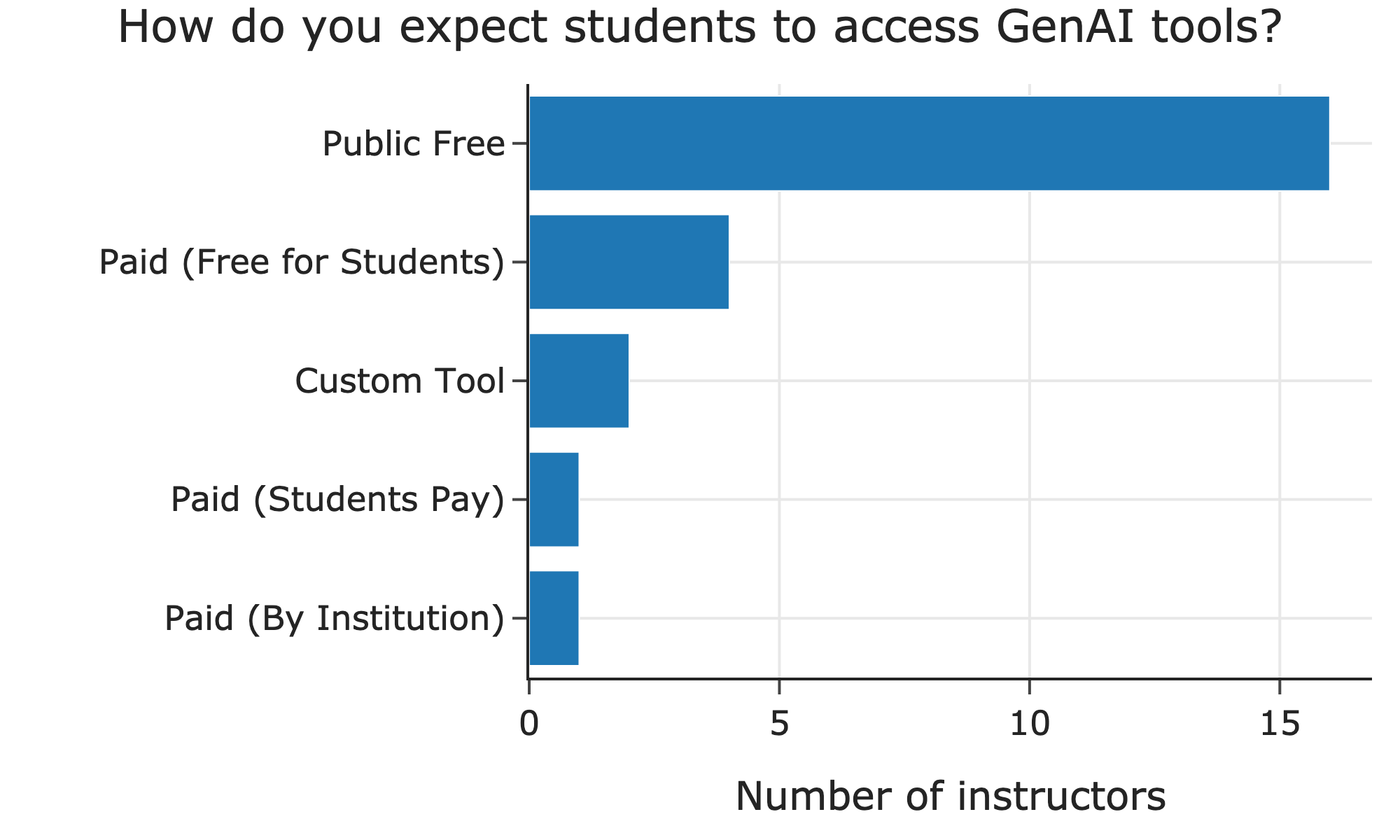}
    \caption{Instructors preferred to use tools that did not require students to pay (ES-15).}
    \label{fig:q15}
\end{figure}

To explore classroom use of GenAI in more depth, we asked instructors how they expect students to access GenAI tools (ES-15). As shown in \autoref{fig:q15}, when students were expected to use GenAI tools, instructors incorporated a tool that was free to use, like a free publicly available tool (16 out of 27, 59.3\%) or a paid tool that allowed students to use it for free (4 out of 27, 14.8\%).

In response to a question asking which types of GenAI tools instructors incorporated (ES-16), all educators reported that they used standard industry GenAI tools like ChatGPT and Copilot (27 out of 27, 100\%), while a few instructors also used customized tools that the instructors had created themselves (2 out of 27, 7.4\%).

Question ES-17 provided additional insight, finding instructors primarily incorporated GenAI tools to teach students about using GenAI tools (20 out of 27, 74.1\%) and as an educational content generator for teaching material (16 out of 27, 63.0\%). Other educators used GenAI for feedback, correcting student work, validating quality of assignments, and to support grading, as shown in \autoref{tab:educator-useInCourse}.

\begin{table}[htb]
    \centering
    \caption{Ways in which instructors incorporated GenAI (ES-17).}
    \begin{tabularx}{\linewidth}{X r}
        \toprule
        \textbf{Answer} & \textbf{Count} \\
        \midrule
        to teach students about using GenAI tools & 20 \\
        as educational content generator for teaching material & 16 \\
        to automatically provide feedback to students using a custom tool & 5 \\
        to support the correction of student work & 4 \\
        to validate the quality of assignments & 3 \\
        to support grading & 2 \\
        \bottomrule
    \end{tabularx}
  \label{tab:educator-useInCourse}
\end{table}

The interviews revealed that while many instructors are using generative AI to provide feedback to student submissions or increase access to student help-seeking, many are still not directly using it in class to support their teaching. What we did find still seems rather rudimentary and is one piece of this that we expect will continue to evolve. First, some instructors are using generative AI to help students understand bias in the models and planned in-class activities around it. Others will use it to generate code or proofs and then have students attempt to pick it apart, determining if it is correct or incorrect, and why. For example, one instructor detailed an in-class activity on generating code and then having students write tests as a first homework assignment. Finally, several educators mentioned using generative AI coding tools in upper-division courses or courses with heavy software development tasks as a way to prepare students for industry.

Regarding bias in the models, one instructor said:
\begin{quote}
    ``\textit{And then these models can perpetuate bias because they're trained on data that's biased. And we try to look at some examples of that. It does seem like students are a little bit surprised about some of these bias issues. I think it's important to talk about how these tools can have bias in them.}''
\end{quote}

When talking about an upper division course that an instructor teaches, they said: 
\begin{quote}
    ``\textit{So for upper division classes, where the point is coding as a professional would: Hell yeah, like we'll [use it for] everything. It'll be Gen AI from beginning to end.}''
\end{quote}

Regarding using it for large software development project courses, one instructor said:
\begin{quote}
     ``\textit{I'm about to teach a software development course that I've taught a lot before. Well, AI and generative AI is a big part of software development. So I'm going to be working on teaching generative AI as a tool, as part of the software development process as well. And so working with students on: okay, what can it do? How do we use it effectively and integrate it as a tool? Just like we have IDEs and debuggers and linters and everything else.}''
\end{quote}






In the interviews, the participants were predominantly mentioning using the GenAI tools to directly interface with the students and help them with their work. A common theme was tweaking such tools as to provide meaningful learning experience.

\begin{quote}
``\textit{I added the functionality of having an instructor be able to specify keywords or concepts that we didn't want to have show up in a response to be able to kind of better tailor the responses for the courses.}''
\end{quote}

\noindent Another theme was using the GenAI tools in specific contexts to support students while not interfering with their learning.

\begin{quote}
``\textit{So it is something we've built into an office hour queuing system. So we have a website students log into when they want to join an office hour. Just join the office hour queue and get TA support. And when the students type in what their question is, they're gonna ask the TA here, or why they're in office hours, then to our tool and it basically takes and adds some context to the question. And says this question is from a student with this level of knowledge, and who knows these languages and so on. And give them an answer, and don't provide a complete working program.}''
\end{quote}

\noindent Finally, the participants often mention they would use the GenAI tools to create learning materials or brainstorm lesson plans.

\begin{quote}
``\textit{And so, we've used it, me and [CS teacher] have used it. And so is our other teacher a little bit. I think we're probably the only ones, but it's it's exclusively to help us with our lesson planning. It's phenomenal.}''
\end{quote}

\subsection{RQ4-Motivations}
To better understand the educators' motivations behind their decision to prevent, tolerate or actively integrate GenAI tools, we asked three questions in the respective survey:
\begin{itemize}
    \item ES-9 -- Why don't you allow GenAI tools in your courses?
    \item ES-18 -- Why have you incorporated GenAI tools into your recent course? 
    \item ES-20 -- Why have you not incorporated GenAI tools (e.\,g., actively integrating it into the curriculum or exercises) into your recent courses? 
\end{itemize}


The reasons for not allowing the use of GenAI are very diverse (ES-9): 
39.3\,\% (11 out of 28) of the statements are about didactic aspects and curricular approaches, with a common theme being that students should first acquire a solid foundation in programming. 
A further 21.4\,\% (6 out of 28) name tasks or topics in which students can use GenAI and how they should document their use. 
17.9\,\% (5 out of 28) of the statements express an attitude towards the students: students would misuse GenAI, and they do not (yet) have the skills to use GenAI or to understand and interpret the results. 
It is also interesting to note some statements (10.7\,\%; 3 out of 28) reflecting negativity towards GenAI in general; for example, that GenAI is not capable of solving the course assignments anyway. Another concern is that ``\textit{GenAI will almost certainly be criminalized (as stealing code)}'', or that only humans should write code, not an AI.
Organizational and legal aspects were mentioned in only 7.1\,\% (2 out of 28) of the statements.

The following quote from the survey summarizes a general view of these educators on the use of GenAI tools in their courses:
\begin{quote}
``\textit{Introductory CS students need to learn basic skills before they can move on. It is a similar situation to calculators in K-12, where usage is not generally permitted in the primary grades where basic skills and number sense are being taught.}''
\end{quote}

Educators also provided their motivation and rationale for incorporating GenAI tools into their recent courses (ES-18), which asked: ``Why have you incorporated GenAI tools into your recent course?''. 24 open-ended responses were collected, and a thematic analysis by two of the authors revealed the following reasons:

\begin{enumerate}[leftmargin=*]
    \item \textbf{Preparation for industry and career readiness}. Some educators believe students will use GenAI tools in their  career, so it is a matter of career-readiness to successfully use these tools: ``\textit{I believe that nearly everyone will use AI assistants like Copilot and ChatGPT to program in the future. I want to prepare my students for their careers and hence wish to train them in how to use those tools.}'' Another related response is: ``\textit{In work life, students will use AI, too. So, we try to work under `normal' work conditions in the course.}''
    \item \textbf{Ethical use and responsibility}. Another reason educators mentioned was their responsibility as a teacher. Due to the sheer potential and assumed impact of GenAI, but also its bias and challenges, educators think it is responsible to teach about GenAI:  ``\textit{Because to not do so would be irresponsible, and because GenAI engages students.}''. Another educator highlights respective concerns: ``\textit{I have observed the technology's usage with disastrous learning outcomes, easily shown in a few iterations of two different courses where students did very well on homework that is code-oriented and were unable to do even the most basic things on the same topic in written or oral context.}''
    \item \textbf{Efficiently supporting student learning}. In some cases, educators use LLMs to compensate for a lack of teaching resources, or to generate teaching and learning material. For example, ``\textit{For scalability reasons (my teaching team is too small for the large quantity of students we have). To support students in their moment of need and better support their learning.}'' Several educators mentioned how they themselves use GenAI: ``\textit{It makes me faster at generating materials, like brainstorming the setup for exam questions.}''
    \item \textbf{Adapting to recent technological trends}. Another motivating factor for educators to use GenAI tools was to educate themselves about recent technologies. For example, ``\textit{I wanted to get a personal sense of the capabilities of models, and to then showcase effective ways to use it in the class should students elect to use it.}''
    For courses with an AI focus, integrating GenAI was a given: ``\textit{I teach an introduction to artificial intelligence. Not using GenAI tools would be a great disservice to the students.}''
\end{enumerate}

We were also interested in why educators do not incorporate GenAI into their courses (ES-20). We identified five areas of underlying motivations within 55 statements which we categorized and summarized. 
\begin{enumerate}[leftmargin=*]
    \item \textbf{Lack of educator resources}. 24 of the statements were related to the educators.  
    Educators cited a lack of time, lack of skills, or lack of didactic competency in particular. 
    \item \textbf{Students' motivation and abilities}. Another important aspect mentioned by 10 educators was students' perceived skills and motivation. 
    For example, educators assume that students already know and use GenAI, that they teach themselves how to use it anyway -- but also that students lack the skills to use it properly. 
    \item \textbf{Doubts regarding usefulness}. Attitudes towards GenAI also appear to be a motivating factor in 8 responses. 
    Educators are cautious about the technology and its benefits, especially in basic programming education. 
    \item \textbf{External factors}. Institutional factors such as a lack of support, unresolved legal issues or privacy concerns were also mentioned in 6 responses. 
    \item \textbf{Other reasons}. The remaining 7 statements 
    concerned general teaching practice issues, such as not teaching a suitable course at the moment or explaining the use of GenAI to students but not integrating the tools further. 
\end{enumerate}


In the interviews, the participants often stressed the importance of designing and providing solutions that focus on education. A common concern was that general tools are focused on providing complete solutions as opposed to helping students learn.

\begin{quote}
``\textit{The tools are really good at giving complete working program functions. But what we want with it, with a student who has a code problem is some kind of hint to get them going and get them unstuck. So a lot of times the tools give too much help.}''
\end{quote}

\noindent Another motivation for solutions that are educationally-focused is centered around potential over-reliance of students on these always available assistants.

\begin{quote}
``\textit{they can become very reliant on just asking questions as soon as they get stuck. And I think the idea of being able to guide students much in the way that a human teaching assistant would. I mean, if you go and ask a TA for help. They would typically try and use a kind of Socratic method to maybe ask a question back at you to guide you to a solution rather than just giving you the answer.}''
\end{quote}

Another common theme often mentioned by participants was that GenAI tools often provide a significant help to the instructors, easing their work load and allowing them to focus their attention where it is needed the most.

\begin{quote}
``\textit{The idea was that basically I get another teaching assistant, another person in the room that can give the students tailored feedback. I do think there's a small subset of students that actually prefer getting feedback from the AI. Another goal of mine has been to increase the breadth of students that are getting through introductory CS courses.}''
\end{quote}

A very important motivation for using GenAI in computing education often mentioned by the participants was providing immediate (timely) help to the students. This would otherwise not be possible in many contexts.

\begin{quote}
``\textit{in the age of AI for me philosophically, and technologically, this is strictly a net positive, because we now have the ability to provide students with far more real time and iterative feedback, not only for our undergraduates on campus at Harvard and Yale but open courseware audience around the world, who never had access to TA's office hour sections, and that human support structure, and they still don't have a human support structure, but arguably an increasingly good approximation thereof.}''
\end{quote}

On a related note, another person said,

\begin{quote}
``\textit{[We] leverage GenAI to provide help at scale, which is one of the things we're always trying to do and modify how we teach computer science to better help more people succeed. That's always my goal. Help more people. My diversity of people succeed in computing and find interest in it, you know. Use it to do interesting things.}''
\end{quote}

Some participants explained their motivation along the lines of students needing to encounter GenAI tools during their studies because they will be using them in practice.

\begin{quote}
``\textit{I'm kind of pushing them hard to adopt the view that it's almost academic malpractice to not teach people how to write with these things, because once you're out in practice you're going to have access to these and your competitor [...] will have access to these things. And so you need to know how to use them responsibly, how to best use them. How to check them for errors like, what types of things can go wrong with this, etc, etc.}''
\end{quote}

Among the participants studying GenAI in computing education, a common motivation was the need to understand how these are changing the classroom:

\begin{quote}
``\textit{I think one main point for me would be, how the introduction of GenAI into introductory programming courses are affecting students, learning and also affecting the concept that the students need to learn and then how the problem can be this become the solution 
itself, how we can utilize the affordances of GenAI. For helping students with learning what they need to learn.}''
\end{quote}


\subsection{RQ5 -- Tools}

As mentioned in Section \ref{sec:rq2rq3}, all educators reported using standard industry GenAI tools like ChatGPT and Copilot (ES-16, 27 out of 27, 100\%) via the online survey. Only two instructors also used customized tools that they had created themselves (2 out of 27, 7.4\%).

Interviews revealed a varying range of tools from AI-powered assistants to specific applications designed to enhance learning experiences in programming and other courses. One notable category of tools is Retrieval-Augmented Generation (RAG), which not only provides information but does so within the context of a given course. Another category of tools tracked student interactions and completion of the course assessment elements, which helps educators understand how students are learning so that they can adapt course curriculum and delivery accordingly. These tools initially were used and deployed in computing courses but now are being increasingly adopted in other disciplines such as business and economics.

The usage and impact of these tools are considerable. GenAI tools are extensively used in writing problem sets, assisting with assignments, and providing explanations and help in coding tasks. There are tools specifically developed to facilitate functionalities for asking questions, explaining code, writing code, and fixing code. A common trend is to put guardrails in place so these tools cannot directly return generated code. Many of these tools are built on top of existing AI models and are trained on specific course material.
\textit{
\begin{quote}
    ``So it's essentially like a sandbox version of ChatGPT. Now, that doesn't mean, I think, that it's completely safe. But I think that what it means is that the responses that students get from the model are going to be based only on the material that the instructor has trained the model on.''
\end{quote}
}

An interesting observation by educators who deployed and used GenAI assistants in their courses is the noticeable reduction in the use of office hours by students following the deployment of these tools, indicating a shift in how students seek help and interact with course material. 

In terms of design and feedback mechanisms, the tools are crafted to be simple, avoiding complex options for students. 
\textit{
\begin{quote}
    ``I think AI tools in CSed should be designed to ensure users remain cognitively engaged in their coding tasks, preventing situations where users can offload the necessary cognitive efforts onto the AI without thought, while also not causing frustration.''
\end{quote}
}

The responses provided by these tools are also simplified, and students have the ability to offer feedback on the responses they receive. This feedback loop is critical for continuous improvement.
\textit{
\begin{quote}
    ``If the student doesn't know what the AI is doing and how it's coming up with something, they cannot debug this help seeking tool.''
\end{quote}
}

The use of GenAI in education extends beyond direct student support. GenAI tools are employed to analyze student data, including test case-driven auto-graders, knowledge component-based tutors, and adaptive feedback mechanisms.

\subsection{RQ6 -- Perceived Outcomes}

In the interviews, instructors shared perceived outcomes of GenAI use that they consider to be both negative and positive. The negative perceptions mostly revolve around students using GenAI to write code for them. This has raised concern in multiple ways. First, there is concern that the students are able to circumvent work by having the AI do the work for them. Multiple instructors have reported a sizable increase in academic dishonesty violations in the last few semesters. 
\textit{
\begin{quote}
``The bar to cheating is lower because of these GenAI tools.''
\end{quote}
}

Second, it is perceived that this use of GenAI is resulting in lack of actual learning. For example, a scenario shared by professors is that students will use GenAI to complete open assignments and therefore not actually understand the learning objectives. Professors are saying this scenario causes students to be unprepared for assessments. 

\begin{quote}
\textit{``I know I don't think we've ever had such large numbers of students who go into a test and simply can't do anything.''}
\end{quote}

In some cases, it appears that students think they are learning by using the AI but they are mistaken. Even when using GenAI in ways supported by instructors, it can be problematic for students. Instructors shared that it is very easy for students to be over-trusting of the AI-generated suggestions and waste their time going down incorrect paths. In some extreme cases, students have been discouraged from continuing their education due to the fear that AI will make their careers obsolete.

However, there are still many positive perceived outcomes shared by instructors. First, students who use GenAI are able to create work that is more complicated and of a higher quality than they would without assistance. This includes both coding projects and writing assignments. Used as a tutor, AI can help students by explaining difficult concepts, spotting logical errors, and even providing better compiler error messages than the actual compiler. Use of GenAI in this capacity has even been perceived to lower the load on office hours held by instructors and teaching assistants. 

The interviews also revealed some larger structural changes reported by instructors. One common theme is the idea that the very concept of being a programmer will change. It is predicted that students will graduate with fewer fundamental coding skills, but with the ability to produce more advanced work. Some instructors predict a branching into two fields: computer scientists and conversational programmers. Conversational programmers will focus much less on writing code and much more on understanding how to write code and use AI tools to generate code for them. One professor went so far as to say that this shift in computing education is impactful on the same scale as the invention of the printing press.

\subsection{RQ7 -- Industry Usage}
\label{sec:rq7}

In this section, the perspectives of industry developers towards GenAI tools are presented. Moreover, we compare the developers' views and their use of GenAI with educators' beliefs of the industry's use of GenAI. 


The majority of surveyed developers (31 out of 39, 79.5\,\%) use GenAI tools in their professional role in developing software (DS-1). Eight developers claimed not to use GenAI. Two of these stated that they were concerned about ethical issues as their main reason not to use it (DS-14). The reasons ``\textit{My company does not let me}'', ``\textit{I do not believe they will help me code better}'', ``\textit{Haven't gotten around to it}!'' and ``\textit{I tried it and did not work for my needs}'' were stated once each. Three developers did not provide a main reason why they do not use GenAI. Elaborated reasons were entered by three developers. One developer stated that they did not receive the ``needed code'', maybe they were trying on ``not so standard'' problems (DS-15). Other reasons were mentioned in the open-ended response field: ``\textit{In theory AI generated code could be a derivative work}'' and ``\textit{it's not allowed to use GenAI tools but we expect it will be in the near future}'' (cf. \autoref{sec:rq1-policies}).

\begin{figure*}
    \centering
    \begin{subfigure}[b]{0.45\textwidth}
        \centering
        \includegraphics[width=\textwidth]{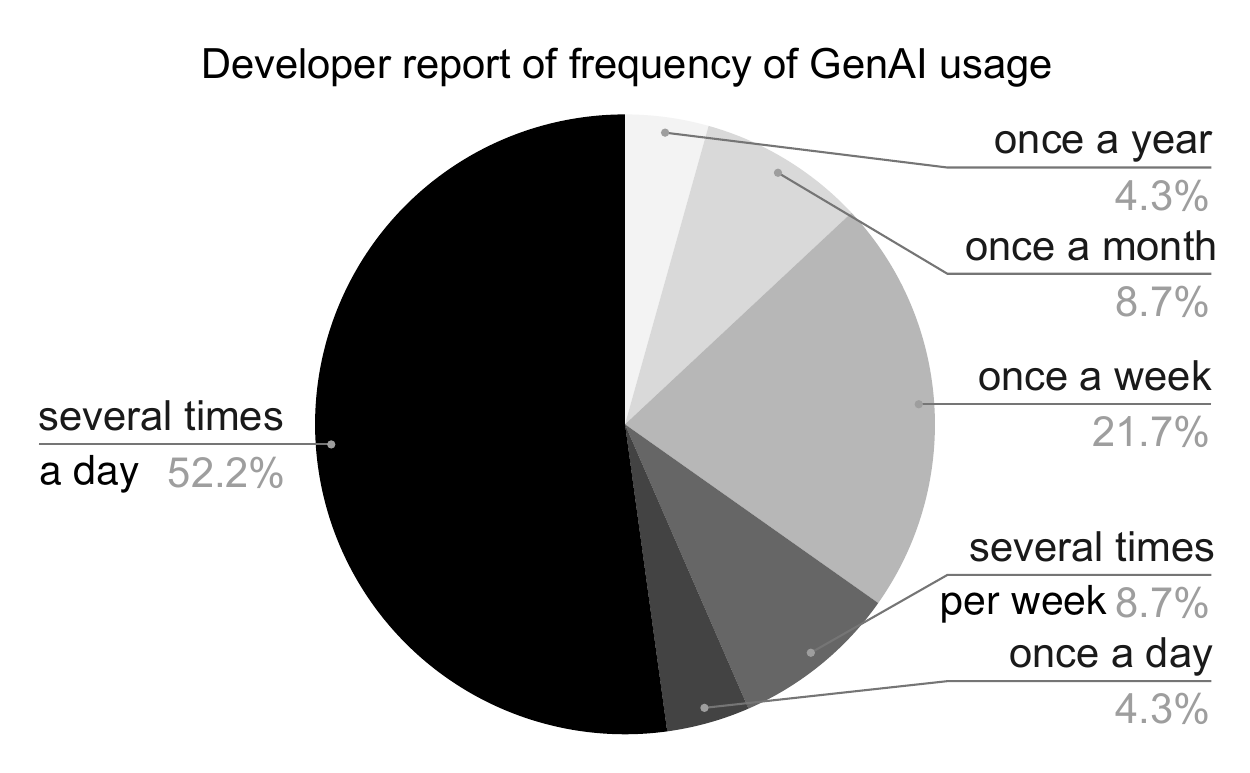}\\
    \end{subfigure}
    \hfill
    \begin{subfigure}[b]{0.45\textwidth}
        \centering
        \includegraphics[width=\textwidth]{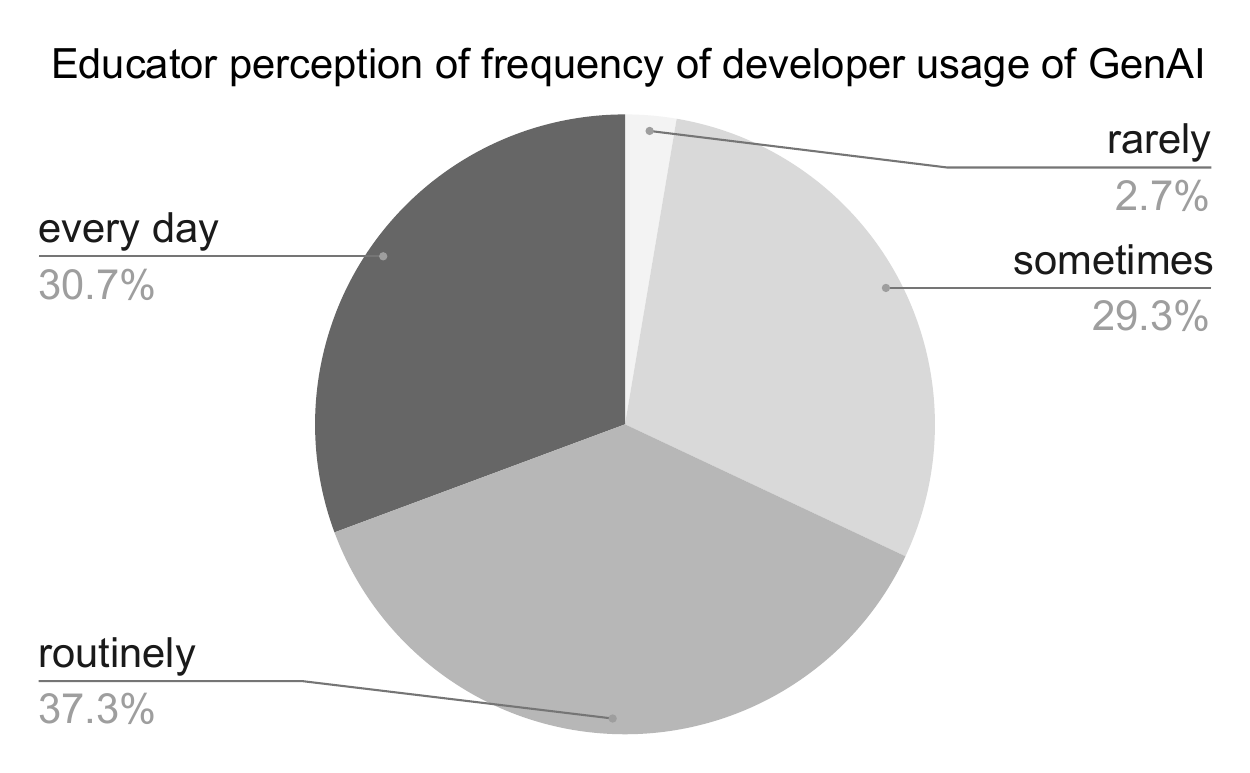}\\
    \end{subfigure}
    \caption{Left: Frequencies of developers (n=23) reporting on their use of GenAI tools as part of their professional role. Right: Perspective of educators (n=76) about the frequent use of GenAI tools by professional developers.}
    \label{fig:survey-devs-eds:howoften}
\end{figure*}

In the following, we present the results of the responses from developers who are users of GenAI (n=31).
We asked developers about the frequency of their GenAI use in their professional role (DS-2). Most of the developers in our survey reported that they use GenAI at least once a day, with 52.2\,\% reporting using it several times a day and 4.3\,\% using it once a day (see~\autoref{fig:survey-devs-eds:howoften}). A similar question (ES-7) was provided to educators, asking them to estimate how often they believe professional developers use GenAI  in their work (see~\autoref{fig:survey-devs-eds:howoften}). 
Many educators in our survey expect professional developers to use GenAI at least routinely (36.4\,\%). 
But the percentage of developers in our survey reporting to use GenAI every day is 56.5\,\% (sum of responses to ``once a day'' and ``several times a day''). This indicates that educators' expectations seem to underestimate the use of GenAI tools by professional developers.

Asked to select the types of tools being used (DS-3), developers reported that autocompletion tools are the most frequently used type (52\,\%, 16 out of 31), followed by Chatbots (48\,\%, 15 out of 31). Nine developers (29\,\%) did not provide any answer to question DS-3.

We asked developers to identify the tasks for which they use GenAI from a preset list of programming tasks (DS-5). Eight developers did not answer this question. On average the developers selected 4.4 tasks (median 4).
The most often selected tasks were \textit{generating code} and \textit{autocompleting code} (both selected by 17 developers, 70.8\,\%). Furthermore, the tasks \textit{modifying code} and \textit{creating documentation/comments} were selected by the majority. The least frequently selected task is \textit{modeling algorithms}, i.\,e., one developer selected this option.
Only one other task, ``research'', was mentioned for \textit{Other, please specify} by one developer.

Just prior to the closed-ended DS-5, we provided an open-ended question asking the developers to describe how they use GenAI tools in their professional work (DS-4). It was intentionally placed there to avoid influencing the developers by the options offered in DS-5. The open-ended question was answered by 17 developers mostly on a high abstraction level and was coded using the existing options of DS-5 as a starting point. In general, the explicitly named tasks match the selected tasks in DS-5, but most developers selected more tasks in response to DS-5 than they had selected in DS-4. On average, 2.6 tasks were tagged in the open-ended answers to DS-4 and 4.4 tasks were selected in the closed-ended question DS-5. 

In the following, some interesting examples of the open-ended responses are provided: 
Two developers explicitly stated that they use GenAI for repetitive tasks, or to ``\textit{automate boring stuff (i.\,e., code that no human should ever write)}.''
Three developers stated they use GenAI for generating code or examples for languages they are not comfortable or less familiar with. A common theme was to use GenAI to avoid reading documentation (mentioned 4 times) or to ``\textit{get help with undocumented features}'' (mentioned once). 
For the task \textit{modifying code}, refactoring or cleaning code were mentioned explicitly by two developers.
Interestingly, one developer stated they use GenAI to get ``\textit{code reviewed to make it more clear, concise and maintainable}'' but did not select \textit{modifying code} and one developer claimed to use GenAI to ``\textit{explain errors or exceptions for frameworks that I rarely use}'' but did not select ``Debugging.''
Further notable examples that did not fit into the existing categories were ``\textit{Generating workshop material},'' helping ``\textit{with unusual tasks (e.\,g., I recently had to convert timestamps stored in a funny text format into a number of minutes 
 \dots\ in Excel)}'', ``\textit{summarizing large text}'', and ``\textit{writing quality updates}''. The latter two seem to be focused on prose text instead of program code.

\begin{figure*}
    \centering
    \includegraphics[width=0.95\linewidth]{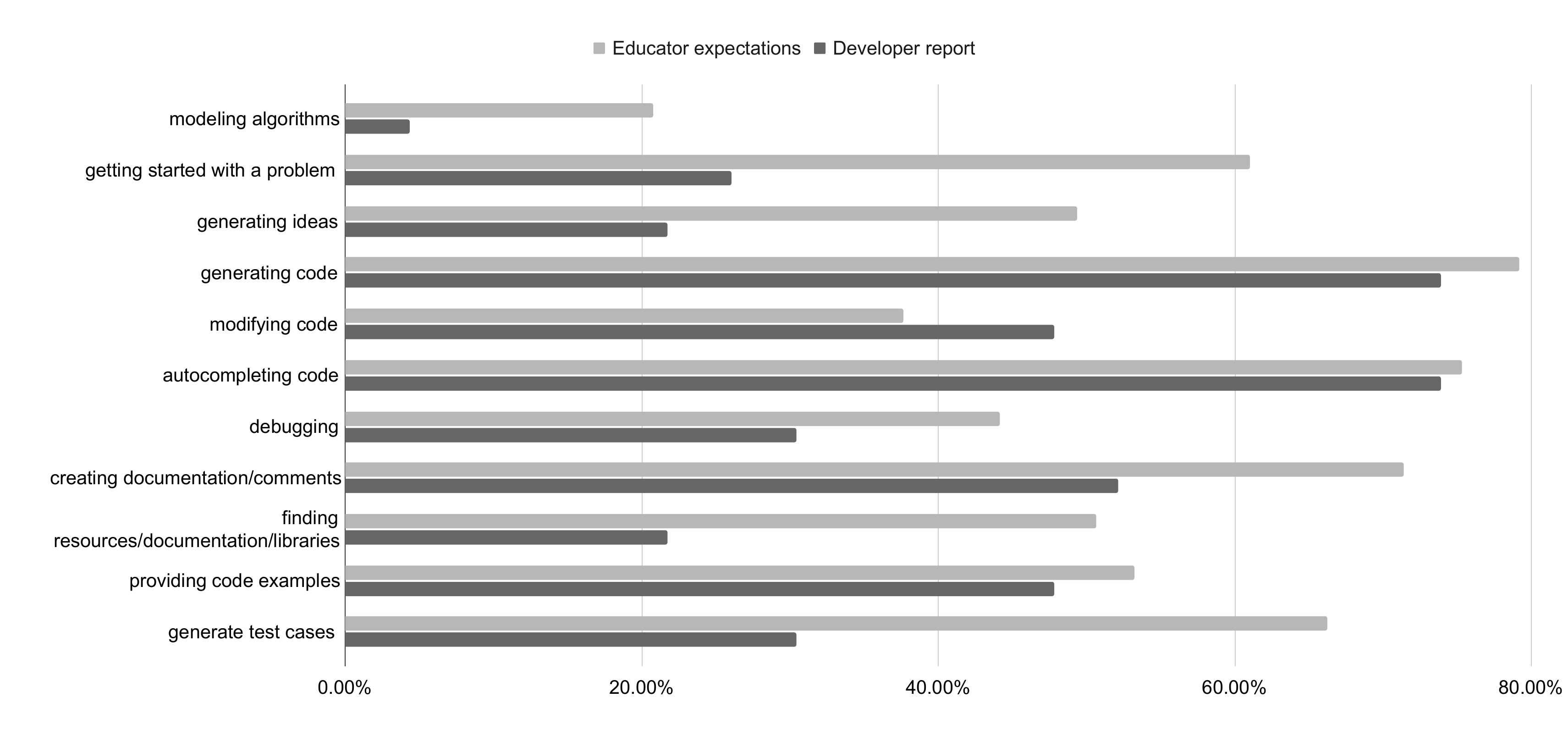}
    \caption{Comparison of tasks developers use GenAI for and educators think developers use it for.}
    \label{fig:survey:usage}
\end{figure*}

We asked educators to report their perceptions of which tasks developers use GenAI tools for (ES-8). 
74 educators answered this question. 
Educators, on average, selected 6.3 tasks that they assume developers use GenAI for (median 6). Most frequently, the educators selected \textit{generating code} (79.2\,\%), followed by \textit{autocompleting code} (75.3\,\%), and least frequently \textit{modeling} \textit{algorithms} (20.8\,\%). Only the tasks \textit{modeling} \textit{algorithms, modifying code, generating ideas}, and \textit{debugging} were not selected by the majority of the educators. Five educators selected \textit{Other, please specify}. The following statements by educators should be noted: ``\textit{Most of the ndustry [sic!] people I know are just experimenting, not using it for official tasks}'', ``\textit{all of the above}'', ``\textit{exploring alternatives, explaining code, explaining downsides/upsides of approaches}'', and ``\textit{developers I've talked use ai differently depending on their context}.'' 
The comparison of the results from educators and developers is summarized in \autoref{fig:survey:usage}.

The frequencies of educators estimating developers' GenAI usage and actual developers' GenAI usage seem to approximately match for the tasks \textit{generating code}, \textit{autocompleting code},
 and \textit{providing code examples}.
There is a particularly higher percentage of educators who identified \textit{getting started with a problem, modeling algorithms, generating ideas, generating test cases}, and \textit{finding resources/documentation/libraries}. 
Interestingly, more developers selected \textit{modifying code} than educators. In all other cases, the expectation of the educators were higher.

\begin{figure}[htb]
    \centering
    \includegraphics[width=0.95\linewidth]{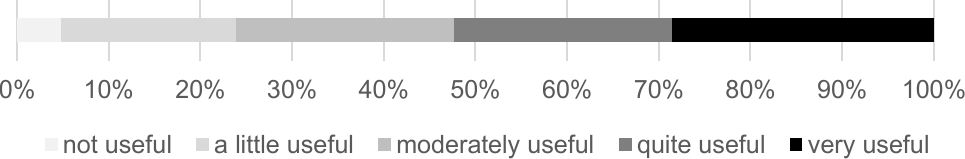}
    \caption{Ratings of the usefulness of GenAI as reported by the developers (N=21)}
    \label{fig:survey-devs:usefulness}
\end{figure}

\autoref{fig:survey-devs:usefulness} shows the ratings of the usefulness of GenAI as reported by the developers (n=21, DS-6). Almost all developers who answered (95\,\%, 20 out of 21) find using GenAI at least a little useful. Six developers out of 21 (29\,\%) find it very useful and only one developer did not find it useful. 

At the same time, the majority of developers (81\,\%, 17 out of 21) who answered the subsequent question (DS-7) think GenAI makes their work more efficient. Three developers (14\,\%) saw no change, and only one developer found GenAI to make their work much less efficient.


When asked about the harmfulness of GenAI (n=21, DS-8), the majority of our surveyed developers found it \textit{not harmful} (62\,\%, 13 out of 21), one third \textit{a little harmful} (33\,\%, 7 out of 21), and one developer \textit{moderately harmful} (5\,\%). Six of the developers who found GenAI (a little) harmful elaborated on the harmfulness (DS-9). Two themes emerged from the thematic analysis. The first one was \textbf{wrong results or bugs in the code} (even for simple cases, n=5) and the second represents \textbf{concerns regarding code quality} (which is reported to be ``\textit{generally worse than humans}'', n=1). To present more specific examples, one developer reported on an issue with an invalid SQL statement and another developer said that GenAI invented features.
 A developer summarized that every line of code needs to be rechecked -- even though errors may only occur in rare cases.



\subsection{RQ8 -- Competencies}
\label{sec:competencies}

\subsubsection{Educators' Perspectives}

The majority of educators (57 out of 76) who participated in the survey believe the skills needed to create software have changed after the advent of GenAI (ES-3).  Among educators who have changed their courses to integrate GenAI, 25 out of 27 believe the skills to create software have changed.  In contrast, among educators who have not changed their courses to integrate AI, 32 out of 49 believe the skills have changed.

Examining educators' opinions based on whether they believe the skills to create software have changed, we found that among faculty who do not believe the skills have changed, only 2 out of 19 have changed their courses to incorporate GenAI.  However, among faculty who believe the skills have changed, 25 out of 57 have changed their courses to incorporate GenAI.

Question 4 of the educator survey (ES-4) was given to faculty who responded ``No'' to whether the skills needed to write software have changed.  The question was ``Please elaborate on your last response why skills have not changed. (open-ended)''. Two members of the working group performed a thematic analysis of the 16 open ended responses on \textbf{why skills have not changed} uncovering the following themes:

\begin{itemize}[leftmargin=*]
    \item \textbf{Same Skills/Shifting Importance.} The largest group of respondents, seven, believe that the skills are the same. However, the importance of these skills has changed. For example, multiple participants reported believing that understanding code is still crucial, but 
    code evaluation has become more important. ``\textit{Software skills remain a mixture of design, implementation, static analysis, and debugging. Changing the balances does not change the core skills involved.}''  Their responses seem to indicate that evaluating code generated by AI is not different than the code evaluated in the past.
    \item \textbf{Accelerating skilled programmers.} Two respondents believe that GenAI will only serve to speed up programmers who are already skilled on their own. ``\textit{GenAI is only a tool which can speed up the development process for those already skilled and knowledgeable.}''
    \item \textbf{Know everything.}  One participant believe that programmers still need to know everything at all levels below them regardless of GenAI: ``\textit{Using gen ai is a lot like being a team lead. The team leads and project managers still need the skills of the programmers under them}''.  However, it was unclear how deep they felt these skills should go.
    \item \textbf{Don't know yet.}  One participant believe that GenAI is changing too rapidly for us to know if or how these skills will change. ``\textit{Too little time to know (from my limited point of view)}''.
\end{itemize}


Question 5 on the educator survey was given to participants who replied ``Yes'' to the question of whether the skills to write software have changed with the advent of GenAI (ES-5).  Question 5 asked: ``In what ways do you think the skills needed to create software have changed with the introduction of GenAI tools?''. Two members of the working group performed open-coding on the 50 open-ended responses followed by a thematic analysis to uncover the following themes related to \textbf{why educators think skills have changed}:

\begin{itemize}[leftmargin=*]
    \item \textbf{Less writing, more reading code.} The most common responses from participants were that reading and understanding code is more important (21 participants): ``\textit{It's becoming more important to be able to read code that you have not written}'' and writing syntactically correct code from scratch is less important (11 participants), ``\textit{There is less need to write code.}''
    \item \textbf{Higher-level skills are needed.} Fifteen participants said that writing software with GenAI makes higher level skills more important.  Examples of higher level skills included creativity, problem understanding, problem specification, problem decomposition, software design, and software architecture. ``\textit{The focus needs to be more on the algorithmic thinking and problem solving, not the actual syntax/coding}'', and ``\textit{Problem decomposition becomes essential in a way it wasn't in the past.}''
    \item \textbf{Evaluating and fixing code.} Eleven participants reported that evaluating code (testing) from the GenAI is an important skill and nine participants said debugging the code from GenAI is important as well: ``\textit{Code testing and debugging, if not previously emphasized, should be now.}''
    \item \textbf{Prompting as a new skill.} Seven participants reported that being able to prompt the GenAI to receive desired code is a new and important skill: ``\textit{There are new skills like prompting AI.}''
\end{itemize}


The educators who believe skills have changed (ES-6) chose \textit{problem understanding} as the most important skill for programming with GenAI. They see \textit{reading code} as the second most important competency and \textit{problem decomposition} as the third most important.


Only three people who allowed GenAI in their classes changed any of the learning objectives of their recent courses (ES-19); six people reported not changing them at all.

\subsubsection{Developers' Perspectives}

In question 10 of the developer survey (DS-10), we asked the developers whether the competencies required to professionally develop software are changing due to the availability of GenAI tools. 21 developers responded:  12 of them (57\,\%) indicated a \textit{slight change}, 4 of them a \textit{moderate change}, and 5 of them \textit{no change}.

The 16 participants who recognized a change in the competencies in DS-10 were of particular interest for the follow-up questions DS-11 and DS-12. We received 11 open-ended responses to DS-11, which asked for the \textbf{new relevant competencies to professionally develop software} with GenAI tools. Within these responses, we identified several themes. It should be noted that some of these themes had occurred within the responses of educators.  
\begin{itemize}[leftmargin=*]
    \item \textbf{Same skills/Shifting importance.} Not necessarily new competencies, but shifting focus on understanding of concepts, e.\,g., ``\textit{there's a bigger need to understand underlying concepts}''
    \item \textbf{Accelerating skilled programmers.} Successfully use GenAI tools to increase productivity, at least for simple tasks, e.\,g., ``\textit{leverage GenAI tools to improve productivity}'', ``\textit{using the speed for solving simple/generic problems.}''
    \item \textbf{Knowing about general limitations.} Know the strengths and weaknesses of GenAI tools, e.\,g.,  ``\textit{accounting for hallucinations from GenAI.}''
    \item \textbf{Evaluating GenAI use and its output.} Evaluate, if it is appropriate to (not) use GenAI tools for a specific problem, e.\,g., ``\textit{knowing the sorts of problems that GenAI can be useful to solve is important}'', ``\textit{Being able to determine if autocompletions are actually useful.}''
    \item \textbf{Prompting as a new skill.} Prompting GenAI to generate useful and appropriate responses, e.\,g., ``\textit{prompting AI with appropriate context to receive relevant/valid responses.}''
    \item \textbf{Meticulousness.} Paying attention to detail (being meticulous) when checking generated code, e.\,g., ``\textit{I need to carefully examine the output of Copilot [...] and I often have to make small edits to its output.}''
\end{itemize}

The second competency-related follow-up question for developers (DS-12, n=9) asked for \textbf{competencies that were perceived as no longer or less relevant} to professionally develop software with GenAI tools. We identified the following themes in the responses: 
\begin{itemize}[leftmargin=*]
    \item \textbf{Know everything.} Know syntax and other elements of programming languages by heart, e.\,g., ``\textit{Less need for specific knowledge; eg language specific nuances.}''
    \item \textbf{Know and use other sources.} Using other external resources, e.\,g., ``\textit{searching stack overflow or other non-official help or documentation (official docs still has some usage).}''
    \item \textbf{Tolerating high levels of frustration.} Being purpose-driven or persistent despite frustration, e.\,g., ``\textit{The competence not to be easily frustrated if the first hit on Google does not provide the answer to the problem.}''
    \item \textbf{None.} None of the previous acquired competencies are perceived as irrelevant e.\,g., ``\textit{I see no competencies that are less relevant}'', ``\textit{It's a good starting point, nothing else.}'' 
\end{itemize}

Interestingly, when asked about what advice developers would give to novice programmers regarding the use of GenAI tools (DS-13 and DS-16, n=18), many responses related to the responses to DS-11. The analysis of the open-ended responses revealed the following themes, which repeat some of the competencies mentioned before: 
\begin{itemize}[leftmargin=*]
    \item \textbf{Evaluating GenAI use and its output.} Evaluate, if it is appropriate to (not) use GenAI tools for a specific problem, e.\,g., ``\textit{Find the right use cases for GenAI but don't trust it. Verification is crucial.}'', ``\textit{Don't simply trust them}.''
    \item \textbf{Meticulousness.} Paying attention to detail (being meticulous), e.\,g., ``\textit{always read back over the code the ai recommends: sometimes it is not entirely correct}''
    \item \textbf{Prompting as a new skill.} Prompting GenAI to generate useful and appropriate responses, e.\,g., ``\textit{learn how to speak AIs language but also learn enough domain knowledge that lets you adequately explain your needs to an AI.}''
\end{itemize}

The recommendations by developers are somewhat divided between using GenAI tools to learn and get comfortable with them -- and not using them:
\begin{itemize}[leftmargin=*]
    \item \textbf{Resist using GenAI.} Do not use GenAI tools as a novice programmer, e.\,g., ``\textit{Don't.}'', ``\textit{Please take the time to learn to be an algorithmic thinker yourself. Try the problem on your own }'', ``\textit{Learn the basics of programming without GenAI.}''  
    \item \textbf{Rarely use GenAI.} Use GenAI, but rarely, e.\,g., ``\textit{learn to work without them as much as you can}'', ``\textit{There's value in not needing it}'', ``\textit{Don't lean on it too heavily}.''
    \item \textbf{Critically use GenAI.} Use GenAI tools, but always be critical of the output
    ``\textit{Use it to start out, but still need to read over the code that you write to understand it.}'', ``\textit{Use it to learn languages (explain what existing code is doing) but know that it's nowhere near perfect.}''
    \item \textbf{Excel at using GenAI.} Use GenAI tools, and become more productive, e.\,g., ``\textit{Start using it early and get comfortable with leveraging it as a way to code faster.}''
\end{itemize}

It should be noted though that the last comment was the exception. The majority of the comments highlighted the need for a critical use of GenAI tools, and many comments were in favor of learning the basics of programming without GenAI to avoid over-reliance early on respective tools. 

The interviews further reveal that educators are grappling with how GenAI tools are reshaping students' competencies in computing education. One of the most commonly discussed shifts is the reduced emphasis on writing code from scratch and the growing importance of reading and understanding pre-existing code (or code written by GenAI tools). Educators believe that as GenAI tools provide students with ready-made solutions, the ability to critically evaluate and modify this code becomes more essential. Furthermore, higher-level cognitive skills, such as problem decomposition, task specification, and computational thinking, are emphasized as essential competencies that students must develop to navigate the evolving landscape of AI-assisted programming. Several interviewees also highlight the necessity of integrating communication skills, such as explaining one's code, into course objectives to ensure that students genuinely understand what they create rather than merely generating functional code.

\begin{quote}
    ``\textit{I think potentially, students will start knowing less about how computers work. Right? As these tools evolve, we're potentially having students that come in with less knowledge at that level, but more capability at higher levels.}''
\end{quote}

\begin{quote}
    ``\textit{I think with these tools, people are going to spend more time reading because they're gonna get help. They're gonna get assistance from these tools. It's gonna give them code. People are gonna need to read it and understand it.}''
\end{quote}

Moreover, prompt engineering is emerging as a new core competency. Educators are increasingly aware that these models can do much of the "heavy lifting" in coding, but students must learn how to ask the right questions and interpret AI-generated outputs to succeed. Some express concerns that students might bypass key stages of the problem-solving process, such as task decomposition, which could lead to gaps in their understanding. While many educators see these shifts as an opportunity to teach more advanced and meta-cognitive skills, there is also a concern that students may lose touch with foundational computing concepts. Ultimately, the consensus is that GenAI is pushing educators to rethink their course objectives and better prepare students for a world where interacting with AI is a critical part of the programming process.

\begin{quote}
    ``\textit{At the end of the day, we want students to practice task decomposition. So how should AI be used? Hopefully, it should help them develop those skills.}''
\end{quote}

\subsection{RQ9 -- Equity}


As a part of the survey study, educators were asked whether they teach at an institution that serves a minority population in their country (ES-26). 23.7\,\% (18 out of 76) of educators report teaching at an institution serving a minority population. However, 30.3\,\% (23 out of 76) are not sure whether it applies to their institution (one participant did not answer this question). Due to these relatively low response rates and insecurity among the participants, we did not analyze the data any further with regard to GenAI teaching practices and instructional approaches and whether and how they differ depending on the educators' institutions.


In the interviews, educators raised significant concerns regarding the equity implications of GenAI tools in education. Many pointed out that access to these tools is often tied to financial resources, both at the individual and institutional levels. For instance, students from wealthier backgrounds or universities may have access to premium tools such as Microsoft Copilot, while others may have limited or no access. This disparity could exacerbate existing inequalities, as students with better tools may accelerate their learning, leaving behind those with fewer resources. 

\begin{quote}
``\textit{ChatGPT-like technologies are not a silver bullet to help the poor students get better.}''
\end{quote}

The issue extends beyond financial access; educators also highlighted that not all students possess the meta-cognitive skills required to use these tools effectively. Those who are more capable of self-regulation may benefit from AI tools, while students lacking these skills might misuse them, producing correct outputs without learning the underlying processes. 

\begin{quote}
``\textit{[N]ot all students are equipped with the Metacognitive skills to use these sorts of unconstrained tools productively.}''
\end{quote}


Additionally, linguistic and cognitive barriers were flagged as potential equity concerns. Non-native English speakers and students with lower literacy skills might struggle with AI-generated responses, which are often complex to read and comprehend. As such, AI tools could exacerbate difficulties in comprehension and critical thinking. 

\begin{quote}
    ``\textit{For students whose first language isn't English, they may struggle more with interpreting the AI-generated responses, which could create another layer of disadvantage.}''
\end{quote}

Furthermore, students with disabilities, particularly those relying on assistive technologies like text-to-speech tools, may face challenges in interacting with GenAI. 

\begin{quote}
    ``\textit{we'll have students with various disabilities in classes who will be watching their peers use it in some senior-level software engineering or capstone course, and they'll be charging forward much, much faster to do this. But like, we'll have text-to-speech issues for people who have motor impairments who can't type where they're doing speech to text to do text entry for some large language model, or they have a speech impediment. And so the speech recognition for the model doesn't work very well, and then they can't actually just get precise enough prompts into the system.}''
\end{quote}

While some educators noted efforts to mitigate these issues, such as providing institutional licenses or designing tools for broader access, the overall sentiment was that AI tools, at present, risk amplifying existing inequities rather than alleviating them.

\subsection{RQ10 -- Future}



The future of GenAI in education, particularly in computer science, is seen as both an opportunity and a challenge. Interviewees generally agree that AI will continue to reshape the landscape, but the extent of its influence remains uncertain. 
\begin{quote}
    ``\textit{I think if we really were honest with ourselves about new learning outcomes and wanting students to demonstrate proficiency. Working with these models. It would be good if we actually had some assessments where they were allowed to do that.}''
\end{quote}

Many educators foresee GenAI becoming an integral part of the classroom, assisting in various ways, from automating tedious tasks such as grading and creating assessments, to dynamically analyzing student code and providing real-time feedback. Some interviewees suggested that AI could cluster student submissions to provide more tailored feedback, saving time for instructors while offering more personalized learning experiences. 
\begin{quote}
    ``\textit{One interesting project\dots is to make a conversational agent that takes students' programs as input and then ensures they have learned what they need to learn.}''
\end{quote}

However, the consensus is that while these advancements have the potential to greatly improve education, careful planning and continuous adaptation are necessary to ensure that AI is used effectively and equitably.

Several interviewees also envision a future where learning outcomes and assessments shift to accommodate AI's growing role. Instead of focusing purely on coding, students might be evaluated on how well they interact with AI tools to solve complex problems, emphasizing prompt engineering and problem decomposition. The challenge lies in balancing AI's capabilities with the need to preserve essential skills, such as critical thinking and deep learning, which are at risk if students rely too heavily on AI. Many interviewees expressed hope that AI would free up educators to focus on higher-level teaching tasks, fostering stronger teacher-student relationships and creating more meaningful, engaging learning environments. At the same time, they acknowledged that this future is still unfolding, and much of the educational community is cautiously experimenting with how best to integrate these new tools.





%
%
%
%
\section{Discussion}
\label{sec:discussion}

In this section, we discuss some of the most interesting findings of the present study in an integrated manner. The discussion is thus led by themes and findings, and not by the research methods used to gather them.

\textbf{Changing Competencies.}
Several interesting insights emerged from the educator survey. The majority of faculty (77\,\%) believe the skills to write software have changed because of the advent of LLMs, but less than half (36\,\%) have changed their courses to include GenAI. Perhaps those who feel the skills have changed, but have not changed their courses to include GenAI, have changed their courses in other ways. Regardless, an increasing number of instructors are adopting (or at least not banning) GenAI in their courses compared to prior studies~\cite{prather2023wgfullreport, hou2024effects}. In the present study, we see about 30\,\% of educators actively integrating GenAI, and 75\,\% acknowledge it in their class. It is not a total change but a clear trend.

We also found that educators who are changing their courses to incorporate LLMs are focusing on the new competencies, such as prompting (ES-5, see \autoref{sec:competencies}). Based on the results of the literature review, there are differences in how well generative AI can support different tasks in computing courses. For example, studies that used it for code comprehension reported more positive findings compared to studies that used it for hint generation or code writing.

Lastly, recognizing that LLMs can solve take-home assignments, faculty are allocating more course grades on proctored assessments and creating more, or new, proctored assignments.

In the educator survey, educators are divided about how competencies are changing. Most educators report that they believe the skills to write software are either changing or shifting in some way as a result of LLMs and GenAI. In contrast, a minority of educators report believing either that there has been no change to competencies, or that students still need to develop certain programming competencies and skills before they should start to work with/use GenAI tools. Educators who see a shift in competencies feel we are shifting away from code writing to code reading; toward higher level skills like creativity, problem understanding, problem specification, problem decomposition, and software architecture; and toward evaluating/testing and debugging. 

\textbf{Reasons for Disallowing GenAI.}
Educators disallowing the use of GenAI provide two main reasons, (1) their own (at least perceived) lack of skills and competencies regarding the concept and use of GenAI, and (2) their attitude that the students would not have the necessary skills and should, therefore, learn the basics of programming first without any support of GenAI. Further research is needed to determine whether this attitude is pedagogically sound or whether using GenAI can make it easier for beginners to learn programming --- and if so, in what ways it may even enhance learning (these ways may relate to the shift in competencies discussed earlier). 

\textbf{Gap between Educators Expectations and Developers Actual Use of GenAI.}
We further noted some differences between the use of GenAI tools among developers and how educators think developers are using GenAI tools. More educators expected developers to use GenAI for certain tasks than developers reported. This applied to the following tasks: Modeling algorithms, getting started with problem solving, generating ideas, debugging, creating documentation/comments, finding resources/documentation/libraries, providing code examples, and generation of test cases (cf. \autoref{fig:survey:usage}). Modifying code was the only task where developers exceeded educators' expectations of GenAI use. It is thus important for computing education to regularly connect to industry practices and to not lose track of the competencies needed in the workplace.
Despite these differences regarding expectations and actual use, the survey results alone show that GenAI tools are indeed used by developers for a great number of different tasks.

\textbf{Developers' Use of and Trust in GenAI Tools.}
The survey with professional software developers further revealed that GenAI is used for getting started, debugging, code cleanup, and not having to read documentation (cf. \autoref{sec:rq7}, question DS-4). 

Even though only 39 professional developers took part in our survey, the results are in line with larger surveys, e.\,g., the StackOverflow 2024 survey with professional developers and their AI usage (see~\cite{stackoverflowstudy2024}, conducted in May 2024, filtered for professional developers). It should be noted that the number of responses greatly varied among questions which is why we also provide the absolute numbers of responses in the following. In the StackOverflow survey, about 63\,\% of about 46,000 professional developers state that they use GenAI (vs. 79.5\,\% in our survey) and about 23\,\% state that they do not use GenAI and do not plan to do so in the future. About 83\,\% of about 28,500 developers see an increase in productivity and 58\,\% a greater efficiency. This number matches the 81\,\% of developers in our survey who see an increase of efficiency when using GenAI tools.

The StackOverflow survey further shows that developers are split about whether they trust AI output: about 41\,\% have (some) trust in AI and about 31\,\% have (some) distrust in the accuracy. Furthermore, about 45\,\% of the developers in the StackOverflow study think AI tools are bad or very bad on complex tasks. Both aspects are also reflected in comments in our survey. Hence, programming skills are still required to spot such issues and creating complex software. The top ethical concerns that the majority of developers stated in the StackOverflow survey are ``misinformation and disinformation in AI results'' (79\,\%) and issues with ``source attribution'' (65\,\%). This may contradict our result that about 62\,\% of the developers in our study found GenAI \textit{not harmful}, but may also be caused by the way the ethical concerns were collected in the StackOverflow survey (i.\,e., via a closed question).

\textbf{Equity Concerns.}
With standard industry tools being used by all educators who participated in our survey, we are concerned about how these tools may negatively impact equity among students. For example, we cannot neglect the cost of using GenAI tools. Considering a standard three- or four-year program (undergraduate degree in Europe or the US) and a monthly cost of 20 US dollars per month, this may amount to about one-thousand US dollars. Access alone may thus be another financial burden for students with an already tight budget. 

According to the educator survey (ES-15), educators seem to expect students to use the free version of standard industry tools. However, we did not ask them whether they have any knowledge of students using the paid versions regardless, and how that may have had an impact on their outcomes or exam results. When being used for take-home tasks or assignments, educators have no control whatsoever on students' tool use. Even in the classroom, some students will have the resources to access GenAI easily via their smartphone, while others will not. For this reason, we recommend educators to carefully consider how and for which tasks they incorporate GenAI tools from an equity lens to not disadvantage any of their students. In related work, it has been shown that students with various prior knowledge and education use GenAI tools differently in introductory programming classes, and they have varying success rates~\cite{kiesler2024novice}. Similarly, we are only beginning to understand the persisting inequities between GenAI technologies and the accessibility needs of people with disabilities~\cite{alshaigy2024forgotten}.

\textbf{Need for Educators' Professional Development.}
An important finding of the educator survey is that many educators have not yet considered the potential risks of GenAI in depth or the benefits that GenAI tools can bring to their courses or students. On the one hand, educators stated that they have not yet familiarized themselves with GenAI, usually due to a lack of time, and therefore do not incorporate or allow GenAI in their courses. On the other hand, they seemed to have fixed attitudes about whether GenAI is beneficial to students or not. If we assume that they have hardly worked with GenAI, this attitude is based less on facts than opinions. This lack of familiarity with GenAI tools may also help explain why some of the responding educators had no clear ideas about using GenAI in their courses in a pedagogically meaningful way. The following quote in response to survey question ES-20 represents this challenge:
\begin{quote} 
    ``\textit{I haven't yet investigated the Generative AI tools. I'm observant and beginning to research and use the tools currently myself first to try to understand the realm of what can be done and how it can be integrated. I want to see how to scale the integration of the tools. I want to preserve the integrity of learning but still keep new and developing technologies viable because they are now in society, but I need examples for the classroom.}''
\end{quote} 
    
Another consequence of not having explored GenAI tools is that educators may not be fully aware of the risks and potential harm, meaning they are unable to discuss it critically with their students. These barriers to adoption should be addressed more intensively by professional development courses and pedagogical training opportunities at all levels of computing education.

\textbf{Future Developments.} 
As we consider the future, we also speculate on potential use cases for generative AI (GenAI) tools. Currently, most GenAI tools primarily rely on text-based interactions. However, current GenAI tools can already generate additional modalities, such as images, audio, and video. A future tool might, for example, be able to analyze a plot created by a student and identify if certain data points have been omitted.

We also expect that GenAI tools will become more personalized. A GenAI tool tailored to a specific course could recognize that students are developing new skills as they progress and adjust its generated results accordingly, avoiding advanced language features until they have been introduced by the instructor. Additionally, a GenAI tool personalized for an individual student might identify specific content areas where the student needs further assistance and direct them to relevant course materials to address any misconceptions, or even generate such personalized materials on the fly, targeted towards the individual student's educational needs (see also \cite{logacheva2024evaluating,sarsa2022automatic,keuning2024goodbye}).

\section{Threats to Validity}
\label{sec:threats}
The present work has some limitations that need to be considered. In the following, we outline the threats to validity of the respective methodologies, the data and its analysis and interpretation.

\subsection{Systematic Literature Review}

Regarding the systematic literature review, it should be noted that we did not check every available database in the context of computing education. However, we are confident that the applied methodology suffices in representing the state-of-the-art literature. Due to the cut-off date, some publications appearing around that date may also be missing, as arXiv, for example, can have delays in the process due to its moderation system. Another limitation related to the arXiv search is that it only allowed the title and abstract to be used for the search.  

Related to the analysis on the nature of findings, we acknowledge that there is likely publication bias, which should be taken into account. Studies with positive findings could more likely be accepted for publication (or written up by the authors in the first place), which could inflate the number of included studies that reported positive results.

\subsection{Educator and Developer Surveys}
We invited the participants of both surveys through several mailing lists to draw a broad sample of educators and computing professionals. 
However, it is possible that some educators on these mailing lists are more involved in CS education than their peers and, given the acute impact of GenAI over the past few year(s), may have spent more time thinking through the implications of GenAI. To mitigate this, we also encouraged recipients of our emails to share the survey with others. Moreover, we noted that many potential respondents dropped out of the survey after seeing the consent form. It might have had a discouraging effect as participants had to sign a form to proceed.

Regarding the developer survey, it is possible that the respondents are not typical software developers, as they could be more interested in developments in academia and the development of tools.
Moreover, we recognize that we are computing education researchers who do not have access to developers within large tech companies worldwide, which explains the low number of (full) responses. 
To mitigate this issue, we correlated our results with other research initiated by industry.

A limitation that applies to both surveys is that of self-reporting, meaning participants may have exaggerated, omitted information, or expressed thoughts they believe are socially desired. To somewhat mitigate this limitation, we included both open-ended and closed response options, and triangulated the survey data with the data gathered via the interviews with educators.

\subsection{Interview Study}

Regarding the interviews with educators, it should be noted that the sample may not be indicative of the computing education community. For example, interviewees were recruited in English, and the interviews were also conducted in English only. 

Another limitation is, again, related to self-reporting, which may involve subjective representations, exaggerations, omissions, or socially desirable statements.

%
%
%
%
\section{Conclusions}
\label{sec:conclusions}

The present working group report aimed to address two overarching goals: (1) identifying how and why instructors incorporate GenAI tools into their teaching, and (2) outline how the competencies and skills in software developments changed and will change further in the future due to GenAI. To address these goals, we applied a mixed-methods design comprising a systematic literature review (SLR), a survey for computing educators and developers in the software industry, and semi-structured interviews with computing educators.

\subsection{Systematic Literature Review}
The SLR focused on the reported evidence of GenAI in CER, more specifically on the types of class interventions used and whether findings have been positive or not. We then chose the search strings and databases (ASEE PEER, arXiv, Scopus, ACM Digital Library, and IEEE Xplore).
After filtering of the papers based on inclusion and exclusion criteria, the main characteristics of the papers was extracted. This lead to the final set of 71 papers that were included in the literature review. The findings of the literature review suggest that thus far, generative AI has mostly been studied in unsupervised conditions, such as having students use it for homework. Generative AI has mostly been used for writing code, code comprehension, automatic hints and generating learning resources. Most commonly, students were not instructed on how to use generative AI, and were directed to use general tools such as ChatGPT. However, we did find that there are many custom tools available. These often include some sort of pedagogical guardrails aimed to make the use of generative AI more productive for learning.

Related to the nature of the findings, we found that studies that used custom tools that included some instructor scaffolding (e.\,g., pedagogical guardrails) more often reported positive findings. However, this was only the case when students were not given explicit instructions or guidance on how to use generative AI. Thus, based on the findings of the literature review, it can be recommended that instructors should either guide students on how to use generative AI if the tools used are general purpose (such as ChatGPT or Gemini), or alternatively use custom tools that include scaffolding (such as guardrails that, e.\,g., guide the model to provide more educationally appropriate responses).

\subsection{Educator and Developer Views}
We designed a survey for educators and software developers to capture perspectives on competencies and skills required for future graduates of computing. Precisely, we wanted to gather the educator's perspective on using and teaching the use of GenAI tools to their students, which tools are used, policies that need to be considered or developed, and their motivations to integrate GenAI (RQ1-RQ6). Moreover, we wanted to identify the impact of GenAI on students' competencies (RQ8), equitable conditions for learning (RQ9), and future perspectives (RQ10).  

Another goal was gathering the industry perspective (RQ7) of generative AI usage, so we developed a survey emphasizing the experiences and reflections of software developers regarding GenAI tools. This includes both their use pattern but also potential harm being done by using GenAI. In addition, we also compared educator perspectives on industry usage with industry reports. The main conclusions related to the educators' and developers' perspectives are summarized in the grey box.

\begin{figure}[h!]
\small
\begin{custombox}{Key takeaways from our results include}
\begin{itemize}[leftmargin=1em]
    \item \textbf{GenAI impacts educators and developers:}  80\,\% of developers use GenAI tools in their professional roles (DS-1) with those not using it citing either ethical concerns or company limitations in its use. At the same time, 75\,\% of educators acknowledge that the skills to program are changing as a result of GenAI (ES-3) and 30\,\% of educators are integrating it into their classes (ES-2). In contrast, only 22\,\% of educators are explicitly disallowing GenAI use (ES-1). Given the wide availability of GenAI tools for students, we are heartened that instructors are adapting to the new reality and not attempting to ban what effectively cannot be banned.
    \item \textbf{GenAI changes assessments:} Educators are increasing the weighting of exams (ES-22), exploring alternative invigilated exam techniques (including oral exams), and are emphasizing assessing the process of learning over correctness of answers.
    \item \textbf{GenAI changes programming competencies:}  The majority of surveyed educators (75\,\%) believe program competencies have changed as a result of GenAI (ES-3). Of the remaining 25\,\%, many felt that the skills to program have remained the same, however, the importance of some skills have shifted (ES-4). Educators believe code reading has become more important than writing code from scratch, and that higher level skills like code testing, problem decomposition, problem understanding, and debugging have become more essential (ES-4, ES-5). When given options to rank, educators chose \textit{problem understanding}, \textit{reading code}, and \textit{problem decomposition} as the most important skills for programming with GenAI. Developers pointed out a similar shift (DS-10), but added the need to \textit{critically evaluate GenAI use and its output}, \textit{writing prompts}, and \textit{meticulousness} as new relevant competency components.
    \item \textbf{Need to train students to use GenAI in industry:}  GenAI tools are almost ubiquitous in industry. Educators need to train students for industry so they are successful in their career. To achieve this goal, it is crucial to keep an open eye on recent developments regarding GenAI and job ads, and cooperate with industry partners so that educators and institutions can align their curricula, competency expectations, and study programs~\cite{kiesler2023industry,valstar2020aquantitative,kiesler2023socially}. 
    \item \textbf{Guide students on how to use GenAI or use custom tools:} Based on the results of the literature review, studies reported more positive results when students were provided with guidance on how to use GenAI tools or used custom tools that include pedagogical guardrails. 
    \item \textbf{Teach (educators) and students about GenAI challenges:} Critically using GenAI is crucial according to educators and developers. It is therefore important to teach students about the limitations and issues of current GenAI tools (e.\,g., ethical aspects, biases, plagiarism, etc.) so they can make informed decisions. For that to happen, educators need to receive relevant training. Our educator survey showed that those educators who did not integrate GenAI cited the lack of time, skills, pedagogical training, support, or resources as reasons for non-use (ES-20).
\end{itemize}
\end{custombox}
\end{figure}

To elaborate on the educator perspective, we further used se\-mi\-struc\-tured interviews with tool creators and experienced educators studying or using GenAI on how they developed or integrated respective tools. The interviews addressed the research questions regarding actual student outcomes (RQ6), changing competencies (RQ8), and how GenAI shapes the future of computing education (RQ10). We found that educators are already seeing both the negative and positive impacts of GenAI from student use of these tools. Despite intentional planning for integration into their courses, instructors and noticing a large influx in both cheating and student unpreparedness for exams. However, students who use these tools responsibly are able to accomplish more than traditionally possible, especially in introductory courses. Instructors also highlighted the importance of GenAI tools in expanding assistance to students through help-seeking tools, allowing them to ask questions without the social risk of doing so in front of peers. While some might lament these uses as a degradation of student knowledge, many instructors who are already using GenAI in their courses see it as something that deprecates certain previously required competencies and enables new ones. Whether or not that's true, most instructors shared the belief that competencies are currently in flux and that computing education is rapidly shifting to meet the occasion.


\section{Future Work}
\label{sec:futurework}
Building upon the presented results, there are several pathways for future work. For example, we could replicate the systematic literature review in other domains beyond computing education and compare the CER community's perspective to those of other domains, e.\,g., teacher education, other engineering disciplines, or even less technical domains. The same applies to the survey and interviews with educators. It would be interesting to identify differences and overlaps in how educators across domains integrate and perceive GenAI tools in their teaching practices. 

The systematic literature review shows that research has focused on developing GenAI tools to help computer science students with tasks like writing code, understanding code, and receiving feedback and hints for programming problems. These tools are a natural extension of existing GenAI tools like ChatGPT, which excel at solving problems in introductory computer science courses but struggle with guiding students to find answers on their own without giving away the solution entirely. Looking forward, we should consider how GenAI could address other challenges in learning to program. One possible area is debugging. Expert tutors often guide students to use debugging techniques like print statements, debuggers, or searching Stack Overflow. A key question is how future GenAI tools can effectively incorporate best practices, and how they can actually support learning processes, scaffolding, or, for example, mastery learning. The role of human tutors is also important as GenAI tools become more capable of using teaching methods like the Socratic method without giving away solutions. This raises the question of what effective teaching looks like for instructors and how we should train them to help students in a GenAI-supported environment.

Another crucial aspect of future work is the impact of GenAI tools on the job market, and, for example, the qualifications of graduates. Are there going to be fewer entry-level software engineering jobs? To date, we know from studies that CS students and developers are using GenAI at a great scale, and even educators are increasingly integrating it into their courses. It is thus important to regularly investigate and align these recent developments.

\begin{figure}[h!]
    \centering
    \includegraphics[width=0.99\linewidth]{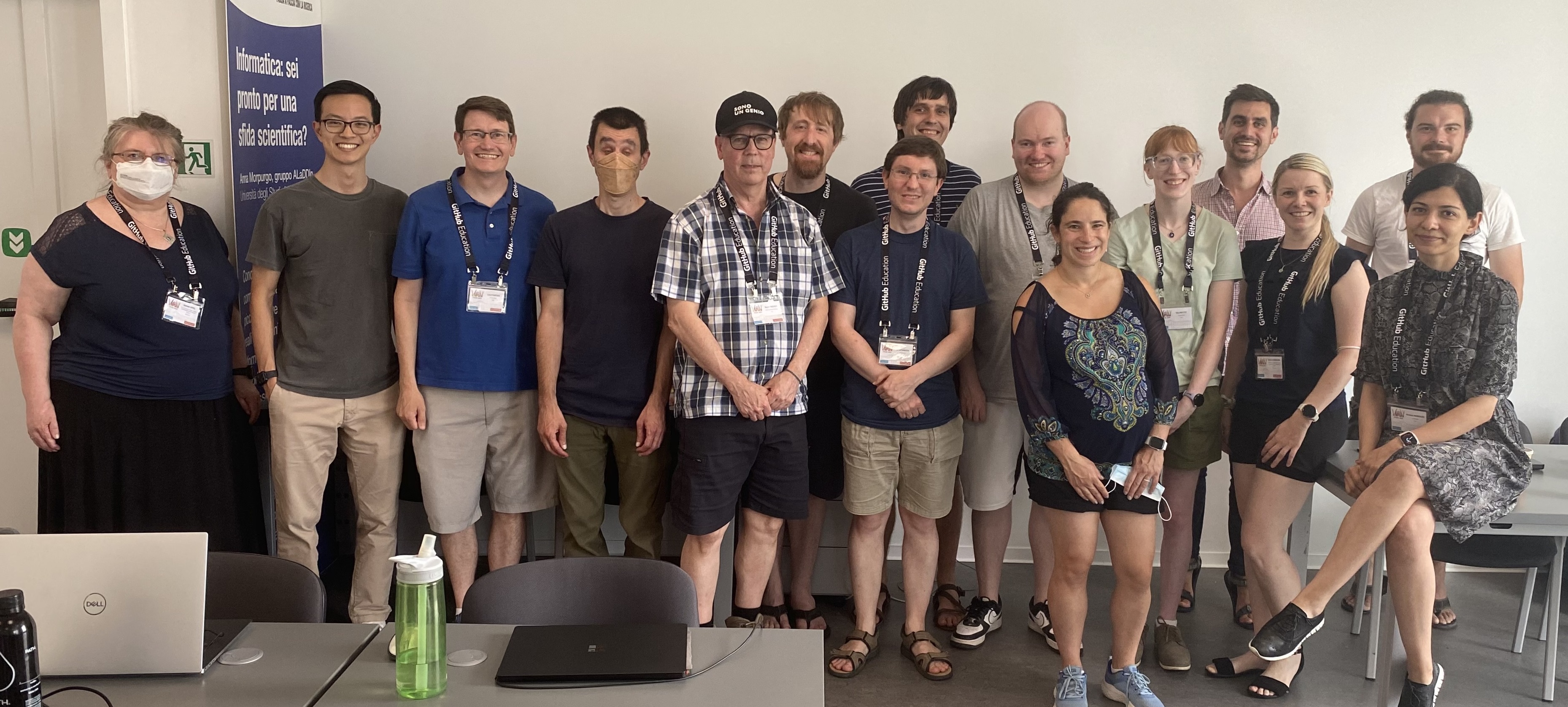}
    \caption{Working Group 09 in Milan, Italy.}
    \label{fig:enter-label}
\end{figure}

\subsection*{Data Availability}
The data of the systematic literature review is published on OSF: \url{https://osf.io/wpxjb/?view_only=cebf366db7f5423792b39de754972400}.

\begin{acks}
This research was supported by the Research Council of Finland (Academy Research Fellow grant number 356114) and in part by the U.S. National Science Foundation IUSE Award \#2417374.

We also acknowledge the support by the Google Award for Inclusion Research Program. 

Finally, we want to say ``Thank You'' to all of our interview and survey participants. In particular, we acknowledge those educators who participated in our interviews who agreed to be named (in alphabetical order by surname):
\begin{itemize}
    \item Bita Akram
    \item Kevin D. Ashley
    \item Brett Becker
    \item Yan Chen
    \item Paul Denny
    \item Barbara Ericson
    \item Majeed Kazemitabaar
    \item Amy Ko
    \item Lauren Margulieux
    \item Tereza Novotna
    \item Wesley Oliver
    \item Ray Pettit
    \item Brad Sheese
\end{itemize}
We acknowledge and thank the participation of several others who participated in the interviews who requested to remain anonymous.
\end{acks}

\balance
\bibliographystyle{ACM-Reference-Format}
\bibliography{sample-base,litreview-included}

\newpage
\appendix
\section*{Appendix}
\section{Educator Survey Questions}
\label{app:appendix-edusurvey}
\small
\begin{enumerate}
    \item[Q1] Do you explicitly disallow students to use GenAI tools for your computing
courses (within the last 12 months)?
    \begin{itemize}
        \item[\Circle] Yes, I explicitly disallow students to use GenAI tools (enables Q9--11)
        \item[\Circle] No, I do not explicitly disallow students to use GenAI tools (condition on Q20)
    \end{itemize}

    \item[Q2] Are you incorporating GenAI tools (e.\,g., actively integrating it into the curriculum or exercises) into your recent courses (within the last 12 months)? 
    \begin{itemize}
        \item[\Circle] Yes (enables Q12--Q19)
        \item[\Circle] No (condition on Q20)
    \end{itemize}

    \item[Q3] Do you believe the skills to create software have changed after the advent of GenAI tools? 
    \begin{itemize}
        \item[\Circle] Yes (proceed with Q5)
        \item[\Circle] No (proceed with Q4)
    \end{itemize}

    \item[Q4] Please elaborate on your last response why skills have not changed. (open question, depends on Q3=No) 

    \item[Q5] In what ways do you think the skills needed to create software have changed with the introduction of GenAI tools? (open question, depends on Q3=Yes)
    
    \item[Q6] When using GenAI tools to create (parts of) software, which skills become the
most important? (please drag and order your top 3)? (depends on Q3=Yes)
    \begin{itemize}
        \item[\Square] Integration skills
        \item[\Square] Testing
        \item[\Square] Reading code
        \item[\Square] Problem decomposition
        \item[\Square] Problem solving
        \item[\Square] Modifying code
        \item[\Square] Prompt engineering
        \item[\Square] Problem understanding
        \item[\Square] Developing algorithms
        \item[\Square] Debugging
        \item[\Square] Understanding error messages
        \item[\Square] Other, please specify (open question)
    \end{itemize}

    \item[Q7] How often do you believe professional software engineers are using GenAI tools as part of their professional role?
    \begin{itemize}
        \item[\Circle] Never
        \item[\Circle] Rarely
        \item[\Circle] Sometimes
        \item[\Circle] Routinely
        \item[\Circle] Everyday
    \end{itemize}    

    \item[Q8] Please select the tasks or contexts you think industry professionals are using GenAI tools.
    \begin{itemize}
        \item[\Square] Modeling algorithms
        \item[\Square] Getting started with a problem
        \item[\Square] Generating ideas
        \item[\Square] Generating code
        \item[\Square] Modifying code
        \item[\Square] Autocompleting code
        \item[\Square] Debugging
        \item[\Square] Creating documentation/comments
        \item[\Square] Finding resources/documentation/libraries
        \item[\Square] Providing code examples
        \item[\Square] Generate test cases
        \item[\Square] Other, please specify: (open question)
    \end{itemize} 
\vspace{1.5cm} 
    \item[----] Block condition (for Q9--Q11): depends on Q1=Yes
    \begin{enumerate}[leftmargin=.7cm]
    \item[Q9] Why don't you allow for GenAI tools use in your courses?

    \item[Q10] Are you doing anything to prevent GenAI tools' use in your course?
    \begin{itemize}
        \item[\Circle] No
        \item[\Circle] Yes (enables Q11)
    \end{itemize}

    \item[Q11] If yes, what are you doing? (open question, depends on Q10=Yes)
    \end{enumerate}

    \item[----] Block condition (for Q12--Q19): depends on Q2=Yes
    \begin{enumerate}[leftmargin=.7cm]
    \item[Q12] We ask you to think of a recent course (within the last 12 months) that you teach that is most influenced by GenAI tools, and respond to the following questions based on this course. Please enter the course name. (open question)
    
    \item[Q13] Select the size of the recent course that you teach that is most influenced by GenAI tools:
    \begin{itemize}
        \item[\Circle] 1--10
        \item[\Circle] 11--25
        \item[\Circle] 26--50
        \item[\Circle] 50--100
        \item[\Circle] 101--250
        \item[\Circle] 250+
    \end{itemize}  

    \item[Q14] Who uses (or is expected to use) GenAI tools in your course(s)?
    \begin{itemize}
        \item[\Circle] Only instructor/instructional staff (proceed with Q16)
        \item[\Circle] Students (proceed with Q15)
        \item[\Circle] Both (proceed with Q15)
    \end{itemize}

    \item[Q15] If students are allowed to use GenAI tools, how do you expect students to access them (depends on Q14=Students or Q14=Both)
    \begin{itemize}
        \item[\Square] N/A
        \item[\Square] Publicly free version
        \item[\Square] Paid version (free for students)
        \item[\Square] Paid version (students are expected to pay)
        \item[\Square] Paid version (paid by institution)
        \item[\Square] Custom tool available to everyone 
    \end{itemize} 
    
    \item[Q16] Which type of GenAI tools are you incorporating into your recent course that is most influenced by GenAI tools? 
    \begin{itemize}
        \item[\Square] Standard industry tools, e.g., ChatGPT, Copilot
        \item[\Square] Customized tools created by others
        \item[\Square] Customized tool created by myself
    \end{itemize} 
    
    \item[Q17] In what ways have you incorporated GenAI tools into your recent course that is most influenced by GenAI tools?
    \begin{itemize}
        \item[\Square] To support grading
        \item[\Square] To automatically provide feedback to students using a custom tool
        \item[\Square] To support the correction of student work
        \item[\Square] To teach students about using GenAI tools
        \item[\Square] As educational content generator for teaching material
        \item[\Square] To validate the quality of assignments
        \item[\Square] Other, please specify: (open question)
    \end{itemize} 

    \item[Q18] Why have you incorporated GenAI tools into your recent course? (open question)

    \item[Q19] Have you changed any of the learning objectives of your recent course based on the capabilities of GenAI tools?
    \begin{itemize}
        \item[\Circle] Yes 
        \item[\Circle] No
    \end{itemize}
    \end{enumerate}


    \item[Q20] Why have you not incorporated GenAI tools (e.g., actively integrating it into the curriculum or exercises) into your recent courses (within the last 12 months)? (open question, depends on Q1=No and Q2=No)


    \item[Q21] Please describe any changes you have made to your teaching approaches in courses you are teaching as a result of GenAI tools. (open question)

    \item[Q22] Please describe any changes you have made to your assessment approaches in courses you are teaching as a result of GenAI tools. (open question)
\end{enumerate}

\par 
\noindent
\textbf{Demographic Questions for all participants:}
\begin{enumerate}
    \item[Q23] In which country are you employed? (select from dropdown list with all countries of the world)

    \item[Q24] How would you characterize your institution? 
    \begin{itemize}
        \item[\Circle] Primary
        \item[\Circle] Secondary
        \item[\Circle] 2-year college (Associates)
        \item[\Circle] Vocational School
        \item[\Circle] College (bachelor's degree granting)
        \item[\Circle] University (graduate degree granting)
        \item[\Circle] Other, please specify: (open question)
    \end{itemize} 

    \item[Q25] Please provide the name of the institution you are currently teaching. (open question)

    \item[Q26] Do you teach at an institution that serves a minority population in your country?
    \begin{itemize}
        \item[\Circle] Yes
        \item[\Circle] No
        \item[\Circle] Unsure/doesn't apply in my country
    \end{itemize}

    \item[Q27] What course/area do you primarily teach or identify with?
    \begin{itemize}
        \item[\Square] CS 1 -- Introduction to Programming
        \item[\Square] Software Engineering
        \item[\Square] Artificial Intelligence (ML/Intelligent Systems)
        \item[\Square] Human-Computer Interaction
        \item[\Square] Networking and Communications
        \item[\Square] Architecture and Organization
        \item[\Square] CS 2 -- Introduction to Data Structures
        \item[\Square] Information Assurance and Security
        \item[\Square] Graphics and Visualization
        \item[\Square] Information Management
        \item[\Square] Software Development Fundamentals
        \item[\Square] Parallel and Distributed Computing
        \item[\Square] Platform-based Development
        \item[\Square] Operating Systems
        \item[\Square] Computational Science
        \item[\Square] Teacher Preparation (for teaching CS to ages 5--18)
        \item[\Square] Programming Languages
        \item[\Square] Systems Fundamentals
        \item[\Square] Discrete Structures
        \item[\Square] Social Issues and Professional Practice
        \item[\Square] Robotics
        \item[\Square] Algorithms and Complexity
        \item[\Square] Other
    \end{itemize}

    \item[Q28] How many years have you been teaching for? (open question)

    \item[Q29] What is the gender you identify yourself with?
    \begin{itemize}
        \item[\Circle] Female
        \item[\Circle] Male
        \item[\Circle] Non-binary or gender diverse
        \item[\Circle] Prefer not to disclose
        \item[\Circle] Prefer to self-describe (open question)
    \end{itemize}    

    \item[Q30] Would you be willing to be interviewed in more detail?
    \begin{itemize}
        \item[\Circle] No
        \item[\Circle] Yes - please enter your email (open question)
    \end{itemize}   

\end{enumerate}
\section{Developer Survey Questions}
\label{app:appendix-devsurvey}
\small
\begin{enumerate}
    \item[Q1] Do you use GenAI tools in your professional role developing software? 
    \begin{itemize}
        \item[\Circle] Yes (proceed with Q2)
        \item[\Circle] No (proceed with Q14)
    \end{itemize}

    \item[Q2] How often do you use GenAI tools as part of your professional role on average?
    \begin{itemize}
        \item[\Circle] Several times a day
        \item[\Circle] Once a day
        \item[\Circle] Several times per week
        \item[\Circle] Once a week
        \item[\Circle] Once a month
        \item[\Circle] Once a year
    \end{itemize}

    \item[Q3] What types of GenAI tools do you use? 
    \begin{itemize}
        \item[\Square] Chatbot (e.g., Chatgpt, Gemini) 
        \item[\Square] Autocomplete code (e.g., Copilot)
        \item[\Square] Other, please specify: (open question)
    \end{itemize}  
    
    \item[Q4] Can you please describe how you use GenAI AI tools in your professional work as a software developer? (e.g., give 2-3 examples of situations in which you apply them) (open question) 
    
    \item[Q5] Please select the tasks or contexts for which you generally use GenAI tools.
    \begin{itemize}
        \item[\Square] Modeling algorithms
        \item[\Square] Getting started with a problem
        \item[\Square] Generating ideas
        \item[\Square] Generating code
        \item[\Square] Modifying code
        \item[\Square] Autocompleting code
        \item[\Square] Debugging
        \item[\Square] Creating documentation/comments
        \item[\Square] Finding resources/documentation/libraries
        \item[\Square] Providing code examples
        \item[\Square] Generate test cases
        \item[\Square] Other, please specify: (open question)
    \end{itemize}
    
    \item[Q6] How not useful or useful do you feel GenAI tools have been to your software development? 
    \begin{itemize}
        \item[\Circle] Not useful
        \item[\Circle] A little useful
        \item[\Circle] Moderately useful
        \item[\Circle] Quite useful
        \item[\Circle] Very useful
    \end{itemize}

    \item[Q7] Have GenAI tools made your software development more or less efficient?
    \begin{itemize}
        \item[\Circle] Much less efficient
        \item[\Circle] Less efficient
        \item[\Circle] No change
        \item[\Circle] More efficient
        \item[\Circle] Much more efficient
    \end{itemize}    
    
    \item[Q8] How not harmful or harmful do you feel GenAI tools have been to your software development? 
    \begin{itemize}
        \item[\Circle] Not harmful
        \item[\Circle] A little harmful
        \item[\Circle] Moderately harmful
        \item[\Circle] Quite harmful
        \item[\Circle] Very harmful
    \end{itemize}  

    \item[Q9] If you consider GenAI tools harmful, please describe a respective situation you have experienced, e.g., what were you doing, what did you expect, why was the use of the GenAI tools harmful and to whom? (open question)
    
    \item[Q10] Did the competencies (i.e., knowledge, skills, dispositions in context of a task) required to professionally develop software change with the availability of GenAI tools?
    \begin{itemize}
        \item[\Circle] No change
        \item[\Circle] Slight change
        \item[\Circle] Moderate change
        \item[\Circle] Extreme change
    \end{itemize}  

    \item[Q11] If you have seen changes, from your experience with GenAI tools, what do you believe are new relevant competencies to professionally develop software with GenAI tools? (open question)

    \item[Q12] If you have seen changes, from your experience with GenAI tools, what do you believe are competencies that are no longer or less relevant to professionally develop software with GenAI tools?

    \item[Q13] What advice would you give to novice programmers regarding the use of GenAI tools? (open question)

    \item[Q14] What is the reason that you do not use GenAI tools for professional software development?
    \begin{itemize}
        \item[\Square] My company does not let me
        \item[\Square] I do not believe they will help me code better
        \item[\Square] I am concerned about ethics issues (e.g., privacy)
        \item[\Square] Other, please specify: (open question)
    \end{itemize}

    \item[Q15] Please feel free to elaborate on your reasoning. (open question)

    \item[Q16] What advice would you give to novice programmers regarding the use of GenAI tools? (open question)
\end{enumerate} 

\par 
\noindent
\textbf{Demographic Questions for all participants:}
\begin{enumerate}
    \item[Q17] In which country are you employed? (select from dropdown list with all countries of the world)

    \item[Q18] What is your job title? 
    \begin{itemize}
        \item[\Circle] Software developer
        \item[\Circle] Product Manager
        \item[\Circle] Research Engineer
        \item[\Circle] Other, please specify: (open question)
    \end{itemize} 
     
    \item[Q19] What is the type of company where you are employed? 
    \begin{itemize}
        \item[\Circle] Start-up (5 engineers or less)
        \item[\Circle] Small software company (10 engineers or less)
        \item[\Circle] Medium software company (50 engineers or less)
        \item[\Circle] Large software company (more than 50 engineers)
        \item[\Circle] Non-profit
        \item[\Circle] Non-software focused company
        \item[\Circle] Government
        \item[\Circle] Research Institute
        \item[\Circle] Other, please specify: (open question)
    \end{itemize} 

\end{enumerate}

\section{Tool Creators Interview Questions}
\label{app:appendix-tool-interview}
\small
\begin{enumerate}
\item[Q1a] Please give us a description of the GenAI tool you created.
\item[Q1b] Who is using it?
\item[Q1c] What did you learn throughout the process of deployment, and what modifications did you make along the way?
\item[Q1d] What are the privacy and security policies required by your institution for the deployment of the tool?
\item[Q2] What outcomes have you seen from usage of the tool?
\item[Q3] What data are you collecting from the tool?
\item[Q4] What learning objectives are you hoping to reinforce through your tool?
\item[Q5] What expertise is required from educators to be able to deploy and use your tools in their courses?
\item[Q6] Please share your future development plans.
\item[Q7a] How do you think AI should be used in CS education to improve teaching and learning?
\item[Q7b]  What tools need to be created?
\item[Q8] Is there anything else you want to discuss about GenAI in your courses?

\end{enumerate}
\section{Educators Studying GenAI Interview Questions}
\label{app:appendix-studying-interview}
\small
\begin{enumerate}
\item[Q1] Describe your understanding of categories of GenAI research in CS education and its landscape.
\item[Q2] What do you hope to learn from studying GenAI in CS education?
\item[Q3] What AI technologies are currently used by educators for what? 
\item[Q4] Have you encountered methods to prevent the negatively perceived aspects of GenAI? If so, please elaborate.
\item[Q5] Have you seen inequities in utilizing GenAI in CS education?
\item[Q6] Have you seen skills/competencies change ever since LLMs emerged? (positively or negatively) And if so, how?
\item[Q7] Where do you think GenAI in CS education is going to take us next?
\end{enumerate}
\section{Educators using GenAI Interview Questions}
\label{app:appendix-using-interview}
\small
\begin{enumerate}
\item[Q1] What course(s) are you using GenAI in?
\item[Q2a] How are you using GenAI tools in your courses? What's the rationale/why?
\item[Q2b] Are AI tools included in the syllabus (or planned for the next one)?
\item[Q2c] Are you transparent with students about the use of GenAI in the course?
\item[Q3] Have you changed the learning objectives of course(s)? How?
\item[Q4a] What were the outcomes (positive/negative) of GenAI use in your courses so far?
\item[Q4b] Have you seen any equity issues due to GenAI? If so, can you provide an example? Are you generally concerned about these issues?
\item[Q5] Are you planning on expanding the use of GenAI in your course(s)? What about others at your institution?
\item[Q6] Is there anything else you want to discuss about GenAI in your courses?
\item[Q7] Is there a particular memorable moment in your course regarding GenAI?
\end{enumerate}
\section{Extraction Survey Questions}
\label{app:appendix-extraction}
\small
\textbf{Include paper?}
    \begin{itemize}
        \item Description: This paper has been originally classified for inclusion. However, now that you have read the paper in detail if you think this is incorrect (and the paper does not meet the inclusion criteria) then there is no need to extract the data.
        \item Options:
        \begin{itemize}
            \item Yes
            \item No, too short
            \item No, K-12
            \item Not genAI
            \item Not computing education
            \item No intervention
            \item No empirical human evidence
        \end{itemize}
    \end{itemize}

\noindent
\textbf{Bibtex entry (note, use authorYEARword format)}
    \begin{itemize}
        \item Description: Please provide a complete bibtex entry -- if possible, please use the entry from the ACM DL as this is typically very complete.
    \end{itemize}
\noindent
\textbf{Data source}
    \begin{itemize}
        \item Description: A ``Supervised'' study is a study that is conducted in a research lab (i.e. a highly controlled environment), while an ``Unsupervised'' study is a study that is conducted in a less restricted environment (such as online or where participants are not supervised).
        \item Options:
        \begin{itemize}
            \item Supervised study (lab study, observations, etc.)
            \item Unsupervised study (no human overseeing use)
            \item Other
        \end{itemize}
    \end{itemize}
\noindent
\textbf{Author affiliation (type)}
    \begin{itemize}
        \item Options:
        \begin{itemize}
            \item Academic
            \item Industry
        \end{itemize}
    \end{itemize}
\noindent
\textbf{Author affiliation (country)}
    \begin{itemize}
        \item Description: Comma separated list of institutions' countries
    \end{itemize}
\noindent
\textbf{Human participants (country)}
    \begin{itemize}
        \item Description: If data is collected from human participants, provide a comma separated list of countries of where participants were located (if not explicitly mentioned, put ``unclear'')
    \end{itemize}
\noindent
\textbf{Human participants (level)}
    \begin{itemize}
        \item Options:
        \begin{itemize}
            \item Tertiary education (e.g. college, university)
            \item Informal education (e.g. MOOCs)
            \item K-12 (in addition to tertiary)
            \item Other
        \end{itemize}
    \end{itemize}
\noindent
\textbf{Number of human participants from whom data was collected (if available, type into other)}
    \begin{itemize}
        \item Options:
        \begin{itemize}
            \item Unclear
            \item Other
        \end{itemize}
    \end{itemize}
\noindent
\textbf{Description of participants}
    \begin{itemize}
        \item Description: A copy-paste (or paraphrased) description of participants from whom data is collected which may be useful to a more detailed thematic analysis. This information can often be found at the beginning of a Methods section.
    \end{itemize}
\noindent
\textbf{How do the authors motivate the work?}
    \begin{itemize}
        \item Description: A copy-paste (or paraphrased) description of the motivation for the work as expressed by the authors.
    \end{itemize}
\noindent
\textbf{What LLM / tool is used?}
    \begin{itemize}
        \item Options:
        \begin{itemize}
            \item GPT-3
            \item GPT-3.5
            \item GPT-4
            \item GPT-4o
            \item Codex
            \item GitHub Copilot
            \item Gemini
            \item Claude
            \item Multimodal model
            \item Open-source model
            \item Unclear
            \item Other
        \end{itemize}
    \end{itemize}
\noindent
\textbf{What are the explicit research questions / research goals / hypotheses in the article?}
    \begin{itemize}
        \item Description: A copy-paste (or paraphrased) description of the RQs, goals, hypotheses
    \end{itemize}
\noindent
\textbf{What programming languages are involved in the study?}
    \begin{itemize}
        \item Options:
        \begin{itemize}
            \item Java
            \item Python
            \item C
            \item C++
            \item Not programming language focused
            \item Other
        \end{itemize}
    \end{itemize}
\noindent
\textbf{How does the article evaluate the data collected?}
    \begin{itemize}
        \item Options:
        \begin{itemize}
            \item Qualitatively
            \item Quantitatively
        \end{itemize}
    \end{itemize}
\noindent
\textbf{Quality assessment}
    \begin{itemize}
        \item Description: An assessment of the research ``quality''. For the last question, on threats to validity / limitations: code as ``Yes'' if there is an explicit (sub)section, ``Vague'' if they are mentioned as part of some other section (i.e. Discussion, Conclusions), or ``No'' if they are not mentioned.
        \item Sub-questions:
        \begin{itemize}
            \item Is there a clearly defined research question/hypothesis?
            \begin{itemize}
                \item Yes
                \item No
                \item Vague / Unclear
            \end{itemize}
            \item Is the research process clearly described?
            \begin{itemize}
                \item Yes
                \item No
                \item Vague / Unclear
            \end{itemize}
            \item Are the results presented with sufficient detail?
            \begin{itemize}
                \item Yes
                \item No
                \item Vague / Unclear
            \end{itemize}
            \item Are threats to validity / limitations addressed in an explicit (sub)section?
            \begin{itemize}
                \item Yes
                \item No
                \item Vague / Unclear
            \end{itemize}
        \end{itemize}
    \end{itemize}
\noindent
\textbf{What is the contribution / what are the key results of the article?}
    \begin{itemize}
        \item Description: Provide a short summary of the main findings.
    \end{itemize}
\noindent
\textbf{How Instructors Incorporate Generative AI into Teaching Computing?}
    \begin{itemize}
        \item Description: Provide information on how generative AI was incorporated into teaching in the article (if applicable).
    \end{itemize}
\noindent
\textbf{And why do they incorporate GenAI tools that way?}
    \begin{itemize}
        \item Description: Provide information on why generative AI was incorporated into teaching in the article (if applicable).
    \end{itemize}
\noindent
\textbf{How have the expectations towards skills in software development changed with the use of Generative AI?}
    \begin{itemize}
        \item Description: Provide any potential changes to expectations towards skills in software development that could result from this work or were discussed in this work. When answering this question, please feel free to provide some of your own commentary -- it is fine to mention changes which the paper prompted you to think about, even if they aren't explicitly mentioned by the paper authors.
    \end{itemize}
\noindent
\textbf{Which computing competencies are required in the future?}
    \begin{itemize}
        \item Description: Provide anything interesting for computing competencies of the future that could result from this work or were discussed in this work. When answering this question, please feel free to provide some of your own commentary -- it is fine to mention changes which the paper prompted you to think about, even if they aren't explicitly mentioned by the paper authors.
    \end{itemize}
\noindent
\textbf{Limitations of the study}
    \begin{itemize}
        \item Description: Add any important limitations of the study (potentially mentioned by the authors, or just noticed by you).
    \end{itemize}
\noindent
\textbf{Additional notes}
    \begin{itemize}
        \item Description: Can be used to note any interesting aspects of the paper or anything else relevant that isn't captured in the extraction fields above.
    \end{itemize}

\end{document}